\documentclass[12pt]{article}

\ifx\pdfoutput\undefined
\usepackage[dvips,bookmarks]{hyperref}
\else
\usepackage{hyperref}
\fi
\hypersetup{colorlinks=false,bookmarksopen,bookmarksnumbered,citecolor=blue,
   pdfstartview=FitH}

\usepackage{latexsym}
\usepackage{amssymb,amsfonts,amsmath}
\usepackage{graphicx} 
\usepackage{indentfirst}
\usepackage{bbm}
\usepackage{amssymb}
\usepackage{verbatim}
\usepackage{amsmath, amsthm,amssymb}
\usepackage{mathrsfs}
\usepackage{hyperref}
\usepackage{amsfonts}
\usepackage{dsfont}
\usepackage{slashed, tensor}
\usepackage{booktabs}
\usepackage{graphicx}
\usepackage{mathrsfs}

\parskip=\medskipamount

\newcommand {\cA}{{\cal A}}
\newcommand {\cB}{{\cal B}}
\newcommand {\cC}{{\cal C}}
\newcommand {\cD}{{\cal D}}
\newcommand {\cE}{{\cal E}}
\newcommand {\cF}{{\cal F}}
\newcommand {\cG}{{\cal G}}

\newcommand {\cK}{{\cal K}}
\newcommand {\cL}{{\cal L}}
\newcommand {\cM}{{\cal M}}
\newcommand {\cN}{{\cal N}}
\newcommand {\cO}{{\cal O}}

\newcommand {\cR}{{\cal R}}
\newcommand {\cS}{{\cal S}}
\newcommand {\cT}{{\cal T}}

\newcommand {\cV}{{\cal V}}
\newcommand {\cW}{{\cal W}}

\def\a{\alpha}
\def\b{\beta}
\def\c{\chi}
\def\d{\delta}
\def\e{\epsilon}
\def\f{\phi}
\def\g{\gamma}
\def\G{\Gamma}

\def\j{\psi}
\def\k{\kappa}
\def\l{\lambda}
\def\m{\mu}

\def\o{\omega}
\def\p{\pi}
\def\q{\theta}
\def\r{\rho}
\def\s{\sigma}
\def\t{\tau}

\def\x{\xi}
\def\z{\zeta}
\def\D{\Delta}
\def\F{\Phi}
\def\J{\Psi}
\def\L{\Lambda}
\def\O{\Omega}

\def\S{\Sigma}
\def\U{\Upsilon}

\def\tr{{\rm tr}}

\def\ri{{\rm i}}
\def\re{{\rm e}}

\newcommand{\ve}{\varepsilon}                            
\newcommand{\cDB}{{\bar\cD}}                            

\newcommand{\ab}{{\a\b}}

\newcommand{\pa}{\partial}                           
\newcommand{\hf}{\frac12}

\newcommand{\vf}{\varphi}

\newcommand{\be}{\begin{equation}}
\newcommand{\ee}{\end{equation}}
\newcommand{\bea}{\begin{eqnarray}}
\newcommand{\eea}{\end{eqnarray}}
\newcommand{\non}{\nonumber}
\newcommand{\ba}{\begin{array}}
\newcommand{\ea}{\end{array}}

\newcommand{\bm}[1]{\mbox{\boldmath$#1$}}

\def\double #1{#1{\hbox{\kern-2pt $#1$}}}

\renewcommand{\Bar}{\overline}

\newcommand{\de}{{\nabla}}
\newcommand{\deb}{{\bar{\nabla}}}

\newcommand{\bbD}{{\mathbb {D}}}

\newcommand{\bsubeq}{\begin{subequations}}
\newcommand{\esubeq}{\end{subequations}}

\newcommand{\eps}{{\ve}}

\newcommand{\rd}{\mathrm d}
\numberwithin{equation}{section}

\newcommand{\RM}{R(M)}
\newcommand{\RD}{R(\mathbb D)}
\newcommand{\RN}{R(N)}
\newcommand{\RJ}{R(J)}
\newcommand{\RS}{R(S)}
\newcommand{\RK}{R(K)}

\begin{document}

\begin{titlepage}
\begin{flushright}
June, 2015 \\
\end{flushright}
\vspace{5mm}

\begin{center}
{\Large \bf 
Higher derivative couplings and  massive supergravity
in three dimensions
}
\\ 
\end{center}

\begin{center}

{\bf
Sergei M. Kuzenko${}^{a}$, Joseph Novak${}^{a}$ and\\
Gabriele Tartaglino-Mazzucchelli${}^{ab}$
} \\
\vspace{5mm}

\footnotesize{
${}^{a}${\it School of Physics M013, The University of Western Australia\\
35 Stirling Highway, Crawley W.A. 6009, Australia}}  
~\\
\vspace{2mm}
\footnotesize{
${}^{b}${\it Instituut voor Theoretische Fysica, KU Leuven,\\
Celestijnenlaan 200D, B-3001 Leuven, Belgium}
}
\vspace{2mm}
~\\
\texttt{joseph.novak,\,gabriele.tartaglino-mazzucchelli@uwa.edu.au}\\
\vspace{2mm}

\end{center}

\begin{abstract}
\baselineskip=14pt

We develop geometric superspace settings to construct 
arbitrary higher derivative 
couplings (including $R^n$ terms)
in three-dimensional supergravity theories with $\cN\leq 3$ 
by realising them as conformal supergravity coupled to certain compensators.
For all known off-shell supergravity formulations, we construct supersymmetric 
invariants with up to and including four derivatives. As a warming-up exercise, 
we first give a new and completely geometric derivation of such invariants in $\cN=1$ supergravity. Upon reduction to components, they agree with those given in 
arXiv:0907.4658 and arXiv:1005.3952. 
We then carry out a similar construction 
in the case of $\cN=2$ supergravity for which there exist two minimal 
formulations that differ by the choice of compensating multiplet: 
(i) a chiral scalar multipet; (ii) a vector multiplet. For these 
formulations all four derivative invariants are constructed in 
completely general and gauge independent form.
For a general supergravity model  (in the $\cN=1$ and minimal $\cN=2$ cases)
with curvature-squared and lower order terms,  we derive the superfield equations 
of motion, linearise them about maximally  supersymmetric backgrounds and 
obtain restrictions on the parameters that lead to models for massive supergravity. 
We use the non-minimal formulation for $\cN = 2$ supergravity 
(which corresponds to a complex linear compensator) to construct 
a novel consistent theory of massive  supergravity. 
In the case of $\cN = 3$ supergravity, we employ the off-shell formulation with 
a vector multiplet as compensator to construct for the first time various 
higher derivative invariants.  These invariants may 
be used to derive models for $\cN = 3$ massive supergravity. 
As a bi-product of our analysis, we also present superfield equations 
for massive higher spin multiplets in (1,0),  (1,1) and (2,0) anti-de Sitter superspaces.

\end{abstract}

\vfill

\vfill
\end{titlepage}

\newpage
\renewcommand{\thefootnote}{\arabic{footnote}}
\setcounter{footnote}{0}

\tableofcontents



\allowdisplaybreaks

\section{Introduction}

Higher-derivative gravity has attracted attention,
on and off,
for over half a century. Interest in such theories 
was spurred on in the early 1960s when it was noticed \cite{UDW,DeWitt}
that  the renormalization of divergences in quantum field theories in curved spacetime
requires higher-derivative counterterms containing the curvature tensor squared.
A decade later it was established \cite{Stelle} that adding
 the higher-derivative structures
$R^{ab}R_{ab}$ and $R^2$ to the Einstein-Hilbert (EH) Lagrangian 
leads to a renormalizable theory in four spacetime dimensions, 
the price for renormalizability being unphysical ghost modes in the theory.
Furthermore, 
an important development took place in 1980 when Starobinsky
proposed his (nowadays famous) model of inflation \cite{Starobinsky}
obtained by complementing the  EH Lagrangian
with a term proportional to the scalar curvature squared. 

In three dimensions (3D),  consistent models for massive gravity can be constructed
by making use of certain higher-derivative extensions of the EH action. 
One such extension was proposed more than thirty years ago
\cite{DJT} and is known as topologically massive gravity (TMG). 
This model is obtained by adding a Lorentzian  Chern-Simons term 
(which is cubic in derivatives of the gravitational field) to the EH action. 
The resulting theory does not preserve parity, 
is ghost-free and propagates a single 
massive state of helicity $\pm2$, where the sign depends on that of the 
Lorentz Chern-Simons term. More recently, 
a parity-preserving model for 3D massive gravity has been proposed   
\cite{BHT1} (see also \cite{BHT2}). It is obtained by combining the ``wrong sign'' EH Lagrangian 
with  a fourth-order term 
$m^{-2}(R^{ab}R_{ab} -\frac{3}{8} R^2)$, which introduces a mass parameter $m$.
The resulting theory, dubbed ``new massive gravity''  (NMG), proves 
to be unitary \cite{Nakasone:2009bn,Deser:2009hb, Ohta12} 
 (unlike its 4D predecessor \cite{Stelle}) and 
it propagates two massive states of both helicities $\pm 2$ in a Minkowski vacuum.\footnote{It has 
been claimed that NMG is power-counting renormalizable \cite{Oda:2009ys}. However, this statement is 
incorrect as shown in \cite{MuneyukiOhta12}.}
Further generalisations of NMG are also possible. First of all, 
one may consider a hybrid parity-violating model which  
interpolates between TMG and NMG \cite{BHT1} and is known as ``general massive gravity'' (GMS). Its specific feature is that the $\pm 2$ helicity states 
have different masses $m_\pm $. Furthermore, adding a cosmological term 
(in the spirit of cosmological TMG \cite{Deser,LSS,CDWW}) leads to cosmological 
GMG \cite{BHT1}. 
It turns out that all of these 3D models for massive gravity admit 
supersymmetric extensions.

Topologically massive supergravity (TMSG) with $\cN=1$  was introduced in \cite{DK}
and its cosmological extension followed in \cite{Deser}.
The off-shell versions of cosmological TMSG  theories were presented in \cite{KLRST-M}
for $\cN=2$ and in  \cite{KN14} for $\cN=3$ and $\cN=4$.  
The off-shell $\cN=1$ supergravity extensions of the models for massive gravity 
proposed in \cite{BHT1} were given in \cite{Andringa:2009yc} 
(see also \cite{BHRST10}), while the $\cN=2$ 
case was studied in a recent paper \cite{BOS14}. 

The constructions in \cite{Andringa:2009yc,BHRST10} and \cite{BOS14}
made use of component techniques.\footnote{It should be mentioned that 
the superspace formalism to derive all the $\cN=1$ invariants
given in \cite{Andringa:2009yc,BHRST10} has been available since 1979 
\cite{BG,Siegel,GGRS}.  However, the questions posed and 
answered in \cite{Andringa:2009yc,BHRST10} 
had not been asked by the authors of \cite{BG,Siegel,GGRS}. 
In principle, the off-shell formulation for 3D $\cN=1$ supergravity 
proposed in 1978 \cite{HT} is also perfectly suitable  for 
the explicit construction of the invariants 
given in \cite{Andringa:2009yc,BHRST10}.}   
Such techniques are quite adept for deriving supergravity-matter systems with 
at most two derivatives. However, they can become rather 
involved 
when it comes to constructing higher-derivative couplings such as 
supersymmetric extensions of the curvature squared terms.\footnote{The 
supersymmetric extensions of $R^3$ terms in 3D $\cN=1$ supergravity 
were constructed in \cite{Bergshoeff:2014ida}.} 
For instance, the Ricci squared invariant in $\cN = 2$ 
supergravity with a chiral compensator was 
only given at the bosonic level in \cite{BOS14}. Moreover, the component 
formalism does not seem to provide a clear approach to 
higher derivative invariants with more than four derivatives. For 
this reason it is worth looking for alternative approaches.

There exist fully-fledged superspace formulations 
for off-shell 3D $\cN$-extended conformal 
supergravity \cite{KLT-M11,BKNT-M1}, of which  \cite{KLT-M11}
is a gauged-fixed version of \cite{BKNT-M1}.
The ${\rm SO}(\cN)$ superspace approach of \cite{KLT-M11} 
has been used to construct general 
off-shell supergravity-matter couplings for $1\leq \cN \leq 4$.  
The conformal superspace of \cite{BKNT-M1} has been applied to provide 
a universal construction of the conformal supergravity actions 
for $1\leq \cN \leq 6$ \cite{BKNT-M2,KNT-M} (for each $\cN$, the conformal supergravity
action is a locally supersymmetric Lorentzian Chern-Simons term required to formulate 
TMSG). Off-shell versions for 3D Poincar\'e and anti-de Sitter (AdS) supergravity 
theories naturally follow by coupling conformal supergravity to conformal 
compensators, see \cite{KT-M11} for the complete description of the $\cN=2$ case.
In this paper we show that all the supergravity invariants 
required for the construction of the massive supergravity models
proposed in \cite{Andringa:2009yc,BHRST10,BOS14}
naturally originate within the superspace approaches of \cite{KLT-M11,BKNT-M1}.
In particular, the construction of four-derivative invariants 
in 3D $\cN=2$ supergravity is analogous to that in 4D $\cN=1$ supergravity 
\cite{Theisen}.
We also construct, for the first time,
curvature squared invariants in $\cN=3$ supergravity. 

Before turning to the technical aspects of this work, we would like to 
make several comments concerning $\cN=2$ supergravity.\footnote{
For early works 
on off-shell 3D $\cN=2$ supergravity, see \cite{RvN,ZP88,NG93,HIPT}.
In on-shell 3D $\cN=2$ supergravity, the matter couplings were studied in 
\cite{dWNT,IT, DKSS}.
} 
There are three off-shell formulations for 3D $\cN=2$ Poincar\'e and
AdS supergravity theories \cite{KLT-M11,KT-M11}:
(i) type I minimal; (ii) type II minimal; and (iii) non-minimal. 
They differ by the structure of the conformal compensators employed. 
Type I minimal supergravity is a 3D analogue of the old minimal 
formulation
for 4D $\cN=1$ supergravity \cite{old} (see \cite{GGRS,Ideas,WB} for reviews).
Type II minimal  supergravity is a 3D analogue of the new minimal 
formulation\footnote{Unlike the new minimal formulation for 4D $\cN=1$ 
supergravity, the type II  minimal formulation is suitable to describe AdS supergravity, 
which is a unique feature of three dimensions.}
for 4D $\cN=1$ supergravity \cite{new} (see \cite{Ideas,GGRS} for reviews).
The non-minimal supergravity theories are analogues of  
 the following 4D $\cN=1$ theories: (i) non-minimal supergravity
 without a cosmological term \cite{non-min,SG};
and (ii)  non-minimal AdS supergravity \cite{BK11dual}.
As shown by Ach\'ucarro and Townsend \cite{AT}, in three dimensions 
$\cN$-extended AdS supergravity exists in several incarnations. 
They were called the  $(p,q)$ AdS supergravity theories  
where the  non-negative integers $p \geq q$ are such that 
$\cN=p+q$.   It was demonstrated in \cite{AT} that these theories  
are naturally  associated with the 3D AdS supergroups
$\rm OSp (p|2; {\mathbb R} ) \times  OSp (q|2; {\mathbb R} )$.
There are two off-shell realisations for (1,1) AdS supergravity \cite{KT-M11}, 
which are the type I theory with a cosmological term and  
the non-minimal AdS theory. 
There is only one off-shell realisation for (2,0) AdS supergravity \cite{KT-M11}, 
which is the type II theory with a cosmological term. Strictly speaking, 
the terminology $(p,q)$ AdS supergravity should be used only for supergravity theories
with a cosmological term. In the literature, however, 
the names (1,1) and (2,0) supergravity theories are also used for the type I and type II 
minimal formulations. 

This paper is organised as follows. 
Section \ref{section2} is devoted to the description of $\cN = 1$ supergravity models.
In sections  \ref{section3}, \ref{section4} and \ref{section5},
$\cN = 2$ supergravity models with a chiral compensator, 
with a real linear compensator and 
a complex linear compensator
are presented, respectively. 
In all the sections \ref{section2}--\ref{section5}
special attention is given to those models that describe massive supergravity. 
In section \ref{section6} we construct new invariants in $\cN = 3$ supergravity with 
a vector multiplet compensator. A discussion of our results and 
concluding comments are given in section \ref{section7}.

We have also included a few technical appendices. In appendix \ref{geometry} we 
summarise the essential details of conformal superspace for $\cN \leq 3$. Appendices
\ref{prepotentialN=1} and \ref{prepotentialN=2} are devoted to 
prepotential deformations for $\cN = 1$ and $\cN = 2$ supergravity.


\section{$\cN = 1$ supergravity models} \label{section2}

The construction of $\cN = 1$ supergravity models in three dimensions
can be performed using the conventional superspace formalism of 
\cite{BG,Siegel,GGRS}.
It makes use of a curved superspace $\cM^{3|2}$ parametrized by real bosonic ($x^m$) 
and real fermionic ($\q^{\mu}$) coordinates
$z^{M}=(x^m,\q^{\mu})$, where  $m=0,1,2$ and $\mu=1,2$.

\subsection{Conventional superspace} \label{subsection2.1}

The superspace geometry is described in terms of covariant derivatives of the form
\bea
\cD_{A}&=& (\cD_a, \cD_\a )= E_{A}-\O_A~.
\eea
Here the vector fields $E_A=E_A{}^M 
\pa/\pa z^M$ define the inverse vielbein,
and 
\bea
\O_A=\hf\O_{A}{}^{bc} M_{bc}=-\O_{A}{}^b M_b=\hf\O_{A}{}^{\b\g} M_{\b\g}
\eea
is the Lorentz connection.
The Lorentz generators with two vector indices 
($M_{ab}= -M_{ba}$), with one vector index ($M_a$)
and with two spinor indices ($M_{\a\b} = M_{\b\a} $) 
are related to each other by the rules:
$M_a=\hf \ve_{abc}M^{bc}$ and $M_{\a\b}=(\g^a)_{\a\b}M_a$.
The Lorentz generators act on the covariant derivatives  as follows:
\bea
&{[}M_{\a\b},\cD_{\g}{]}
=\ve_{\g(\a}\cD_{\b)}~,\qquad
{[}M_{ab},\cD_c{]}=2\eta_{c[a}\cD_{b]}~.
\label{generators}
\eea
In the notation of \cite{KLT-M11}, 
the   covariant derivatives obey the following (anti-)commutation relations:
\bsubeq \label{cdaN=1-con}
\bea
\{ \cD_\a , \cD_\b \} &=& 2 \ri \cD_{\a\b} - 4 \ri \cS M_{\a\b} \ , \\
\left[ \cD_{\a \b} , \cD_\g \right] &=& 
- 2 \eps_{\g(\a} \cS \cD_{\b)} + 2 \eps_{\g(\a} \cC_{\b) \d \r} M^{\d\r}
\non\\
&&
+ \frac{2}{3} \big( \cD_\g \cS M_{\a\b} 
- 4 \cD_{(\a} \cS M_{\b) \g} \big) \ , \\
\left[\cD_a , \cD_b\right] &=& 
-\frac{\ri}{2} \eps_{abc} (\g^c)^{\a\b} \Big\{
\cC_{\a\b\g} \cD^\g
+ \frac{4 }{3} \cD_\a \cS \cD_\b
- \cD_{(\a} \cC_{\b\g\d)} M^{\g\d} \non\\
&&+  \frac{2 }{3} (\cD^2 -6\ri \cS)\cS 
 M_{\a\b}
 \Big\}
 \ ,
\eea
\esubeq
where $\cS$ and $\cC_{\a\b\g}$ are related to each other by  the Bianchi identity
\bea 
\cD^\g \cC_{\a\b\g} = -\frac{4\ri }{3} \cD_{\a\b} \cS \ .
\label{2.444}
\eea

Practically all supergravity actions 
(with the action for conformal supergravity being a notable exception)
may be 
realized as 
invariants of the form\footnote{In $\cN$-extended superspace we 
use the notation $\rd^{3|2\cN} z := \rd^3 x \rd^{2\cN} \q$.
The $\cN = 1$ supergravity measure, $\rd^{3|2} z \, E $,
 is imaginary.}
\be S = \ri \int \rd^{3|2} z \, E \, \frak{L}(\cT , \cD \cT  , \cD^2 \cT  , \cdots ) \ ,
\qquad E^{-1} ={\rm Ber}(E_A{}^M)~,
\ee
where $\cT$ schematically represents the torsion components appearing in the 
covariant derivative algebra \eqref{cdaN=1-con}. 
Various choices for $\frak{L}$ lead to different supergravity models. 
As far as the higher-derivative supergravity invariants are concerned, 
the important observations are: 
(i) the top component of $\cS$ gives a scalar curvature 
contribution; and (ii)  a linear in $\q$ component 
of $\cC_{\a\b\g}$, $\cD_{(\a} \cC_{\b\g\d)}$, contains 
the traceless part of the Ricci curvature tensor. Therefore, choosing $\frak L \propto ( \cD^\a \cS )\cD_\a \cS $
leads to a supersymmetric completion of  the scalar curvature squared, 
while $\frak L \propto  \cC^{\a\b\g} \cC_{\a\b\g} $ produces the Ricci tensor squared
along with some other contributions. 

As is well known, gravity in $d>2$ dimensions can be realized 
as a Weyl invariant dynamical system describing conformal gravity coupled to 
a conformal compensator \cite{Deser70,Zumino}.
It is also well known that similar formulations exist for various supergravity 
theories. Such formulations are useful for certain applications,
including the component reduction of supergravity models.
It is especially suitable when the conformal supergravity action is a sector of 
the complete action of the theory under consideration.
In three dimensions, $\cN=1$ conformal supergravity can be described using the 
above curved superspace setting by requiring an additional gauge symmetry
known as super-Weyl invariance. 
The algebra of covariant derivatives \eqref{cdaN=1-con} is invariant under 
super-Weyl transformations \cite{ZP88,ZP89,LR-brane} 
of the form\footnote{Only infinitesimal 
super-Weyl transformations were given in \cite{ZP88,ZP89,LR-brane}.}
\bsubeq \label{N=1sW}
\bea
\cD'_\a &=&
\re^{\hf \s} (\cD_\a + \cD^{\b}\s M_{\a\b})
~,
\\
\cD'_a
&=&
\re^\s \Big( 
\cD_a + \frac{\ri}{2} (\g_a)^{\a\b} (\cD_\a \s) \cD_\b
+ \eps_{abc} (\cD^b \s) M^c \non\\
&&- \frac{\ri}{8} (\g_a)^{\a\b} (\cD^\g \s) \cD_\g \s M_{\a\b}
\Big)
~,
\eea
with the parameter $\s$ being a real unconstrained superfield. 
The corresponding transformation of the torsion superfields is
\bea
\cS' &=&
 \frac{\ri}{2} \re^{\frac{3}{2} \s} \big(\cD^2 -2\ri \cS\big)\re^{-\hf \s} 
  ~,
\qquad
\cC'_{\a\b\g}= 
- \hf \re^{\hf\s}\big( \cD_{(\a\b} \cD_{\g)} -2\cC_{\a\b\g}\big)\re^\s
~.
\eea
\esubeq
Every supergravity-matter action can be made super-Weyl invariant 
by coupling the fields to a conformal compensator $\vf$, which is 
a nowhere vanishing scalar superfield with the super-Weyl transformation law
\bea
 \vf' =\re^{\hf \s} \vf~.
\eea
Applying a finite super-Weyl transformation allows one to choose
the gauge $\vf =1$, in which the super-Weyl invariant action reduces to the original one. 

The super-Weyl 
invariance \eqref{N=1sW} is intrinsic to conformal supergravity. 
The action for $\cN=1$ conformal supergravity\footnote{The action 
 for $\cN=1$ conformal supergravity was originally constructed in components  
using the  superconformal tensor calculus \cite{vanN85}.}
does not depend on $\vf$ and is given by \cite{KT-M12}
\bea
S_{\rm CSG}
&=& -2 \int \rd^{3|2} z \, E \, \Omega^{\a\b\g} 
\cG_{\a\b\g} 
\non\\
&& + \frac{2}{3} \int \rd^{3|2} z \, E \, \Big\{ \tr (\Omega^\a \Omega^\b \Omega_{\a\b} - 2 \cS \Omega^\a \Omega_\a) 
- \cS \Omega^{\a\b}{}_\a \Omega^\g{}_{\b\g} \Big\} \non\\
&&+32 \ri  \int \rd^{3|2} z \, E \, \cS^2 \ ,
\label{N=1CSA}
\eea
where we have used the matrix notation $\O_A = (\O_{A\, \b}{}^\g )$ 
and introduced the tensor 
\be 
\cG_{\a\b\g} := \cC_{\a \b\g} - \frac{4}{3} \eps_{\a (\b} \cD_{\g)} \cS ~, 
\qquad \cD^\a \cG_{\a \b\g} = 0~.
\label{Gtensor}
\ee
This tensor will 
be used for later considerations.
Modulo an overall coefficient, the structures in the first and second lines of \eqref{N=1CSA} 
are uniquely 
fixed by the condition of invariance under the local Lorentz transformations 
\bea
\d_K \O_A{}^{bc} = K_A{}^D \O_D{}^{bc} - \cD_A K^{bc}~.
\eea
The last term in \eqref{N=1CSA} is uniquely fixed by requiring invariance 
under the super-Weyl transformations. 
Separate sectors of the superfield action \eqref{N=1CSA}
 had appeared long ago \cite{GGRS,ZP88,ZP89}, 
 but the complete action was given only in \cite{KT-M12}.
 

\subsection{The superconformal setting}

Off-shell $\cN$-extended conformal supergravity in three dimensions 
can be realized in superspace \cite{BKNT-M1} 
as a gauge theory of the superconformal group  ${\rm OSp}(\cN|4, {\mathbb R})$.
This formulation, known as conformal superspace, is briefly reviewed 
in appendix \ref{geometry}. It is the most powerful  approach to derive off-shell 
conformal supergravity actions \cite{BKNT-M2,KNT-M}. 
In the $\cN=1$ case, the Weyl invariant formulation
for conformal supergravity sketched above originates from conformal 
superspace by partially fixing certain local symmetries, see 
\cite{BKNT-M1} for the details. 
Therefore, it is quite natural to carry out our subsequent analysis in 
conformal superspace; all results may be recast in the conventional superspace formalism 
by imposing the gauge conditions required. 
One of the advantages of conformal superspace
is that it improves the complexity in performing component reduction.

Within the conformal superspace setting, the conformal compensator 
$\vf$ has to be a primary superfield of dimension 1/2, 
\be 
\mathbb D \varphi = \hf \varphi \ , \qquad K_A \vf =0~.
\ee
We define the component fields of $\varphi$ as follows\footnote{The component fields of the conformal supergravity multiplet
were elaborated in \cite{BKNT-M2}.}
\bea
l := \varphi| \ , \quad \l_\a := \ri \nabla_\a \varphi| \ , \quad S := \frac{\ri}{2} \nabla^2 \varphi| \ ,
\label{2.99}
\eea
where the bar-projection \cite{WZ2, WB, GGRS} 
of a superfield $V(z) = V(x,\theta)$ is defined in the standard way $V| := V(x, \theta)|_{\theta = 0}$. Here we have introduced the 
operator
\be \nabla^2 := \nabla^\a \nabla_\a \ .
\ee

Using $\varphi$ one can deform the covariant derivatives of conformal superspace 
$\nabla_A$ to 
new covariant derivatives $\mathscr D_A$
that are dimensionless and take primary 
superfields to primary ones. 
This procedure is very much like the one adopted 
in \cite{KugoU,BKO89, Butter09, BK11dual, BN12} to construct Weyl invariant covariant derivatives. We define the new covariant derivatives as follows:
\bsubeq
\bea 
\mathscr D_\a &:=& \frac{1}{\varphi} \big( \nabla_\a - 2 \nabla^\b \ln \varphi \ \! M_{\a\b} - 2 \nabla_\a \ln \varphi \ \! \bbD \big) \ , \\ 
\mathscr D_a &:=& \frac{\ri}{4} (\g_a)^{\a\b} \{ \mathscr D_\a , \mathscr D_\b \} - 2 \mathscr S M_{a} \ ,
\eea
\esubeq
where we have introduced the dimension zero primary superfield
\be \mathscr S := \frac{\ri}{2 \varphi^3} \nabla^2 \varphi \ . \label{SN=1}
\ee
Note that $\varphi$ is covariantly constant with respect to $\mathscr D_A$, $\mathscr D_A \varphi = 0$. 
When acting on a primary superfield,
the covariant derivatives $\mathscr D_A$  satisfy the algebra
\bsubeq \label{cdaN=1}
\bea
\{ \mathscr D_\a , \mathscr D_\b \} &=& 2 \ri \mathscr D_{\a\b} - 4 \ri \mathscr S M_{\a\b} \ , \\
\left[ \mathscr D_{\a \b} , \mathscr D_\g \right] &=& 
- 2 \eps_{\g(\a} \mathscr S \mathscr D_{\b)} + 2 \eps_{\g(\a} \mathscr C_{\b) \d \r} M^{\d\r}
\non\\
&&
+ \frac{2}{3} \big( \mathscr D_\g \mathscr S M_{\a\b} 
- 4 \mathscr D_{(\a} \mathscr S M_{\b) \g} \big) \ ,
\eea
\esubeq
where we have introduced
\bea
\mathscr C_{\a\b\g} &:=& - \frac{1}{2 \varphi} \nabla_{(\a\b} \nabla_{\g)} \frac{1}{\varphi^2} \ . \label{CN=1}
\eea

The algebra \eqref{cdaN=1} formally coincides with 
\eqref{cdaN=1-con}. In fact, we can relate the 
superconformal framework presented above to the one 
of conventional 
superspace by gauge fixing the additional symmetries. The 
conformal boosts and $S$-supersymmetry transformations can 
be fixed by imposing the gauge condition
\be B_A = 0 \label{gaugeFixN=1-1} \ ,
\ee
which reduces conformal superspace to conventional superspace 
via the degauging procedure of \cite{BKNT-M1}. The composites \eqref{SN=1} 
and \eqref{CN=1} become
\bsubeq \label{2.188}
\bea
\mathscr S &=& \frac{\ri}{2 \varphi^3} \big(\cD^2-2\ri\cS\big) \varphi  \label{SN=1_CSS} \ , \\
\mathscr C_{\a\b\g} &=& - \frac{1}{2 \varphi}\big( \cD_{(\a\b} \cD_{\g)}-2\cC_{\a\b\g} \big)\frac{1}{\varphi^2} 
 \ . \label{CN=1_CSS}
\eea
\esubeq
One can then use the super-Weyl transformations to impose the 
gauge condition
\be \varphi = 1 \ . \label{gaugeFixN=1-2}
\ee
One can see that in the above gauge the composites \eqref{SN=1} and \eqref{CN=1} 
coincide with the torsion components $\cS$ and $\cC_{\a\b\g}$ of conventional superspace.

We will make use of the composites \eqref{SN=1} and \eqref{CN=1} to construct supergravity invariants 
as superspace integrals. 
Superspace actions have the form
\be S = \ri \int \rd^{3|2} z \, E \, \cL \ , 
\ee
where $\cL = \bar \cL$ is a real primary superfield of dimension 2,
\be 
\bbD \cL = 2 \cL \ , \quad K_A \cL = 0 \ .
\ee
Using our constructions one may consider general actions of the form
\be 
S = \ri \int \rd^{3|2} z \, E \, \varphi^4 
\frak{L}(\mathscr T , \mathscr D \mathscr T  , \mathscr D^2 \mathscr T  , \cdots )
\ ,
\label{2.211}
\ee
where $\frak{L}$ is a dimensionless superfield constructed out of the torsion components $\mathscr T$ of the 
covariant derivatives $\mathscr D_A$.
In particular, one can in principle construct general higher derivatives couplings.
In this section, we will focus our attention on actions containing at most curvature squared terms. 

In order to reduce the superspace actions to 
components we make use of the following 
component reduction formula \cite{BCGLMP04}: 
\bea \label{N=1APrinc}
S &=& \ri \int \rd^{3|2} z \, E \, \cL
~,\non\\
&=&  - \frac{\ri}{4} \int \rd^3x \, e \, \Big{\{}
 \nabla^2  - \ri (\g^a)^{\a\b} \psi_{a \a} \nabla_\b 
- \hf \eps^{abc} (\g_a)_{\a\b} \psi_b{}^\a \psi_c{}^\b 
\Big{\}} \cL| \ ,
\eea
where $e := \det \big(e_m{}^a\big)$. Here the component vielbein $e_m{}^a$, 
its inverse $e_a{}^m$ and the gravitino field $\j_a{}^\b$ are defined by
\bsubeq
\bea e_m{}^a &:=& E_m{}^a| \ , \quad 
e_a{}^m e_m{}^b = \d_a^b \ , \quad e_m{}^a e_a{}^n = \d_m^n \ , \\
\psi_a{}^\b &:=& e_a{}^m \psi_m{}^\b \ , \quad
\psi_m{}^\a := 2 E_m{}^\a| \ .
\eea
\esubeq
In what follows, we will mostly be interested in the bosonic sectors of
locally supersymmetric actions, although by using the previous results it is straightforward to derive 
the full component actions.

Applying the component reduction formula, eq. \eqref{N=1APrinc}, to our supergravity models, one will often find 
the appearance of terms such as $\nabla^a \nabla_a \varphi|$ and $\nabla^a \nabla_a \nabla^b \nabla_b \varphi|$. 
Such terms are of significance because they involve scalar curvature and Ricci curvature squared contributions. In 
general one finds for a  primary scalar superfield $\f$ of dimension $\D$ and lowest component $f:=\f|$ the following 
results:
\bsubeq \label{compBoxResultsN=1}
\bea 
\nabla^a \nabla_a \f| &=& 
\Big({D}^a {D}_a
+\frac{\D}{4}{R}\Big) f
+{\rm fermion \ terms}
 \ , 
\label{2.14a} \\
\nabla^a \nabla_a \nabla^b \nabla_b \f| &=& 
{D}^a
\Big({D}_a{D}^b{D}_b f
+\frac{\D}{4}({D}_a{R}) f
+\frac{\D}{4}{R} {D}_a f \non\\
&&\qquad +(2\D-1)\Big({R}_{ab}{D}^bf
-\frac{1}{4}{R}{D}_af\Big)
\Big)
\non\\
&&
+\big(2\D-1\big){R}^{ab}{D}_{a}{D}_{b}f
-\frac{(\D-3)}{4}{R}{D}^{a}{D}_{a}f
\non\\
&&
+\D\big(2\D-1\big){R}^{ab}{R}_{ab}f
-\frac{\D(9\D-7)}{16}{R}^2f
+{\rm fermion \ terms}
 \ .~~~~~~~~~
 \label{2.14b}
\eea
\esubeq
Here we have introduced the covariant derivative
\be 
{D}_a = e_a{}^m \Big(
\partial_m - \hf \omega_m{}^{bc} M_{bc}
- b_m \mathbb D \Big) \ ,
\ee
where the 
Lorentz connection $\omega_m{}^{bc}$
and dilatation connection $b_m$ are defined as component 
projections of their corresponding superspace connections,
\be
\omega_m{}^{bc} = \Omega_m{}^{bc}| \ , 
\quad b_m = B_m| \ .
\ee
The scalar curvature $R$ is constructed from the Lorentz curvature ${R}_{ab}{}^{cd}$
as follows
\be
{R} = {R}_{ab}{}^{ab} \ ,
\ee
where the Lorentz curvature ${R}_{ab}{}^{cd}$ is given by
\be
{R}_{ab}{}^{cd}
= 2 e_a{}^m e_b{}^n \big(
\partial_{[m} \omega_{n]}{}^{ab}
- 2 \omega_{[m}{}^{cf} \omega_{n]}{}_{f}{}^d \big)\ .
\ee
In the cases we consider we will only need to make use of the results \eqref{compBoxResultsN=1} 
in the gauge where $\phi = 1$.
For a more detailed discussion of the component results and conventions, 
the reader is referred 
to \cite{BKNT-M2}.

It should be mentioned that at the component level the gauge conditions \eqref{gaugeFixN=1-1} and \eqref{gaugeFixN=1-2} 
corresponds to setting
\be l = 1 \ , \quad \l_\a = 0 \ , \quad b_m = 0 \ . \label{compGaugeFixN=1}
\ee
Here the first gauge condition
fixes the dilatations, the second fixes the $S$-supersymmetry 
transformations and the last fixes the conformal boosts.
We also point out that the top component $S$ of $\vf$, eq. \eqref{2.99}, 
does not vanish in the gauge $\vf =1$. These gauge conditions are useful in 
deriving component actions corresponding to supergravity invariants.


\subsection{Supergravity invariants} \label{subsection2.3}

We now turn to describing locally supersymmetric invariants
which contribute to massive supergravity actions.  


\subsubsection{The supergravity action}

The standard $\cN=1$ supergravity action with a cosmological term is given by 
\bea
S =
\frac{1}{\kappa} S_{\rm SG} + \l S_{\rm cos}
~,
\eea
where
\bea
S_{\rm SG} &=& 8 \ri \int \rd^{3|2}z \, E \, \varphi^4 \mathscr S \ , 
\label{2.300}
\eea
is the supersymmetric Einstein-Hilbert action, and 
\bea 
S_{\rm cos} &=& \ri \int \rd^{3|2}z \, E \, \varphi^4
\label{2.311}
\eea
is the supersymmetric cosmological term. 
The functional $S_{\rm SG}$ gives rise to the Einstein-Hilbert term
$-\hf R$ once one reduces to components and imposes the gauge conditions 
\eqref{compGaugeFixN=1}. To see this one applies the 
component reduction formula
\eqref{N=1APrinc} to the action \eqref{2.300}. Keeping in mind the expression for 
$\mathscr S$ in terms of the compensator, eq. \eqref{SN=1}, one finds a term involving the component 
projection of
$\nabla^2 \nabla^2 \varphi = - 4 \nabla^a \nabla_a \varphi$. Finally, making use of \eqref{compBoxResultsN=1} 
in the gauge $\varphi = 1$ recovers the Einstein-Hilbert term. The full component action can be similarly 
computed. Here we are primarily concerned with the curvature dependence of our supergravity invariants.


\subsubsection{The $S^n$ invariants}

Keeping in mind the gauge condition \eqref{compGaugeFixN=1}, 
one can construct an invariant 
which contains an $S^n$ term,  with $n$ a positive integer. 
Such a functional is given in terms of $\mathscr S$ as follows:
\be S_{S^n} = \ri \int \rd^{3|2} z \, E \, \varphi^4 \mathscr S^{n-1} \ . \label{N=1S^naction}
\ee
For $n = 1$ and $n = 2$ one recovers the supergravity cosmological term 
\eqref{2.311} and 
the Poincar\'e supergravity action \eqref{2.300}, respectively. 
Similarly to the supergravity action one can check that 
the action \eqref{N=1S^naction} contains the contribution 
$-\frac{(n-1)}{16} {R} S^{n-2}$ 
for $n \geq 2$. 
The corresponding bosonic component action was given in 
\cite{BHRST10}. For $n = 1$ it coincides with the cosmological term, 
while for $n = 2$ it gives the Einstein-Hilbert term.


\subsubsection{The scalar curvature squared invariant}

A functional containing a scalar curvature term ${R}^2$ may be constructed using
\be 
S_{\rm scalar^2} =  
\int \rd^{3|2} z \, E \, \varphi^4 \, (\mathscr D^\a \mathscr S) \, \mathscr D_\a \mathscr S \ . \label{N=1scalarcurvatureaction}
\ee
Upon integration by parts the above functional may be taken to the equivalent form
\be 
S_{\rm scalar^2} =
- \int \rd^{3|2} z \, E \, \varphi \mathscr S \nabla^2 (\varphi \mathscr S) 
- 2 S_{S^4}
\ .
\ee
One can check that the first term in 
$S_{\rm scalar^2} $
gives rise to a scalar curvature squared term,
$- \frac{1}{64} {R}^2$. At the component level a scalar curvature 
squared action was given in \cite{Andringa:2009yc} using different techniques. 
Our curvature squared action \eqref{N=1scalarcurvatureaction} differs from the one in \cite{Andringa:2009yc} by the 
addition of a multiple of the $S^4$ invariant.


\subsubsection{The Ricci curvature squared invariant}

An invariant containing a Ricci squared term, 
${R}^{ab}{R}_{ab}$, is given by
\be 
S_{\rm Ricci^2} 
= -2 \int \rd^{3|2} z \, E \, \varphi^4 \mathscr C^{\a\b\g} \mathscr C_{\a\b\g} \ .
\ee
One can verify this readily by imposing the gauge conditions \eqref{gaugeFixN=1-1} and \eqref{gaugeFixN=1-2} and working in 
conventional superspace. It is not difficult to see that the 
component action will involve a Riemann curvature squared contribution, which leads to the Ricci squared term.
In particular, in the gauge \eqref{compGaugeFixN=1} one finds the curvature squared contribution to be
\be {R}^{ab}{R}_{ab}
-\frac{1}{3} {R}^2 \ .
\ee
One can then show that the combination
\be \label{N=1pureRicci}
S_{\rm Ricci^2}^{\rm (pure)}
:=
S_{\rm Ricci^2} - \frac{64}{3} S_{\rm scalar^2}
\ee
gives a pure Ricci curvature squared invariant without any scalar curvature squared terms.

Remarkably one can write down an alternative invariant in a compact form that also gives rise to a Ricci squared term
\be 
S_{\S} = \int \rd^{3|2} z \, E \, \frac{\S}{\varphi} \ , 
\quad \S := 
\nabla^a \nabla_a
\nabla^2 \frac{1}{\varphi} \ .
\label{N=1altInvariant}
\ee
The dimension-$\frac{5}{2}$ superfield $\S$ can be shown to be primary. Using eq.~\eqref{compBoxResultsN=1}
one can check  that in the gauge \eqref{compGaugeFixN=1} the invariant \eqref{N=1altInvariant} contains the following
curvature squared contribution:
\bea 
{R}^{ab} {R}_{ab}-\frac{23}{64}{R}^2 \ .
\eea
The relative coefficients between the Ricci squared and scalar curvature squared contributions 
exactly coincides with that of the Ricci squared invariant constructed at the bosonic level in \cite{Andringa:2009yc}. 
The result thus appears to coincide with the Ricci squared invariant in \cite{BHRST10} 
up to the addition of a multiple of the $S^4$ invariant. 
The form of the curvature squared terms tells us that the invariant \eqref{N=1altInvariant} may be 
expressed in terms of a linear combination of the Ricci squared invariant $S_{\rm Ricci^2}$, the scalar 
curvature squared invariant $S_{\rm scalar^2}$ and the $S^4$ invariant $S_{S^4}$.

There is another 
linear combination of 
$ S_{\rm Ricci^2} $ and $ S_{\rm scalar^2} $ 
that is worth mentioning. Here we 
make use of the gauges \eqref{gaugeFixN=1-1} and \eqref{gaugeFixN=1-2}, 
 and define the following invariant in conventional superspace
\bea
S_{\rm Ricci^2}^{(\rm YM)} 
&:=& S_{\rm Ricci^2} - \frac{16}{3} S_{\rm scalar^2} \ .
\eea
The reason for the superscript $(\rm YM)$ will become clear shortly. The form of the action allows 
one to write it entirely in terms of the tensor $\cG_{\a\b\g} = \cG_{\a(\b\g)}$, eq. \eqref{Gtensor}, 
which has the property 
$\cD^\a \cG_{\a \b\g} = 0$.
The action reads
\bea
S_{\rm Ricci^2}^{(\rm YM)} &=& - 2 \int \rd^{3|2} z \, E \, \cG^{\a \b\g} \cG_{\a\b\g }
 = 2 \int \rd^{3|2} z \, E \, {\rm tr} \left\{ \cG^\a \cG_\a \right\} \ ,
\eea
where we have defined the Lorentz generator valued superfield $\cG_\a := \cG_\a{}^{\b\g} M_{\b\g}$. The 
form of the action makes clear a striking connection with 
the $\cN = 1$ super Yang-Mills action
\bea
S_{\rm YM} &=& 2 \int \rd^{3|2} z \, E \, {\rm tr} \left\{ \bm G^\a \bm G_\a \right\} \ ,
\eea
with $\bm G_\a$ the field strength of a Yang-Mills multiplet which satisfies the divergenceless condition $\bm \cD^\a \bm G_\a = 0$.
Here $\bm \cD_\a$ is the Yang-Mills group gauge covariant derivative.

The fact that an invariant containing a Ricci squared term may be made to resemble the Yang-Mills 
action is most significant from a component perspective. In particular, in \cite{BHT1} the full component 
action for a supergravity invariant containing a Ricci squared term was efficiently constructed in the 
gauge \eqref{compGaugeFixN=1} by reducing the problem to one of coupling a certain 
Yang-Mills multiplet to supergravity.\footnote{This is often referred to 
as the ``Yang-Mills trick.''} 
The procedure is equivalent to judiciously replacing the Yang-Mills multiplet 
component fields with those of the component fields of $\cG_{\a}{}^{\b\g}$, which transform as a Yang-Mills 
multiplet (with a Lorentz group index) by virtue of $\cD^\a \cG_{\a}{}^{\b\g} = 0$. At the component level one can check that 
the invariant contains the curvature squared contribution ${R}^{ab} {R}_{ab}-\frac{1}{4}{R}^2$.

It is worth mentioning that although we imposed the gauge conditions \eqref{gaugeFixN=1-1} and \eqref{gaugeFixN=1-2} 
it is straightforward to 
restore the compensator. One simply uses the action
\bea
S_{\rm Ricci^2}^{(\rm YM)} &=& - 2 \int \rd^{3|2} z \, E \, \varphi^4 \mathscr G^{\a \b\g} \mathscr G_{\a \b\g } \ ,
\eea
where
\be \mathscr G_{\a}{}^{\b\g } = \mathscr C_{\a}{}^{\b\g} + \frac{4}{3} \d_{\a}^{(\b} \mathscr D^{\g)} \mathscr S \ .
\ee

Although we have restricted our attention here
to curvature squared 
invariants, our approach makes it possible to generate locally supersymmetric 
functionals containing higher powers of the curvature tensor and its covariant derivatives. 
All such invariants are described by actions of the form \eqref{2.211} 
where $\mathscr T$ denotes the primary dimensionless superfields \eqref{2.188}. 
Among the descendants of \eqref{2.188}, the following rank 3 symmetric spinor
\be 
\mathscr W_{\a\b\g} = -\ri \mathscr D^2 \mathscr C_{\a\b\g} - 2 \mathscr D_{(\a\b} \mathscr D_{\g)} \mathscr S
- 8 \mathscr S \mathscr C_{\a\b\g} 
\ee
plays a special role. It is related to 
the super-Cotton tensor \cite{KT-M12}, $W_{\a\b\g}$,  
by the rule
\be 
W_{\a\b\g} = \varphi^5 \mathscr W_{\a\b\g} \ .
\ee
The super-Cotton tensor has been written in terms of $\varphi$ 
to make clear how it is related to the torsion components in conventional 
superspace. It should be kept in mind that it is actually independent of the 
compensator $\varphi$. Given a curved superspace background,
it is conformally flat if and only if $W_{\a\b\g} $ is equal to zero \cite{BKNT-M1}.
 This property explains the geometric meaning of the super-Cotton tensor. 
 The super-Cotton tensor contains the ordinary Cotton tensor as 
the component field $\nabla_{(\a} W_{\b\g\d)}|$ and 
obeys the equation \cite{BKNT-M1}
 \bea
 \nabla^\g W_{\a\b\g} =0~.
 \eea


\subsection{Models for massive supergravity}

Using the invariants constructed in the previous section one can 
build models for massive supergravity. The actions for these models are 
built out of a linear combination of the supergravity invariants together 
with the action for conformal supergravity. In this section we 
analyse the dynamical properties of such theories and derive the 
necessary conditions for massive supergravity.

We begin by considering a general $\cN=1$ supergravity model
described by the action 
\be S = \l S_{\rm cos} + \frac{1}{\kappa} S_{\rm SG} 
+ \mu_0 S_{\cS^3} 
+ \mu_1 S_{\cS^4} 
+ \mu_2 S_{\rm scalar^2}
+ \mu_3 S_{\rm Ricci^2} 
+ \frac{1}{\mu} S_{\rm CSG} \ , 
\label{N=1_SG_model}
\ee
where 
$S_{\rm CSG}$ denotes the $\cN=1$ conformal supergravity action, 
 eq. \eqref{N=1CSA}. In what follows we will assume that $\mu_3 \geq 0$ 
 as in \cite{Andringa:2009yc, BHRST10}.

It is an instructive exercise to derive the equations of motion in the
theory with action \eqref{N=1_SG_model}.
Varying the action \eqref{N=1_SG_model} with respect to 
the compensator $\vf$ leads to the equation
\bea 
0 &=& 
\l 
+ \frac{4}{\kappa} \mathscr S
- \ri \mu_3 {\mathscr C}^{\a\b\g} {\mathscr C}_{\a\b\g} 
- \frac{\mu_1}{2} \mathscr S^3 \non\\
&&+ \frac{\ri}{4} \mu_0 \mathscr D^2 \mathscr S
+ \ri \Big(\frac{3}{8} \mu_1 - \frac{3}{2} \mu_2\Big) \mathscr D^2 \mathscr S^2 
+ \frac{\ri \mu_2}{2} ( \mathscr D^\a \mathscr S )\mathscr D_\a \mathscr S \non\\
&&+ \Big(\mu_2 - \frac{8}{3} \mu_3\Big) \mathscr D^a \mathscr D_a \mathscr S \ . 
\label{gen_S_eq-or}
\eea
There is no contribution proportional to $1/  \m$ since the conformal supergravity action 
is independent of the compensator. 
The  equation of motion for the conformal supergravity prepotential is 
\bsubeq\label{2.466}
\bea
\frac{1}{\mu} W_{\a\b\g} + T_{\a\b\g} = 0 \ , \label{2.466-a}
\eea
where
\bea
\vf^{-5}
T_{\a\b\g} &=& 
\frac{1}{\kappa} \mathbb T^{({\rm SG})}{}_{\a\b\g} + \mu_0 \mathbb T^{({S^3})}{}_{\a\b\g}
+ \mu_1 \mathbb T^{({S^4})}{}_{\a\b\g}
+ \mu_2 \mathbb T^{({\rm sca^2})}{}_{\a\b\g}
+ \mu_3 \mathbb T^{({\rm Ric}^2)}{}_{\a\b\g}
~~~~~~~~~~
\eea
with the right-hand side consisting of the following contributions:
\bea \label{2.522}
\mathbb T^{({\rm SG})}{}_{\a\b\g} 
&=& - 2 {\mathscr C}_{\a\b\g} \ , \\
\mathbb T^{({S^n})}{}_{\a\b\g}
&=& \frac{n}{16} ({\mathscr D}_{(\a\b} {\mathscr D}_{\g)} + 2 {\mathscr C}_{\a\b\g}) \mathscr S^{n-1} \ , \\
\mathbb T^{({\rm sca^2})}{}_{\a\b\g} 
&=& 
2 ({\mathscr D}_{(\a} \mathscr S) {\mathscr D}_{\b\g)} \mathscr S
- \frac{\ri}{4} ({\mathscr D}_{(\a\b} {\mathscr D}_{\g)} + 2 {\mathscr C}_{\a\b\g}) {\mathscr D}^2 \mathscr S \ , \\
\mathbb T^{({\rm Ric^2})}{}_{\a\b\g} 
&=& 
- 4 {\mathscr D}^a {\mathscr D}_a \mathscr C_{\a\b\g}
+ \frac{8 \ri}{3} {\mathscr D}_{(\a\b} {\mathscr D}_{\g)} {\mathscr D}^2 \mathscr S
- 4 \ri {\mathscr C}_{\a\b\g} {\mathscr D}^2 \mathscr S 
+ 4 \ri \mathscr S {\mathscr D}^2 \mathscr C_{\a\b\g}
\non\\
&& + 32 \mathscr S {\mathscr D}_{(\a\b} {\mathscr D}_{\g)} \mathscr S
+ 12 \mathscr C_{(\a\b}{}^\r {\mathscr D}_{\g) \r} \mathscr S
+ \frac{56}{3} ({\mathscr D}_{(\a\b} \mathscr S) {\mathscr D}_{\g)} \mathscr S
\non\\
&&
+ \frac{8}{3} {\mathscr K}_{\a\b\g\d} {\mathscr D}^\d \mathscr S
- 16 \mathscr C_{(\a}{}^{\d\r} \mathscr K_{\b\g) \d \r} 
+ 12 {\mathscr S}^2 \mathscr C_{\a\b\g}
~.
\eea
\esubeq
Here we have denoted
\be \mathscr K_{\a\b\g\d} = \ri \mathscr D_{(\a} {\mathscr C}_{\b\g\d)} \ .
\ee
The equation of motion \eqref{2.466}
can be obtained by using the results in appendix \ref{prepotentialN=1},
which imply 
\be 
\d S_{\rm CSG} = \ri \int \rd^{3|2} z E \d \Psi^{\a\b\g} W_{\a\b\g}~, \qquad 
\d \Big( S - \frac{1}{\mu} S_{\rm CSG} \Big) 
= \ri \int \rd^{3|2} z E \d \Psi^{\a\b\g} T_{\a\b\g}
\ .
\ee
It is seen that  \eqref{2.466} does not involve 
the cosmological constant $\l$. This is due to the fact that 
the cosmological term \eqref{2.311} does not depend on 
the conformal supergravity prepotential. 
It also follows from the analysis in appendix \ref{prepotentialN=1}
that $T_{\a\b\g}$ obeys the conservation equation 
\bea
\nabla^\g T_{\a\b\g} = 0
\eea
provided the compensator is subject to its equation of motion \eqref{gen_S_eq-or}.

By construction, the supergravity equations \eqref{gen_S_eq-or} and  \eqref{2.466} 
are super-Weyl invariant. Upon imposing the gauge \eqref{gaugeFixN=1-1} 
and reducing to conventional superspace, the local super-Weyl symmetry may be 
fixed by imposing 
the gauge
\be
\vf =1~,
\label{2.499}
\ee
which amounts to replacing $\mathscr D_A \to \cD_A$, 
${\mathscr C}_{\a\b\g} \to {\cC}_{\a\b\g} $ and ${\mathscr S} \to \cS$ everywhere. 
The super-Weyl invariance can always be restored by performing an inverse replacement.
The gauge condition \eqref{2.499} will be assumed in what follows.

The supergravity equations of motion 
have maximally supersymmetric solutions.
Maximally supersymmetric backgrounds are specified by the conditions
\be 
\cC_{\a\b\g} = 0 \ , \quad
\cS =  {\rm const} \ , 
\label{2.588}
\ee
which imply that 
the algebra of covariant derivatives 
\eqref{cdaN=1-con} drastically simplifies 
\bsubeq \label{N=1AdSSS}
\bea
\{ \cD_\a , \cD_\b \} &=& 2 \ri \cD_{\a\b} - 4 \ri \cS M_{\a\b} \ , \\
\left[ \cD_{\a \b} , \cD_\g \right] &=& 
- 2 \eps_{\g(\a} \cS \cD_{\b)}  \ , \\
\left[\cD_a , \cD_b\right] &=& -4    \cS^2
 M_{ab}
 \ .
\eea
\esubeq
Such a superspace describes either anti-de Sitter geometry for $\cS \neq 0$ or a flat spacetime for $\cS=0$. 

Let  us look for a maximally supersymmetric background \eqref{2.588}
with $\cS =\cS_0$ which is 
a solution of the supergravity equations of motion. 
In this case we have 
\be 
 W_{\a\b\g} = T_{\a\b\g} =0 \ , 
\ee
and therefore the equation \eqref{2.466-a} is satisfied identically, 
while the equation on the compensator, eq. \eqref{gen_S_eq-or}, 
 becomes algebraic
\be 
0 = \l + \frac{4}{\kappa} \cS_0 - \frac{\mu_1}{2} \cS_0^3~.
\label{Scubic_eq}
\ee
This cubic equation in $\cS_0$
coincides with the one found 
in \cite{Andringa:2009yc, BHRST10}.
The real solutions of this equation (at least one real solution always exists) 
determine the maximally supersymmetric solutions of the 
supergravity theory under consideration. 

One may impose the constraint
\be \mu_2 = \frac{8}{3} \mu_3 \ , \label{GGMSG_condition}
\ee
which reduces the dynamical system \eqref{N=1_SG_model} 
to a six-parameter subclass of models. In general this leads to a propagating 
scalar mode, which is eliminated in the case of generalized massive supergravity (GMSG) via a 
further choice of coefficients.  
At the component level
one finds the equation of motion on the compensator to be
\begin{align} \label{compProjEoM}
\l + \frac{4}{\kappa} S
&
- \frac{3 \mu_0}{2}
 S^2
+ (12 \mu_2 - 5 \mu_1) {S}^3 \non\\
&= - \frac{1}{16} \Big( \mu_0 + (3 \mu_1 - 12 \mu_2) S \Big) {R} + {\rm fermion \ terms} \ ,
\end{align}
where we have used
\be {\cD}^2 \cS| = 6 \ri {S}^2 - \frac{\ri}{4} {R} + {\rm fermion \ terms} \ .
\ee
One can solve eq. \eqref{compProjEoM} for $S$ in terms of the scalar curvature ${R}$. 
At the bosonic level this was worked out in \cite{BHRST10} and leads to an action 
non-polynomial in
${R}$. It is worth mentioning that although we have suppressed 
the fermionic terms for simplicity, it is a straightforward exercise to recover them from eq. \eqref{gen_S_eq-or}.

One may consider perturbations in the supergravity model \eqref{GGMSG_condition} from the 
maximally supersymmetric solution:
\be \cS = \cS_0 + {\D \cS} \ , \quad \cC_{\a\b\g} = {\D \cC}_{\a\b\g} \ .
\ee
The equation of motion on the compensator becomes
\be 
0 = \ri \Big[ \frac{\mu_0}{4} + \Big(\frac{3}{4} \mu_1 - 3 \mu_2\Big) \cS_0 \Big] \cD^2 \D \cS
+ \Big[ \frac{4}{\kappa} - \frac{3 \mu_1}{2} {\cS}_0^2 \Big] \D \cS \ . \label{GMSG_eq_S}
\ee
When the coefficient for $\cD^2 \D \cS$ in eq. \eqref{GMSG_eq_S} does not vanish,
\be \frac{\mu_0}{4} + \Big(\frac{3}{4} \mu_1 - 3 \mu_2\Big) \cS_0  \neq 0 \ , \label{Sdegree1Cond}
\ee
we have the equation
\be \Big(\frac{\ri}{2} \cD^2 - m \Big)\D \cS = 0 \quad \implies 
({\cD}^a {\cD}_a - m^2) \D \cS = 0 \ ,
\ee
where $m$ is given by
\be m = - \frac{8 - 3 \kappa \mu_1 \cS_0^2}{\kappa \Big(\mu_0 + (3 \mu_1 - 12 \mu_2) \cS_0 \Big)} \ .
\ee
Hence in this case $\D \cS$ becomes propagating.

Instead of \eqref{Sdegree1Cond} 
we may impose the condition
\be \mu_0 + (3 \mu_1 - 12 \mu_2 ) \cS_0 = 0 \ .
\ee
and also assume  that 
\be
\frac{4}{\kappa} - \frac{3 \mu_1}{2} {\cS}_0^2 \neq 0~.
\ee
Then eq. \eqref{GMSG_eq_S} gives 
\be
\D \cS = 0~.
\ee 
Linearising the  Bianchi identity \eqref{2.444} about the background chosen 
and taking into account $\D \cS =0$, we obtain the divergenceless condition 
\bea 
\cD^\g \D\cC_{\a\b\g} = 0 \quad \Longrightarrow \quad
\cD^2 \D \cC_{\a\b\g} = 2\ri \cD_\a{}^\d \D \cC_{\b \g \d} +10 \ri 
\cS_0 \D \cC_{\a\b \g} \ .
\label{2.700}
\eea

At this point it is worth remarking on a property of symmetric divergenceless spinors in the 
 AdS background \eqref{N=1AdSSS} chosen. 
In general, given a  
symmetric rank-$n$ spinor, $T_{\a_1 \cdots \a_n} = T_{(\a_1 \cdots \a_n)}$,  
it holds that 
\be 
\cD^\b  T_{\a_1 \cdots \a_{n-1} \b} = 0 \quad \Longrightarrow \quad
\cD^\b \cD^2 T_{\a_1 \cdots \a_{n-1} \b} = 0~.
\ee
The linearised torsion
$\D \cC_{\a\b\g}$ is an example of such a superfield. In particular, the operator 
$\frac{\ri}{2} \cD^2$ preserves the divergenceless condition of $\D \cC_{\a\b\g}$.

Using eq. \eqref{2.700}, the supergravity equation of motion can be written in terms of vector covariant derivatives as follows
\begin{align}
\mu_3 \cD^a \cD_a \D \cC_{\a\b\g}
&+ \Big(2 \mu_3 {\cS}_0 - \frac{1}{2 \mu} \Big) \cD_\a{}^\d \D \cC_{\b\g\d} \non\\
&
+ \Big[ \frac{1}{2 \kappa} - \frac{1}{2 \mu} \cS_0 - \Big(\frac{3}{32} \mu_0 - 7 \mu_3 \Big) \cS_0^2 
- \frac{1}{8} \mu_1 \cS_0^3 \Big] \D \cC_{\a\b\g} = 0 \ .
\end{align}
When $\mu_3 \neq 0$ the equation of motion may be written in the following factorized form
\bea
\Big( \frac{\ri}{2} \cD^2 + m_- \Big) \Big( \frac{\ri}{2} \cD^2 - m_+ \Big) \D \cC_{\a\b\g} = 0 \ ,
\eea
where the constants $m_+$ and $m_-$ are such that
\bsubeq
\bea 
m_+ m_- &=& - \frac{1}{2 \mu_3} \Big{[}
\frac{1}{\kappa}
+ \frac{4}{\mu} \cS_0
- 3 \Big(\frac{1}{16} \mu_0 - 8 \mu_3\Big) \cS_0^2
- \frac{\mu_1}{4} \cS_0^3
\Big{]} \ , \\
m_+ - m_- &=& 8 \cS_0 + \frac{1}{2 \mu_3 \mu} \ .
\eea
\esubeq
The constants $m_+$ and $m_-$ are real when the following inequality is satified:
\be \frac{1}{\kappa} - \frac{1}{8 \mu_3 \mu^2}
- \Big(\frac{3}{16} \mu_0 + 8 \mu_3 \Big) \cS_0^2
- \frac{\mu_1}{4} \cS_0^3 \leq 0 \ .
\ee

The supergravity equations of motion 
have massive solutions in a number of cases. For instance, 
generalized massive supergravity \cite{Andringa:2009yc, BHRST10} is characterised by a negative Einstein-Hilbert term, 
$\kappa < 0$, while the case of new topologically massive supergravity \cite{Andringa:2009yc, BHRST10} is 
characterized by $\kappa \rightarrow \infty$. 
Furthermore, new massive supergravity occurs in the case $\mu_3 \neq 0$ and $\mu \rightarrow \infty$. In this case 
it is straightforward to verify that about a Minkowski background ($\cS_0 = 0$) we have the 
massive equation
\bea
(\partial^a \partial_a - \tilde{m}^2 )\D \cC_{\a\b\g} = 0 \ , \quad \tilde{m}^2 := m_+ m_- = - \frac{1}{2 \mu_3 \kappa} \ ,
\eea
where $\tilde{m}$ is real for a negative Einstein-Hilbert term, $\kappa < 0$.

In the case when $\mu_3 = 0$ and $\mu$ is finite 
we have the equation
\be \Big(\frac{\ri}{2} \cD^2 - \hat{m} \Big) \D \cC_{\a\b\g} = 0 \ ,
\ee
where $\hat{m}$ is given by
\be \hat{m} = - \Big( \frac{\mu}{\kappa} + 4 \cS_0 - \frac{3 \mu_0 \mu}{16} \cS_0^3 - \frac{\mu \mu_1}{4} \cS_0^3 \Big) \ .
\ee
About a Minkowski background it is straightforward to verify 
that
\be (\partial^a \partial_a - \hat{m}^2 )\D \cC_{\a\b\g} = 0 \ .
\ee
Note that when $\mu_0 = \mu_1 = 0$ the above coincides with topologically massive supergravity. 

We have reduced the supergravity models to those considered in \cite{Andringa:2009yc, BHRST10}. The analysis of 
unitarity for such theories may be carried out as in \cite{Andringa:2009yc, BHRST10}.


\section{$\cN = 2$ supergravity models with a chiral compensator} \label{section3}


All known off-shell formulations for 3D $\cN=2$ supergravity \cite{KT-M11,KLT-M11}
can be formulated in conventional superspace
 with  structure group  ${\rm SL}(2,{\mathbb{R}})\times {\rm U(1)}_R$.
This curved superspace $\cM^{3|4}$ is  parametrised by
local bosonic ($x^m$) and fermionic ($\q^\m, \bar \q_\m$)
coordinates  $z^{{M}}=(x^{m},\q^{\mu},{\bar \q}_{{\mu}})$,
where 
the Grassmann variables $\q^{\mu} $ and $\bar \q_{{\mu}}$
are related to each other by complex conjugation:
$\overline{\q^{\mu}}=\bar \q^{{\mu}}$.

\subsection{Conventional superspace} \label{subsection3.1}

The covariant derivatives of conventional $\cN=2$ superspace
$\cD_{{A}} =(\cD_{{a}}, \cD_{{\a}},\bar \cD^\a)$
have the form
\bea
\cD_{{A}}&=&E_{{A}}
-\O_{{A}}
-\ri \,\F_{{A}} J~,
\label{CovDev}
\eea
with $J$ the $R$-symmetry generator acting on the covariant derivatives
as follows:
\bea
{[} J,\cD_{\a}{]}
=\cD_{\a}~,
\qquad
{[} J,\cDB^{\a}{]}
=-\cDB^\a~,
\qquad
{[}J,\cD_a{]}=0~.
\eea

In order to describe $\cN=2$ conformal supergravity, the torsion 
has to obey the covariant constraints proposed in \cite{HIPT}.
The resulting algebra of covariant derivatives is \cite{KLT-M11,KT-M11}
\bsubeq \label{algebra-final}
\bea
\{ \cD_\a , \cD_\b \} &=& - 4 \bar \cR M_{\a\b} \ , \\
\{ \cD_\a , {\bar \cD}_\b \} &=&
- 2 \ri (\g^c)_{\a\b} \cD_c 
- 2 \cC_{\a\b} J
- 4 \ri \eps_{\a\b} \cS J 
+ 4 \ri \cS M_{\a\b}
- 2 \eps_{\a\b} \cC^{\g\d} M_{\g\d} \ ,  \\
{[}\cD_a , \cD_\b {]}
&=& \ri \eps_{abc} (\g^b)_\b{}^\g \cC^c \cD_\g
+ (\g_a)_\b{}^\g \cS \cD_\g 
- \ri (\g_a)_{\b\g} \bar \cR \bar \cD^\g
+ \ri (\g_a)_\b{}^\g \cD_{(\g} \cC_{\d\r)} M^{\d\r} \non\\
&&- \frac{1}{3} (2 \cD_\b \cS + \ri \bar \cD_\b \bar \cR) M_a
- \frac{2}{3} \eps_{abc} (\g^b)_\b{}^\a (2 \cD_\a \cS + \ri \bar \cD_\a \bar \cR) M^c \non\\
&&+ \frac{\ri}{2} \Big( 
(\g_a)^{\a\g} \cD_{(\a} \cC_{\b\g)}
+ \frac{1}{3} (\g_a)_\b{}^\g (8 \ri \cD_\g \cS - \bar \cD_\g \bar \cR)
\Big) J \ ,
\eea
\esubeq
where the ${\rm U}(1)_{R}$ charges of the torsion superfields  $\cR$, $\bar \cR$ and $\cC_{\a\b}$
are $-2$, $+2$ and 0, respectively.
They also satisfy the Bianchi identities
\bea 
\cD_\a \bar \cR &=& 0 \ , \quad
\cD^\b \cC_{\a\b} =- \frac{1}{2} (\bar \cD_{\a} \bar \cR + 4 \ri \cD_{\a} \cS) \ . 
\label{BItypeIC}
\eea

The algebra of covariant derivatives given by \eqref{algebra-final}
does not change under the super-Weyl transformation
\cite{KLT-M11, KT-M11}
\bsubeq  \label{super-WeylN=2}
\bea
\cD'{}_\a&=&\re^{\hf\s}\Big(\cD_\a+\cD^{\g}\s
M_{\g\a}-\cD_{\a} \s J\Big)~,
\\
\cD'{}_{a}
&=&\re^{\s}\Big(
\cD_{a}
-\frac{\ri}{2}(\g_a)^{\g\d}\cD_{\g}\s\cDB_{\d}
-\frac{\ri}{2}(\g_a)^{\g\d}\cDB_{\g}\s\cD_{\d} 
+\ve_{abc}\cD^b \s M^c \non\\
&&~~~~~-\frac{\ri}{2}(\cD^{\g}\s)\cDB_{\g}\s M_{a}
- \frac{\ri}{24} (\g_a)^{\g\d} \re^{- 3 \s} [\cD_\g , \bar \cD_\d] \re^{3 \s}
J
\Big)~,
\eea
which induces the following transformation of the torsion tensors:
\bea
\cS'&=&\re^{\s}\Big(
\cS
+\frac{\ri}{4}\cD^\g\cDB_{\g}\s
\Big)~,
 \label{2.11d} \\
\cC'_{a}&=&
\Big( \cC_a + \frac{1}{8} (\g_a)^{\g\d} [\cD_\g , \bar \cD_\d] \Big) \re^\s
~,
\\
\cR' &=&
- \frac{1}{4} \re^{2 \s} (\bar \cD^2 - 4 \cR) \re^{- \s}
~. 
\label{2.11f}
\eea
\esubeq
Here the  parameter $\s$ is an arbitrary real scalar superfield and we have defined
$\cD^2:=\cD^\a\cD_\a$ and $\cDB^2:=\cDB_\a\cDB^\a$.
The super-Weyl invariance \eqref{super-WeylN=2}
is intrinsic to conformal supergravity. 
For every supergravity-matter system, 
its action is required to be a super-Weyl invariant functional  
of the supergravity Weyl multiplet coupled to certain conformal
compensators, see \cite{KLT-M11,KT-M11} for more details.

There exists an important super-Weyl invariant descendent of the 
torsion components $\cC_{\a\b}$ and $\cS$ that is worth mentioning. 
Using the above super-Weyl transformation laws, one can check that the following
real vector superfield \cite{Kuzenko12}
\be \cW_{\a\b} = - \frac{\ri}{4} [\cD^\g , \bar\cD_\g] \cC_{\a\b}
+ \hf [\cD_{(\a} , \bar\cD_{\b)}] \cS + 2 \cS \cC_{\a\b}  \label{N=2covsuperspaceCotton}
\ee
transforms homogeneously,
\be \cW'_{\a\b} = \re^{2 \s} \cW_{\a\b} \ . 
\ee
The superfield is the $\cN = 2$ super-Cotton tensor 
and it vanishes if and only if 
the curved superspace is conformally flat \cite{BKNT-M1}. Using the Bianchi identities one can find the following equivalent 
form for the super-Cotton tensor \cite{KLRST-M}:
\be \cW_a = - \hf (\g_a)^{\a\b} \cW_{\a\b} = 
\frac{1}{4} (\g_a)^{\a\b} [\cD_{(\a} , \bar\cD_{\b)}] \cS
- \eps_{abc} \cD^b \cC^c
- 2 \cS \cC_a \ .
\ee
The covariant derivative algebra of conformal superspace is expressed entirely in 
terms of the super-Cotton tensor, see appendix \ref{geometry}.


\subsection{Type I minimal supergravity in conventional superspace}

Type I minimal supergravity
makes use of two compensators, a
covariantly chiral scalar $\F$ and its conjugate $\bar \F$.
The chiral compensator is defined to be nowhere vanishing, 
have U$(1)_R$ charge equal to $ -1/2$,  
\bea
\bar \cD_\a \F =0~, \qquad J \F = -\hf \F~,
\eea
and possess the super-Weyl transformation law
\bea
\F' = \re^{\hf \s} \F~.
\eea
In general, the U$(1)_R$ charge of a chiral scalar and its super-Weyl weight 
are equal in magnitude and opposite in sign \cite{KLT-M11}.

The freedom to perform the super-Weyl and local 
U$(1)_R$ transformations allows us to choose a gauge
$\F=1$,
which implies the consistency  conditions
\bea
 \cS=0~, \qquad \F_\a=0~,\qquad \F_{a}= -\cC_{a}~.
 \label{3.88}
\eea
This reduces the structure group from
${\rm SL}(2,{\mathbb{R}})\times {\rm U(1)}_R$ to its subgroup 
${\rm SL}(2,{\mathbb{R}})$.
Instead of imposing the gauge condition $\F=1$ (which completely fixes
the super-Weyl and local U$(1)_R$ freedom), it is more convenient 
to partially fix
the super-Weyl and local U$(1)_R$ symmetry by imposing 
only the conditions \eqref{3.88}. 
The residual super-Weyl and local U$(1)_R$ symmetry is
described  by transformations which are parametrised 
in terms of a covariantly chiral scalar 
parameter $\l$, $\bar \cD_\a \l=0$, and have the form \cite{KT-M11}
\bsubeq \label{sW+U(1)}
\bea
\cD'{}_\a&=&
\re^{\frac{1}{2}(3\bar{\l}-\l)}
\Big(\cD_\a+\cD^{\g}\l M_{\g\a}\Big)~,
\\
\tilde \cD'{}_{a}&=&\re^{\l+\bar{\l}}\Big(
\tilde \cD_{a}
-\frac{\ri}{2}(\g_a)^{\a\b}\cD_{\a}\l\bar \cD_{\b}
-\frac{\ri}{2}(\g_a)^{\a\b} \bar \cD_{\a}\bar{\l}\cD_{\b}
\non\\
&&~~~~~~~
+\ve_{abc}\tilde \cD^b(\l+\bar{\l}\big)M^c
-\frac{\ri}{2}(\cD^{\g}\l)\bar \cD_{\g}\bar{\l}M_{a}
\Big)
~,
\eea
\esubeq
where we have introduced the operator $\tilde \cD_a := \cD_a - \ri \cC_a J$.
The covariant derivatives $(\tilde \cD_a, \cD_\a , \bar \cD^\a )$
do not contain any U$(1)_R$ connection and obey the algebra
\bsubeq \label{3.100}
\bea
\{ \cD_\a , \cD_\b \} &=& - 4 \bar \cR M_{\a\b} \ , \\
\{ \cD_\a , {\bar \cD}_\b \} &=&
- 2 \ri (\g^c)_{\a\b} \tilde \cD_c 
- 2 \eps_{\a\b} \cC^{\g\d} M_{\g\d} \ .
\eea
\esubeq

The above partially gauge fixed geometric setting is completely suitable 
to describe type I supergravity and its matter couplings. However,
it is not an ideal 
formalism for reducing the supergravity actions 
to components.
From the point of view of component reduction, it is advantageous to 
make use of a supergravity formulation with a larger gauge group
than that of conventional superspace.
Such a framework is provided by 
 the $\cN=2$ 
conformal superspace developed in \cite{BKNT-M1}. 
Its important features are: (i) it is well adapted to reducing 
off-shell supergravity-matter actions to components; 
and (ii) conventional superspace is a gauge fixed version of 
conformal superspace.
The salient details of $\cN=2$ conformal superspace are given 
in appendix \ref{geometry}. Below we show how to describe 
type I supergravity in this setting.


\subsection{The superconformal setting for type I supergravity}

In conformal superspace, the compensator is 
a primary nowhere vanishing chiral superfield $\Phi$ of dimension 1/2,
\be 
\bar \nabla_\a \Phi = 0 \ , \quad K_A \Phi = 
0 \ ,
\quad \mathbb D \Phi = \hf \F\ .
\ee
The chirality of $\Phi$ fixes its 
U$(1)_R$ charge,  $ {J} \Phi = - \mathbb D \Phi $.
We define the component fields of $\Phi$ as follows:
\bea \phi &:=& \Phi| \ , \quad \z_\a := \nabla_\a \Phi| \ , \quad \bar {M} = - \frac{1}{4} \nabla^2 \Phi| \ .
\eea

Similar to the $\cN = 1$ case, one can use the compensator $\Phi$ to introduce dimensionless and  U$(1)_R$ neutral covariant derivatives, 
$\mathscr D_A=(\mathscr D_a , \mathscr D_\a, \bar {\mathscr D}^\a)$,  
that take every primary superfield to a primary one. They are defined by
\bsubeq \label{3.166}
\bea 
\mathscr D_\a &:=& \sqrt{\frac{\Phi}{\bar\Phi^3}} \Big( \nabla_\a - \nabla^\b \ln \Phi \ \! M_{\a\b} 
+ \nabla_\a \ln \Phi \ \! {J} - \nabla_\a \ln \Phi \ \! \mathbb D \Big) \ , \\
\mathscr D_a &:=& - \frac{\ri}{4} (\g_a)^{\a\b} \{ \mathscr D_\a , \bar{\mathscr D}_\b\} \ .
\eea
\esubeq
One can explicitly check that
\be 
\mathscr D_A \Phi = 0
\ ,
\ee
which tells us that when acting on primary superfields 
the graded commutator $[\mathscr D_A , \mathscr D_B \}$ 
contains no ${\rm U}(1)_R$ curvature. 

When acting on primary superfields the 
operators
$\mathscr D_A$ satisfy the algebra
\bsubeq \label{3.155}
\bea
\{ \mathscr D_\a , \mathscr D_\b \} &=& - 4 \bar {\mathscr R} M_{\a\b} \ , \quad 
\{ \bar{\mathscr D}_\a , \bar{\mathscr D}_\b \} = 4 \mathscr R M_{\a\b} \ , \\
\{ \mathscr D_\a , \bar{\mathscr D}_\b \} &=& - 2 \ri \mathscr D_{\a\b} 
- 2 \eps_{\a\b} \mathscr C^{\g\d} M_{\g\d} \ ,
\eea
\esubeq
where we 
have introduced 
the primary superfields
\bsubeq \label{GeocompositeN=2}
\bea
\mathscr R &:=& - \frac{1}{4 \Phi^3} \bar \nabla^2 \bar \Phi \ , 
\\
\mathscr C_{\a\b} &:=& - \frac{1}{4} [\nabla_{(\a} , \bar \nabla_{\b)}] \Big(\frac{1}{\Phi \bar \Phi}\Big) \ ,
\eea
\esubeq
which are dimensionless and U$(1)_R$ neutral. The algebra of covariant derivatives \eqref{3.155} formally 
coincides with \eqref{3.100}.

We can relate the superconformal framework to the one of conventional superspace 
by gauge fixing the additional symmetries. We can use the conformal boosts and $S$-supersymmetry
transformations to impose the gauge condition
\be B_A = 0 \ , \label{gaugeFixN=211}
\ee
which degauges conformal superspace to conventional superspace \cite{BKNT-M1}. 
The composites \eqref{GeocompositeN=2} become 
the following super-Weyl invariant objects
\bsubeq
\bea 
\mathscr C_{\a\b} &=&  - \frac{1}{4} 
\big([\cD_{(\a} , \bar \cD_{\b)}]-4\cC_{\a\b} \big)\frac{1}{\Phi \bar \Phi} \ , \\
\mathscr R &=& - \frac{1}{4 \Phi^3} (\bar \cD^2 - 4 R) \bar \Phi 
\ ,
\eea
\esubeq
while the super-Cotton tensor of conformal superspace $W_{\a\b}$ coincides 
with \eqref{N=2covsuperspaceCotton}.
Upon imposing the additional gauge condition
\be \label{gaugeFixN=222}
 \Phi = 1 \ ,
 \ee
the composites \eqref{GeocompositeN=2} coincide with the 
torsion components $\cR$
and $\cC_{\a\b}$. 

It is also worth mentioning 
that one can use the compensator to construct super-Weyl invariant covariant 
derivatives $\cD_A^{^{(\Phi)}}$ in conventional superspace as follows:
\bsubeq
\bea
\cD_\a^{(\Phi)} &=& \sqrt{\frac{\Phi}{\bar\Phi^3}} \Big( \cD_\a - \cD^\b \ln \Phi M_{\a\b}
+ \cD_\a \ln \Phi J \Big) \ , \\
\cD_a^{(\Phi)} &=& - \frac{\ri}{4} (\g_a)^{\a\b} \{ \cD_\a^{(\Phi)} , \cD_\b^{(\Phi)} \} \ .
\eea
\esubeq
It can be checked that these covariant derivatives satisfy the same algebra as \eqref{3.155}.
Unlike the operators $ \mathscr D_A$, eq.  \eqref{3.166}, 
they do not annihilate the compensator $\Phi$.

Many supergravity models may be constructed as  integrals over
 $\cN = 2$ superspace of the form
\bea
S = \int \rd^{3|4}z \, E \,\cL \ ,
\label{N=2action}
\eea
where $\cL$ is a real primary superfield of dimension 1. 
One can also use 
the chiral action principle
\be 
S_{\rm c} = \int\rd^{3|2} z_{\rm c} \, \cE \, \cL_{\rm c} \ , \qquad  
\rd^{3|2} z_{\rm c} := \rd^3 x \rd^2 \q \ ,
\label{N=2chiralaction}
\ee
where $\cL_{\rm c}$ is a primary chiral scalar of dimension 2
and $\cE$ denotes the chiral density\footnote{The 
explicit expression for $\cE$ in terms of the supergravity prepotentials is given in 
\cite{Kuzenko12}.}. 
Every action \eqref{N=2action} can be rewritten as a chiral action 
because of the relation \cite{KLT-M11}
\bea
\int\rd^{3|4}z\, E \,\cL
= -\frac{1}{4} \int\rd^{3|2} z_c \, \cE \, 
\bar \nabla^2 \cL_{\rm c} \ .
\label{N=2Ac-2}
\eea
The chiral action can be reduced to components
using the component reduction formula
\cite{KLRST-M,KN14} 
\be S_{\rm c} = - \frac{1}{4} \int \rd^3 x\, e\, \big[
\nabla^2 
- 2 \ri (\g^a)_{\a\b} \bar{\psi}_a{}^\a \nabla^\b 
- 2 \eps^{abc} (\g_a)_{\a\b} \bar\psi_b{}^\a \bar\psi_c{}^\b
\big] \cL_{\rm c}| \ , \label{compRedActionFromulaN=2}
\ee
where $\j_a{}^\b = e_a{}^m \psi_m{}^\b$ and $\bar \j_a{}^\b = e_a{}^m \bar \psi_m{}^\b$ denote the gravitini. The component fields of the 
Weyl multiplet were defined in \cite{BKNT-M2}. The vielbein $e_m{}^a$, the gravitini $\psi_m{}^\a$ and $\bar \j_m{}^\b$, the 
${\rm U}(1)_R$ gauge field $V_m$ and the dilatation gauge field $b_m$ are defined as the lowest components of 
their corresponding superforms,
\be e_m{}^a := E_m{}^a| \ , \quad \psi_m{}^\a := 2 E_m{}^\a| \ , \quad 
V_m := \Phi_m| \ , \quad b_m := B_m| \ .
\ee
At the component level we will be mainly concerned with bosonic fields.

Using the above results one can construct general actions of the form
\be
S = \int \rd^{3|4}z \, E \, \Phi \bar \Phi \frak{L}(\mathscr T , \mathscr D \mathscr T , \cdots ) \ , \label{generalActionPrincipleTypeI}
\ee
where $\frak{L}$ is a dimensionless superfield constructed out of the torsion components $\mathscr T$ and their
covariant derivatives $\mathscr D_A$. As for $\cN = 1$, we will focus on actions that 
in components involve at most curvature squared terms.

To fix the additional symmetries in our invariants one can make use of the 
gauge conditions \eqref{gaugeFixN=211} and 
\eqref{gaugeFixN=222} which correspond to the following conditions at 
the component level
\be \phi = 
1 \ , \quad \z_\a = 
0 \ , \quad b_m = 0 \ . \label{compTypeIgaugeconditions}
\ee
The first condition fixes the dilatations and U(1)$_R$ transformations, the second fixes the $S$-supersymmetry 
transformations and the last fixes the conformal boosts.

Some useful results for identifying curvature terms in the actions 
we construct 
are given below with the gauge conditions \eqref{compTypeIgaugeconditions} (compare with \eqref{compBoxResultsN=1}):
\bsubeq
\bea
\Box \phi &=& \frac{1}{8} {R} +{\rm fermion~terms}
\ , \\
\Box^2  \frac{1}{\phi}
&=&
{R}^{ab}{R}_{ab}
-\frac{23}{64}{R}^2
-\frac{1}{8}D^aD_a{R}
+{\rm fermion~terms}
~,
\eea
\esubeq
where we have defined
\bsubeq
\bea
\Box \phi &:=& \nabla^a \nabla_a \Phi| \ , \\
\Box^2 \frac{1}{\phi} &:=& \nabla^a \nabla_a \nabla^b \nabla_b \frac{1}{\Phi}|
\eea
\esubeq
and introduced the covariant derivative
\be D_a = e_a{}^m \Big( \partial_m - \hf \omega_m{}^{bc} M_{bc} - \ri V_m J- b_m \mathbb D \Big)
\ .
\ee


\subsection{Supergravity invariants} 

In this subsection we construct supergravity invariants in superspace by making 
use of the composites \eqref{GeocompositeN=2}.


\subsubsection{The supergravity action}

The type I minimal supergravity action with a cosmological term was given in \cite{KLRST-M} in 
conventional superspace. It is straightforward to lift the action to conformal superspace and is given by
\be S = \frac{1}{\kappa} S_{\rm SG} + (\l S_{\cos} + {\rm c.c.}) \ ,
\ee
where
\bea
S_{\rm SG} &=& - 4 \int\rd^{3|4}z\, E \, \bar \Phi \Phi
 \ , \\
S_{\rm cos} &=& 
 \int\rd^{3|2} z\, \cE \, \Phi^4
 \label{Type-I_AdS}
 \eea
 and $\l$ is the cosmological constant, which can be complex 
 in general. The above action contains the Einstein-Hilbert term,
$-\hf{R}$,
since one finds the term 
$-4\phi \Box \bar \phi$ at the component level  in the action. 
The detailed component analysis for type I supergravity can be found in \cite{KLRST-M}.


\subsubsection{The ${R} \, M^n$ invariants}

A locally supersymmetric invariant containing a ${R} \, M^n$ term is given by
\bea
S_{{R} M^n} = \int \rd^{3|4} z \, E \, \Phi \bar \Phi \mathscr R^n = \int\rd^{3|2} z \, \cE \, \mathscr R^{n+1} \Phi^4 \ .
\label{3.20}
\eea
The corresponding invariant at the component level contains a term proportional to
$\Big(M/\phi^3\Big)^n \phi \Box \bar \phi $,
which upon gauge fixing gives rise to the 
term
\bea
\frac{(n+1)}{8} {R} M^{n}
~.
\eea
 The component action at the bosonic level was explicitly given in \cite{BOS14}.


\subsubsection{The scalar curvature squared invariant}

A scalar curvature squared invariant is described by
\bea
S_{\rm scalar^2} = 
- 4 \int\rd^{3|4}z\, E \, \Phi \bar \Phi \mathscr R \bar {\mathscr R}
= - \frac{1}{4} \int\rd^{3|4}z\, E \, \frac{1}{(\Phi \bar \Phi)^2} (\nabla^2 \Phi) \bar \nabla^2 \bar \Phi \ .
\eea
One can show that the above invariant will involve a term proportional to
 $1/(\phi \bar\phi)^2 \Box \phi \Box \bar \phi$,
which gives rise to the scalar curvature squared term $ -\frac{1}{16} {R}^2$ 
at the component level.
One can check that the action 
also contains a $|M|^4$ term. At the component level the explicit bosonic action was explicitly given in \cite{BOS14}.


\subsubsection{The Ricci curvature squared invariant}

An invariant containing a Ricci squared term is given by
\be S_{\rm Ricci^2} = 4 \int\rd^{3|4}z\, E \, \Phi \bar \Phi \mathscr C^{\a\b} \mathscr C_{\a\b} \ .
\ee
The above action also can be seen to contain a scalar curvature squared term at the component level 
by making use of the results in \cite{KLRST-M}. 
It therefore
makes sense to introduce the one parameter family of invariants
\be 
S_\z = 4 \int\rd^{3|4}z\, E \, \Phi \bar \Phi \Big( \mathscr C^{\a\b} \mathscr C_{\a\b} 
+ \z \mathscr R \bar {\mathscr R} \Big) \ ,
\ee
where $\z$ parametrizes the scalar curvature squared contribution.

It should be mentioned that an alternative invariant containing a 
Ricci curvature squared term may be constructed and is given by
\bea
S_{\Xi} = - \frac{1}{8} \int\rd^{3|2} z_c\, \cE \, \frac{\Xi}{\Phi} + \rm {c.c.} \ , \label{BergshoeffAction}
\eea
where we have defined
\be \Xi = \nabla^a \nabla_a {\bar \nabla}^2\frac{1}{\bar \Phi} 
= \nabla^a  {\bar \nabla}^2 \nabla_a \frac{1}{\bar \Phi} 
=  \bar \nabla^2 \nabla^a \nabla_a \frac{1}{\bar \Phi} \ .
\ee
Remarkably, one can check that the superfield $\Xi$ is both chiral 
and primary. It corresponds to the dimension $5/2$ 
 composite constructed at the bosonic level in \cite{BOS14}. The 
 fermionic terms may be recovered by straightforward component reduction of our result. 
 It is also worth noting that the 
 invariant \eqref{BergshoeffAction} can be written in terms of the 
conventional superspace formulation of \cite{KLT-M11} as follows:
 \bea
S_{\Xi}
 &=&
- \frac{1}{4} \int\rd^{3|4}z\, E  \Big\{ \frac{1}{\Phi} \cD^{\a\b} \cD_{\a\b} \frac{1}{\bar \Phi}
 + \frac{11 \cC^{\a\b} \cC_{\a\b}}{4 \Phi \bar \Phi}
 + \frac{6 \cR \bar \cR}{\Phi \bar \Phi} 
 + 8 \cC^{\a\b} \Big( \cD_\a \frac{1}{\Phi} \Big) \bar \cD_\b \frac{1}{\bar \Phi}
 \non\\
 &&~~~
  - \frac{8 \ri}{\Phi^2 \bar \Phi} \cC^{\a\b} \cD_{\a\b}\frac{1}{\bar \Phi} 
 - \frac{13 \ri}{ \Phi} \cC^{\a\b} \cD_{\a\b}\frac{1}{\bar \Phi} 
 - \frac{6 \cS^2}{\Phi \bar \Phi} 
 - \frac{3 \ri}{2 \Phi \bar \Phi} \cD^{\a\b} \cC_{\a\b}
 + \frac{9 \bar R}{2 \Phi} \bar \cD^2 \frac{1}{\bar \Phi} 
 \non\\
 &&~~~
 - \frac{6 \ri}{\Phi \bar \Phi} \cD^\a \bar \cD_\a \cS
 - \frac{6 \bar R \bar \Phi^2}{\Phi^2} \Big( \bar \cD_\a \frac{1}{\bar \Phi} \Big) \bar \cD^\a \frac{1}{\bar \Phi}
  \Big\} 
+ {\rm c.c.}  \label{longConvSuperspaceActionRicciTypeI}
 \eea

 Note that in the gauge where $\Phi = 1$ we have
 \be 
 \cS=0
 ~,~~~~~~
 \cD_{\a\b} \Phi  = - \frac{\ri}{2} \cC_{\a\b}
 \ee
 and the action \eqref{longConvSuperspaceActionRicciTypeI} simply becomes
  \bea
S_\Xi = 4 \int\rd^{3|4}z \, E  \Big( \cC^{\a\b} \cC_{\a\b} - \frac{3}{4} \cR \bar \cR \Big) \ . \label{3.40}
 \eea
 It follows that the action \eqref{BergshoeffAction} coincides with $S_{-3/4}$.
 
 Upon reducing to components and imposing the gauge conditions \eqref{gaugeFixN=211} 
 and \eqref{gaugeFixN=222},
the action \eqref{3.40} gives rise to the following combination of Ricci 
and scalar curvature squared 
terms:
\bea
{R}^{ab}{R}_{ab}
-\frac{23}{64}{R}^2
\ .
\eea
A pure Ricci curvature squared invariant can be identified and is simply given by
\be 
S_{\rm Ricci^2}^{(\rm pure)}
= S_5
=  4\int\rd^{3|4}z\, E \, \Phi \bar \Phi \Big( \mathscr C^{\a\b} \mathscr C_{\a\b} 
+5 \mathscr R \bar {\mathscr R} \Big) 
\ .
\ee

In the above we have restricted our attention to curvature squared terms. However, our 
approach naturally provides a means to address locally supersymmetric functionals containing 
higher powers of the curvature tensor and its covariant derivatives. One can simply consider 
other actions of the form \eqref{generalActionPrincipleTypeI}, which involves covariant derivatives 
of the primary superfields \eqref{GeocompositeN=2}. Amongst the descendants of \eqref{GeocompositeN=2} 
it is worth mentioning the following rank 2 symmetric spinor
\be \mathscr W_{\a\b} = - \frac{\ri}{4} [\mathscr D^\g , \bar{\mathscr D}_\g] \mathscr C_{\a\b} \ .
\ee
which is related to the super-Cotton tensor $W_{\a\b}$ by the rule
\be W_{\a\b} = (\Phi\bar\Phi)^2 \mathscr W_{\a\b} \ .
\ee
The super-Cotton tensor is independent of the compensator $\Phi$ and satisfies the 
condition
\be \nabla^\b W_{\a\b} = 0 \ .
\ee


\subsection{Models for massive supergravity}

The invariants in the previous section are useful building blocks in the construction of 
massive supergravity. In this section we analyse the dynamics of a general supergravity model 
and determine the conditions in which we have massive supergravity.

We consider the supergravity model\footnote{The class of supergravity models considered here 
is more general than those considered in \cite{BOS14} because we have 
allowed some of the coupling constants to be complex.
}
\be 
S = \frac{1}{\kappa} S_{\rm SG} 
+ \mu_2 S_{\rm scalar^2}
+ \mu_3 S_{\rm Ricci^2} 
+ \frac{1}{\tilde{\mu}} S_{\rm CSG}
+ \Big( \l S_{\rm cos} 
+ \mu_0 S_{M^3} 
+ \mu_1 S_{M^4} + {\rm c.c.} \Big) \ . 
\label{N=2_SG_model}
\ee
Here $\l$, $\mu_0$ and $\mu_1$ are allowed to be complex in general.
The action for $\cN=2$ conformal supergravity,  $S_{\rm CSG} $, 
was originally constructed 
in \cite{RvN}. Within the conformal superspace approach \cite{BKNT-M1},
its construction was given in \cite{BKNT-M2}. In what follows we assume $\mu_3 \geq 0$ 
as in \cite{BOS14}.

Varying the action \eqref{N=2_SG_model} with respect to the compensator $\Phi$ 
leads to the equation of motion
\bea \label{TypeIcompensatorEoM}
0 &=& 
4 \l 
- \frac{4}{\kappa} \mathscr R
- 2 \mu_0 \mathscr R^2 - \hf \bar{\mu}_0 ({\bar {\mathscr D}}^2 - 4 \mathscr R) \bar {\mathscr R} 
- 5 \mu_1 \mathscr R^3
- \frac{3}{4} \bar{\mu}_1 (\bar {\mathscr D}^2 - 4 \mathscr R) {\bar {\mathscr R}}^2 \non\\
&&
+ \frac{1}{4} (\mu_3 -\mu_2 )({\bar {\mathscr D}}^2 - 4 \mathscr R) ({\mathscr D}^2 - 4 \bar{\mathscr R}) \mathscr R
+ (\mu_3-2\mu_2 ) \mathscr R ({\bar {\mathscr D}}^2 - 4 \mathscr R) \bar {\mathscr R}
\non\\
&&
- 2 \mu_3 ({\bar {\mathscr D}}^2 - 4 \mathscr R) (\mathscr C^a \mathscr C_a)
~.
\eea
As in the $\cN = 1$ case there is no contribution proportional to $\tilde{\mu}^{-1}$ since the 
conformal supergravity action is independent of the compensator. The equation of motion for the 
conformal supergravity prepotential is
\bea
\frac{1}{\mu} W_{\a\b} + T_{\a\b} = 0 \ ,
\label{111111}
\eea
where $\mu$ is related to $\tilde{\mu}$ by a multiplicative constant 
and the supercurrent 
$T_{\a\b}$ is
\bea (\Phi \bar \Phi)^{-2} T_{\a\b} &=& 
\frac{1}{\kappa} \mathbb T^{({\rm SG})}{}_{\a\b} 
+ \mu_2 \mathbb T^{({\rm sca^2})}{}_{\a\b}
+ \mu_3 \mathbb T^{({\rm Ric}^2)}{}_{\a\b} \non\\
&&+ \Big( \mu_0 \mathbb T^{({M^3})}{}_{\a\b}
+ \mu_1 \mathbb T^{({M^4})}{}_{\a\b} + {\rm c.c.} \Big)
\ ,
\eea
where
\bsubeq
\bea
\mathbb T^{({\rm SG})}{}_{\a\b} 
&=& - \frac{1}{2} {\mathscr C}_{\a\b} \ , \\
\mathbb T^{({M^{n+2}})}{}_{\a\b}
&=& - \frac{n + 1}{8} 
\big( [{\mathscr D}_{(\a} , \bar{\mathscr D}_{\b)}] + 4 {\mathscr C}_{\a\b}
\big) {\mathscr R}^n \ , \\
\mathbb T^{({\rm sca^2})}{}_{\a\b} 
&=& 
- \frac{1}{8} \big( [{\mathscr D}_{(\a} , \bar{\mathscr D}_{\b)}] + 4 {\mathscr C}_{\a\b}
\big) \Big( ({\mathscr D}^2 - 4 \bar {\mathscr R}) {\mathscr R} 
+ ({\bar {\mathscr D}}^2 - 4 {\mathscr R}) {\bar {\mathscr R}} 
-4 {\mathscr R} \bar {\mathscr R}\Big) \non\\
&&- 2 \big({\mathscr D}_{(\a} {\mathscr R} \big){\bar {\mathscr D}}_{\b)} \bar {\mathscr R}
 \ , \\
\mathbb T^{({\rm Ric^2})}{}_{\a\b} 
&=& 
\frac{\ri}{2} [{\mathscr D}^\d , {\bar {\mathscr D}}_\d] \mathscr W_{\a\b}
- \frac{1}{8} [{\mathscr D}_{(\a} , {\bar {\mathscr D}}_{\b)}] ({\mathscr D}^2 \mathscr R + {\bar {\mathscr D}}^2 {\bar {\mathscr R}}) 
+ \frac{3}{2} [{\mathscr D}_{(\a} , {\bar {\mathscr D}}_{\b)}] ({\mathscr C}^{\g\d} {\mathscr C}_{\g\d})
\non\\
&& 
+ \frac{5}{6} {\mathscr C}_{\a\b} ({\mathscr D}^2 {\mathscr R} + {\bar {\mathscr D}}^2 {\bar {\mathscr R}})
+ 2 \ri {\mathscr C}^{\g\d} ({\bar {\mathscr D}}_{(\a} {\mathscr C}_{\b\g\d)} + {{\mathscr D}}_{(\a} {\bar {\mathscr C}}_{\b\g\d)}) 
\non\\
&& 
+ \frac{2 \ri}{3} \big({\bar {\mathscr D}}^\g {\bar {\mathscr R}} \big){\mathscr C}_{\a\b\g}
 + \frac{2 \ri}{3} \big( {{\mathscr D}}^\g {{\mathscr R}}\big) {\bar {{\mathscr C}}}_{\a\b\g} 
 + \frac{20}{9} \big( {\mathscr D}_{(\a} {\mathscr R}\big) \bar{\mathscr D}_{\b)} \bar{\mathscr R} 
  \non\\
 &&
+ 2 {\mathscr C}_{\a\b} {\mathscr C}^{\g\d} {\mathscr C}_{\g\d} 
- 8 {\mathscr C}_{(\a}{}^{\g\d} {\bar {\mathscr C}}_{\b) \g\d} 
 - 8 {\mathscr C}_{\a\b} {\mathscr R} {\bar {\mathscr R}} \ .
 \eea
\esubeq
Here we have defined
\be \mathscr C_{\a\b\g} = - \ri \mathscr D_{(\a} \mathscr C_{\b\g)} \ .
\ee
One can check the supergravity equation of motion \eqref{111111} by making use of
the results for the deformation of the prepotential in appendix \ref{prepotentialN=2}, which imply
\be 
\d S_{\rm CSG} = \int \rd^{3|4} z E \d H^{\a\b} W_{\a\b}
~,~~~~~~\d \Big{[}S-\frac{1}{\tilde\mu}S_{\rm CSG}\Big{]} = \int \rd^{3|4} z E \d H^{\a\b} T_{\a\b}
\ .
\ee
The supercurrent $T_{\a\b}$ obeys the conservation equation
\be \nabla^\b T_{\a\b} = 0
\ee
when the compensator obeys its equation of motion \eqref{TypeIcompensatorEoM}.

The supergravity equations \eqref{TypeIcompensatorEoM} and \eqref{111111} are automatically 
super-Weyl invariant. Upon imposing the gauge \eqref{gaugeFixN=211} 
and reducing to conventional superspace, the local super-Weyl and U(1)$_R$ symmetries
 may be fixed by imposing the gauge
\be \Phi = 1 \ . \label{gaugeCondonlyTypeI}
\ee
This is equivalent to making the replacements $\mathscr D_A \rightarrow \cD_A$, $\mathscr C_{\a\b} \rightarrow \cC_{\a\b}$ 
and $\mathscr R \rightarrow \cR$ everywhere. The super-Weyl invariance can be restored by making the inverse replacement. 
In what follows we will assume the gauge condition \eqref{gaugeCondonlyTypeI}.

We are interested in  maximally supersymmetric solutions 
of the supergravity equations of motion. In type I supergravity
backgrounds, all maximally supersymmetric backgrounds \cite{KLRST-M,Kuzenko15}
are characterised by 
 dimension-1 torsion superfields under the following constraints
\bea
\cS=0~, \qquad \cR \,{\cC}_a
= 0 ~,
\qquad
{\cD}_A  {\cR}= 0~,
\qquad
{\cD}_{A} {\cC}_b=0~.
\label{5.211}
\eea
The complete algebra of covariant derivatives is
\begin{subequations} \label{4.37}
\bea
\{\cD_\a,\cD_\b\}
&=&
-4\bar{\cR}M_{\a\b}
~,~~~~~~
\\
\{\cD_\a,\cDB_\b\}
&=&
-2 \ri (\g^c)_{\a\b} \Big(  \cD_c
-\ri  \cC_{c} J \Big)
+4\ve_{\a\b}\cC^{c}M_{c}
~, \\
{[}\cD_{a},\cD_\b{]}
&=&
\ri\ve_{abc}(\g^b)_\b{}^{\g}\cC^c\cD_{\g}
-\ri(\g_a)_{\b\g}\bar{\cR}\cDB^{\g}
~,\\
{[}\cD_a,\cD_b]{}
&=&4  \ve_{abc}\Big(\cC^c \cC_d
+\d^c{}_d\bar{\cR}\cR 
\Big)M^d ~.
\eea
\end{subequations}

The equations of motion \eqref{TypeIcompensatorEoM} and \eqref{111111} 
simplify significantly for maximally supersymmetric backgrounds where 
we have the conditions
\be
\cC_{a} = 0 \ , \quad 
\cR = \cR_0 = {\rm const} \ . 
\ee
In this case the supercurrent and super-Cotton tensor vanish,
\be T_{\a\b} = W_{\a\b} = 0 \ ,
\ee
which means that \eqref{111111} is identically satisfied while the equation on the 
compensator reduces to
\be 0 = 4 \l 
- \frac{4}{\kappa} \cR_0
- 2 \mu_0 \cR_0^2
+ 2 \bar{\mu}_0 \cR_0 {\bar {\cR}_0}
- 5 \mu_1 \cR_0^3 + 3 \bar{\mu}_1 {\cR_0} \bar{\cR}_0^2
+ 4 \mu_2 \cR_0^2 {\bar {\cR}_0} \ .
\label{Rcubic_eq}
\ee

We now consider perturbations in the model \eqref{N=2_SG_model} about the 
maximally supersymmetric solution:
\be \cR = \cR_0 + {\D \cR} \ , \quad \cC_{\a\b} = \D \cC_{\a\b} \ .
\ee
The equation of motion on the compensator becomes
\bea
0 &=& \Big( - \frac{4}{\kappa} 
- 4 \mu_0 \cR_0
+ 2 \bar{\mu}_0 {\bar \cR}_0
- 15 \mu_1\cR_0^2
+ 3 {\bar \mu}_1 {\bar \cR}_0^2
+ 8 \mu_2 \cR_0 {\bar \cR}_0 \Big) \D \cR \non\\
&&+ \Big(
2 \cR_0 ( {\bar{\mu}}_0  + 3 {\bar\mu}_1 {\bar \cR}_0)
+ 4 \mu_2 \cR_0^2
\Big) \D {\bar \cR} \non\\
&&+ \Big(
- \hf ({\bar{\mu}}_0
+ 3 {\bar{\mu}}_1 {\bar \cR}_0 )
- \mu_2 \cR_0
\Big) {\bar \cD}^2 \D \bar \cR \non\\
&& + (\mu_2 - \mu_3) \cR_0 \cD^2 \D \cR 
- \frac{1}{4} (\mu_2 - \mu_3) \bar\cD^2 \cD^2 \D \cR \ . \label{typeIEoMperturb}
\eea
Thus we see that $\D \cR$ is propagating in general. However one can simplify the equation 
of motion \eqref{typeIEoMperturb} by turning it into an algebraic one by setting
\bsubeq
\bea \mu_2 &=& \mu_3  \  , \label{constTypeI-1} \\
0 &=& {\bar{\mu}}_0
+ 3 {\bar{\mu}}_1 {\bar \cR}_0 + 2  \mu_2 \cR_0 \label{constTypeI-2} \ .
\eea
\esubeq
The generic case is characterized by the condition
\be 
\frac{1}{\kappa}
+ \frac{3}{4} (\mu_1 \cR_0^2 + {\bar{\mu}}_1 {\bar \cR_0}^2)
- 3 \mu_2 |\cR_0|^2
\neq
0
\ ,
\ee
which requires $\D \cR = \D \bar \cR = 0$. Then the supergravity equation of motion reduces to
\bea
&-& \frac{\mu_3}{8} [\cD^\g , \bar{\cD}_\g][\cD^\d , {\bar \cD}_\d] \D \cC_{\a\b}
+ \frac{\ri}{4 \mu} [\cD^\g , {\bar \cD}_\g] \D \cC_{\a\b} \non\\
&+& \Big( \frac{1}{2 \kappa} 
- \frac{3}{2} \mu_1 \cR_0^2
- \frac{3}{2} {\bar\mu}_1 {\bar \cR}_0^2
\Big) \D \cC_{\a\b} = 0 \ , 
\eea
which gives
\be \mu_3 \cD^b \cD_b \D \cC_a
+ \frac{1}{2 \mu} \eps_{abc} \cD^b \D \cC^c
+ \Big( \frac{1}{4 \kappa} + 8 \mu_3 |\cR_0|^2 
- \frac{3}{4} (\mu_1 \cR_0^2 + \bar \mu_1{\bar \cR}_0^2)\Big) \D \cC_a = 0 \ .
\ee
By linearizing the Bianchi identity \eqref{BItypeIC} about the background chosen one can see that 
$\D \cC_{\a\b}$ is divergenceless $\cD^\b \D \cC_{\a\b} = 0$.

In general for a symmetric spinor $T_{\a_1 \cdots \a_n} = T_{(\a_1 \cdots \a_n)}$ 
that is divergenceless, 
\be \cD^\a T_{\a \a_1 \cdots \a_{n-1}} = 
0 \ ,
\ee
one can check the following identity holds in the background chosen
\be 
\cD^\b \cD^\g \bar\cD_\g T_{ \a_1 \cdots \a_{n-1} \b} 
= \bar \cD^\b \cD^\g \bar\cD_\g T_{ \a_1 \cdots \a_{n-1}\b} 
= 0 \ . \label{TypeIfactorisation}
\ee
This implies that the operator $\frac{\ri}{2} \cD^\g \bar\cD_\g$ preserves 
the divergenceless condition of the superfield $\D \cC_{\a\b}$.

In the case $\mu_3 \neq 0$ we may write the equation of motion in 
the following factorized form
\be 
\Big( \frac{\ri}{2} \cD^\g\bar \cD_\g + m_- \Big) \Big( \frac{\ri}{2} \cD^\d \bar\cD_\d - m_+ \Big) \D \cC_{\a\b} = 0 \ ,
\ee
where
\bsubeq
\bea m_+ - m_- &=& - \frac{1}{2 \mu \mu_3} \ , \\
m_+ m_- &=& - \frac{1}{4 \mu_3} \Big( \frac{1}{\kappa}
- 3 \mu_1 \cR_0^2 
- 3 \bar\mu_1 {\bar \cR}_0^2 \Big) \ .
\eea
\esubeq
The constants $m_+$ and $m_-$ are real for
\be \frac{1}{4 \mu^2\mu_3}
 \geq \Big(
 \frac{1}{\kappa}
 - 3 \mu_1 \cR_0^2
 - 3 \bar \mu_1 {\bar \cR}_0^2 \label{TypeISUGRAmassiveinequality}
 \Big) \ .
\ee

The supergravity model \eqref{N=2_SG_model} leads to massive supergravity 
for different choices of parameters.
For instance, we can see from eq. \eqref{TypeISUGRAmassiveinequality} that in a Minkowski background with $\mu_3 > 0$ 
and $\mu = \infty$ we must have either a 
negative Einstein-Hilbert term ($\kappa < 0$) or no Einstein Hilbert term ($\kappa \rightarrow \infty$) for massive supergravity. 

In the case where $\mu_3 = 0$ and $\mu$ is finite we have the equation
\be \Big( \frac{\ri}{2} \cD^\g \bar \cD_\g
- \hat{m} \Big) \D \cC_{\a\b} = 0 \ ,
\ee 
where $\hat{m}$ is given by
\be \hat{m} = - \mu \Big( \frac{1}{\kappa} - \frac{3}{2} \mu_1 \cR_0^2 - \frac{3}{2} \bar\mu_1 \bar \cR_0^2 \Big) \ .
\ee
When $\mu_0 = \mu_1 = 0$ we have topologically massive supergravity.

One should note that the supergravity model \eqref{N=2_SG_model} is more general than the one 
considered in \cite{BOS14} since the model contains 9 real free parameters.\footnote{The 
cosmological constant can be made to be real via a rescaling of the chiral compensator,
$\F(z) \to \re^{\ri \a} \F(z)$.} This leads to 
an important consequence. In contrast to \cite{BOS14} we have shown 
that we can eliminate the degrees of freedom associated with the torsion superfield $\cR$ in any AdS background 
satisfying eq. \eqref{Rcubic_eq} that otherwise propagates 
since $\cR$ is coupled to a product of propagating fields. Its elimination 
can be seen to coincide with removing the contribution from the $R (M+\bar M)^2$ and $R (M - \bar M)^2$ terms in the component action
to the linearized equation of motion. These terms 
contribute to highly non-linear interactions upon imposing the equation of motion on $M$. 
In \cite{BOS14} the 
$R (M+ \bar M)^2$ term 
was eliminated by a choice of constraints, which coincides with $\mu_0 = 0$, eq. \eqref{constTypeI-1} and 
$\mu_1 = - \frac{2}{3} \mu_2$. Imposing these constraints and expanding about a background with $\cR_0 = \bar\cR_0$ 
we see that \eqref{constTypeI-2} is identically satisfied and one recovers the factorisation \eqref{TypeIfactorisation}. However, 
it is important to note that the factorisation holds for the model defined by the constraints \eqref{constTypeI-1} 
and \eqref{constTypeI-2} about any AdS background
satisfying eq. \eqref{Rcubic_eq}.

It is worth mentioning that there exists other 
type I maximally supersymmetric backgrounds \cite{KLRST-M}
defined by the conditions
\be \cR = 0 \ , \quad \cC_a = {\rm const} \ .
\ee
In this case the equation on the compensator reduces to
\be \l = 0 \ ,
\ee
while the supergravity equation of motion fixes $\cC^2 := \cC^a \cC_a$ as 
follows
\be \cC^2 = \frac{1}{8 \kappa \mu_3} \ .
\ee
We do not discuss linearization about this background here. 


\section{$\cN = 2$ supergravity models with a real linear compensator} \label{section4}

The $\cN = 2$ conventional superspace formulation was presented 
in the previous section where it was also shown how to describe type I minimal supergravity 
with the use of a chiral compensator and its conjugate. Type II minimal supergravity, which
makes use of a real linear compensator, can be described similarly with 
conventional superspace. In this section we show how to do this 
and generalise the geometric framework to a superconformal setting.


\subsection{Type II minimal supergravity in conventional superspace}

Type II minimal supergravity makes use of a real linear 
compensator $G$. The compensator $G$ is defined to be nowhere vanishing 
and satisfy the following constraint
\be (\cD^2 - 4 \bar \cR) G 
= 0 
\ .
\ee
The superfield $G$ transforms homogeneously under super-Weyl transformations,
\be G' = \re^\s G \ . 
\ee

Since $G$ is nowhere vanishing the super-Weyl transformations permit us to choose 
a gauge where $G = 1$, which leads 
to the consistency condition
\be \cR = 
0 \ . \label{cR=0cond}
\ee
We may refer to the superspace subject to the above conditions as type II geometry. 
Supergravity models constructed with type II geometry 
are often referred to as $\cN = (2, 0)$ supergravity or type II supergravity. Imposing only the conditions 
\eqref{cR=0cond} and keeping in mind eq. \eqref{2.11f}, one can see that the residual
gauge transformations are generated by the superfield $\s$ subject to the constraint
\be \cD^2 \re^{- \s} 
= 0 \ .
\ee


\subsection{The superconformal setting}

Type II geometry can be used to describe type II supergravity and its matter couplings. However, as mentioned in previous 
sections, it is advantagous to make use of conformal superspace from the point of view of 
component reduction. The more general framework can always be reduced to conventional superspace. Below 
we elucidate the description of type II supergravity in this setting.

\subsubsection{The real linear compensator}

In conformal superspace, the real linear compensator $G$ is a primary nowhere vanishing scalar superfield of 
dimension 1,
\be 
\nabla^2 G 
= 0 \ , 
\quad \bbD G = G \ , \quad K_A G = 0 \ . \label{N=2linearConst}
\ee

The constraint \eqref{N=2linearConst} allows us to express $G$ in terms of a prepotential $\cV$ as follows:
\be 
G = \ri \nabla^\a \bar \nabla_\a \cV \label{N2prerep} \ ,
\ee
where $\cV$ possesses the gauge transformations 
\bea
\d \cV = \L +\bar \L~, \qquad \bar \nabla_\a \L =0~, 
\eea
with the gauge parameter $\L$ being an arbitrary covariantly chiral 
dimensionless scalar.

One can associate with $\cV$ a gauge one-form $V = E^A V_A$ describing the vector multiplet. 
Modulo an exact one-form, we can choose 
the components of $V$ as follows:
\be V_\a = \ri \nabla_\a \cV \ , \quad \bar{V}_\a = - \ri \bar{\nabla}_\a \cV \ , \quad V_a = - \frac{1}{2} (\g_a)^{\a \b} [\nabla_\a , \bar{\nabla}_\b] \cV \ .
\ee
The corresponding gauge-invariant field strength is simply given by $F = \rd V$. 
In the complex basis the field strength is
\bea \label{4.9}
 F = \bar{E}^\b \wedge E^\a  F_{\a \b} + E^\b \wedge E^a  F_{a \b} + \bar{E}^\b \wedge E^a \bar{ F}_{a \b} + \hf E^b \wedge E^a  F_{ab}~, 
\eea 
where
\bsubeq \label{4.10}
\bea
 F_{\a \b} &=& - 2 \eps_{\a\b}  G \ , \\
 F_{a \b} &=& \ri (\g_a)_\b{}^\g \nabla_\g  G \ , \\
\bar{ F}_{a \b} &=& - \ri (\g_a)_\b{}^\g \bar{\nabla}_\g  G \ , \\
 F_{ab} &=& \frac{1}{4} \eps_{abc} (\g^c)^{\g\d} [\nabla_\g , \bar{\nabla}_\d]  G~.
\eea
\esubeq

One can use $G$ to construct dimensionless covariant derivatives that preserve the primary 
property of superfields. They are given by
\bsubeq \label{4.111}
\bea \mathscr D_\a &=& G^{- \hf} \Big( \nabla_\a - (\nabla^\b \ln G) M_{\a\b} 
+ (\nabla_\a \ln G) {J} - (\nabla_\a \ln G) \mathbb D \Big) \ , \\
\mathscr D_a &=& - \frac{\ri}{4} (\g_a)^{\a\b} \{ \mathscr D_\a , \bar{\mathscr D}_\b\} 
+ \ri \mathscr C_a J+ 2 \mathscr S M_a
\ ,
\eea
\esubeq
where we have defined
\bsubeq \label{compositesN=2}
\bea
\mathscr S &:=& - \frac{\ri}{4 G} \nabla^\g \bar \nabla_\g \ln G \ ,  \label{ScompositeN=2} \\
\mathscr C_{\a\b} &:=& - \frac{1}{4} [\nabla_{(\a} , \bar \nabla_{\b)}] \frac{1}{G} \label{CcompositeN=2} \ .
\eea
\esubeq
One can check that $G$ is covariantly constant with respect to $\mathscr D_A$. Furthermore, on primary superfields the 
covariant derivatives $\mathscr D_A$ satisfy the algebra
\bsubeq \label{algebraGcovConf-Super}
\bea \{ \mathscr D_\a , \mathscr D_\b \} &=& 0 \ , 
\\
\{ \mathscr D_\a , \bar {\mathscr D}_\b \} &=& - 2 \ri \mathscr D_{\a\b}
- 4 \ri \eps_{\a\b} \mathscr S {J}
+ 4 \ri \mathscr S M_{\a\b}
- 2 \eps_{\a\b} \mathscr C^{\g\d} M_{\g\d} \ ,
\eea
\esubeq
which formally coincides with the algebra \eqref{algebra-final} with $\cR = \bar \cR = 0$.

General supergravity invariants may be realized in the form
\be 
S =  \int \rd^{3|4} z \, E \, G
\frak{L}(G,\mathscr T , \mathscr D \mathscr T  , \mathscr D^2 \mathscr T  , \cdots )
\ , \label{generalTypeIIaction}
\ee
where $\frak{L}$ is a dimensionless superfield constructed out of the torsion components $\mathscr T$,
their covariant derivatives and the compensator $G$.

We can relate the superconformal framework to the one of conventional superspace 
by gauge fixing the additional symmetries. We can use the conformal boosts and $S$-supersymmetry
transformations to impose the gauge condition
\be B_A = 0 \ , \label{gaugeFixN=21}
\ee
which degauges conformal superspace to conventional superspace. 
The composites \eqref{compositesN=2} become 
the following super-Weyl invariant objects
\bsubeq \label{S+CcompositeN=2}
\bea
\mathscr S &:=&  - \frac{\ri}{4 G} \Big( \cD^\g \bar \cD_\g \ln G + 4 \ri \cS \Big) \ , \\
\mathscr C_{\a\b} &:=& - \frac{1}{4} \Big([\cD_{(\a} , \bar \cD_{\b)}] - 4 \cC_{\a\b} \Big) \frac{1}{G} \ .
\eea
\esubeq
Upon imposing the additional gauge condition
\be \label{gaugeFixN=22}
 G = 1 \ ,
 \ee
the composites $\mathscr C_{\a\b}$ and $\mathscr S$ can be seen to 
coincide with the torsion components $\cC_{\a\b}$ and $\cS$, respectively.

It is worth mentioning that one can use the compensator to construct super-Weyl 
invariant covariant derivatives $\cD_A^{^{(G)}}$ in conventional superspace as follows:
\bsubeq
\bea
\cD_\a^{^{(G)}} &=& 
G^{- \hf} \Big( \cD_\a 
- (\cD^\b \ln G) M_{\a\b} 
+ (\cD_\a \ln G) {J} \Big)
\ , \\
\cD_a^{^{(G)}} &=& - \frac{\ri}{4} (\g_a)^{\a\b} \{ \cD^{(G)}_\a , {\bar{\cD}}^{(G)}_\b\} 
+ \ri \mathscr C_a J+ 2 \mathscr S M_a \ ,
\eea
\esubeq
where $\mathscr C_a$ and $\mathscr S$ are given by eqs. \eqref{S+CcompositeN=2}. 
It can be checked that these covariant derivatives satisfy the same algebra as \eqref{algebraGcovConf-Super}.
Unlike the operators $ \mathscr D_A$, eq.  \eqref{4.111}, 
they do not annihilate the compensator $G$.

It is worth noting that $G \mathscr S$ turns out to 
be proportional to the composite linear multiplet\footnote{This composite 
first appeared explicitly in conventional superspace in \cite{KLRST-M}.}
\be \mathbb G := \ri \nabla^\g \bar \nabla_\g \ln \Big( \frac{G}{\Phi \bar \Phi} \Big) 
= \ri \nabla^\g \bar \nabla_\g \ln G  \ ,
\ee
where $\Phi$ is an arbitrary dimension 1/2 chiral superfield. The composite 
$\mathbb G$ will be useful in the construction of supergravity invariants.

It should be mentioned that one can 
construct other composite multiplets by simply choosing the prepotential $\cV$ of a linear 
multiplet to be built out of $G$. 
It is also possible to engineer composite linear multiplets with the use of a number of real 
linear multiplets $G^{\rm I}$.\footnote{Models with a number of real linear multiplets 
were considered in \cite{KT-M11}.} 
For example, we can construct the following composite linear superfields:
\bea \mathbb G_\cA &=& \ri \nabla^\g \bar \nabla_\g \cA(G^{\rm I}) \ , \quad 
\mathbb G_\cB = \ri \nabla^\g \bar \nabla_\g \ln \Big{[} \frac{\cB(G^{\rm I})}{\Phi \bar \Phi} \Big{]} \ ,
\eea
where $\cA$ is a real homogeneous function of $G^{\rm I}$  of degree zero and $\cB$ is a real homogeneous function of $G^{\rm I}$ of degree 1,\footnote{The composite 
vector multiplets constructed in \cite{BOS14} coincide with $\cB = C_{\rm IJ} G^{\rm I} G^{\rm J}$ with $C_{\rm IJ}$ a 
homogeneous function of $G^{\rm I}$ of degree -1, $G^{\rm I} \frac{\partial}{\partial G^{\rm I}} C_{\rm JK}=-C_{\rm JK}$.}
\be G^{\rm I} \frac{\partial}{\partial G^{\rm I}} \cA = 0 \ , \quad G^{\rm I} \frac{\partial}{\partial G^{\rm I}} \cB = \cB \ .
\ee
We will not make use of the composites $\mathbb G_\cA$ and $\mathbb G_\cB$ in what follows.


\subsubsection{The $BF$ action}

One can perform component reduction of superspace integrals 
by reducing to a chiral subspace and making use of the component reduction 
formula \eqref{compRedActionFromulaN=2}. However, 
many of our type II supergravity invariants can be conveniently rewritten as a $BF$ term 
for a composite linear multiplet. 
The locally supersymmetric $BF$ action can be written as
\bea
S_{BF}&=&
\int\rd^{3|4} z \,E \,\cV\, \bm G
~.
\label{N=2BF}
\eea
Here $\cV=\bar \cV$ is the gauge prepotential of an Abelian vector multiplet, 
and $\bm G$ is a real linear superfield. The $BF$ action reduces in components to \cite{KN14}
\begin{align} \label{BFN=2}
S_{BF} =& -\frac{1}{8} \int \rd^3 x \,e \, \Big(
	\eps^{abc} v_a \bm f_{bc} 
	+ \frac{\ri}{2} \l^\g \bar{\bm \l}_\g + \frac{\ri}{2} \bm \l^\g \bar{\l}_\g
	+ g \bm h + \bm g h \non\\
	&- \hf (\g^a)_{\g\d} \psi_a{}^\g (g \bm \l^\d +  \bm g \l^\d) + \hf (\g^a)_{\g\d} \bar{\psi}_a{}^\g (g \bar{\bm \l}^\d + \bm g \bar{\l}^\d) \non\\
	&- \ri \eps^{abc} (\g_a)_{\g\d} \psi_b{}^\g \bar{\psi}_c{}^{\d} g \bm g  \Big) \ ,
\end{align}
where the component fields are defined by\footnote{The supersymmetry transformations 
of the component fields were given in \cite{KN14}.}
\bsubeq \label{compsN=(2,0)}
\bea
g &:=& G| \ , \quad \l_\a :=  -2 \nabla_\a G| \ , \quad \bar{\l}_{\a} :=  -2 \bar{\nabla}_\a G| \ , \quad h := \ri \bar{\nabla}^\g \nabla_\g G| \ , \\
v_a &:=& e_a{}^m V_m| = e_m{}^a V_a| + \hf \psi_m{}^\a V_\a| + \hf \bar \psi_m{}_\a \bar V^{\a}| \ .
\eea
The component field strength can be constructed as follows
\bea
f_{ab} &:=& F_{ab}| - \psi_{[a}{}^\b F_{b] \b}| - \bar{\psi}_{[a}{}^\b \bar{F}_{b] \b}| - \hf \psi_{[a}{}^\a \bar{\psi}_{b]}{}^\b F_{\a\b} \non\\
&=& \frac{1}{4} \eps_{abc} (\g^c)^{\g\d} [\nabla_\g , \bar{\nabla}_\d]  G| + \frac{\ri}{2} \psi_{[a}{}^\b (\g_{b]})_\b{}^\g {\l}_\g \non\\
&&- \frac{\ri}{2} \bar{\psi}_{[a}{}^\b (\g_{b]})_\b{}^\g \bar{\l}_\g + \psi_{[a}{}^\a \bar{\psi}_{b] \a} g \ .
\eea
\esubeq
The same definitions hold for the component fields of $\bm G$. 
The component fields of the Weyl multiplet are defined as in the type I case.

To fix additional symmetries in our invariants one can make use of the gauge conditions \eqref{gaugeFixN=21}
and \eqref{gaugeFixN=22}, 
which leads to the following gauge conditions at the component level
\be g = 1 \ , \quad \l_\a
= 0 \ , \quad b_m = 0 \ . \label{compTypeIIgauge}
\ee
The first fixes the dilatations, the second fixes the $S$-supersymmetry transformations and the last fixes 
the special conformal boosts. To see what the invariants we construct correspond to at the component level 
we give the following useful results in the gauge \eqref{compTypeIIgauge}  (compare with \eqref{compBoxResultsN=1}):
\bsubeq \label{usefulResultsTypeIIcomp}
\bea \Box g &=& \frac{1}{4} {R} + {\rm fermion \ terms} \ , \\
\Box^2 g &=& {R}^{ab} {R}_{ab} - \frac{1}{8} {R}^2 
+\frac{1}{4}{D}^a{D}_a{R}
+ {\rm fermion \ terms} \ ,
\eea
\esubeq
where we have defined
\bsubeq
\bea
\Box g &:=& \nabla^a \nabla_a G| \ , \\
\Box^2 g &:=& \nabla^a \nabla_a \nabla^b \nabla_b G|  \ .
\eea
\esubeq


\subsection{Supergravity invariants} \label{subsection4.3TypeII}

We will write down actions for various supergravity models by constructing a superspace Lagrangian 
built out of $G$, the composites \eqref{ScompositeN=2} and \eqref{CcompositeN=2}, and 
their $\mathscr D$-covariant derivatives.


\subsubsection{The supergravity action}

The type II minimal supergravity action with a cosmological term was given in \cite{KLRST-M}
in conventional superspace. In conformal superspace it is given by
\be S = \frac{1}{\kappa} S_{\rm SG} + \l S_{\rm cos} \ ,
\ee
where
\bsubeq
\bea
S_{\rm SG} &=&
4\int\rd^{3|4}z \,E \, G \ln \Big( \frac{G}{\Phi \bar \Phi} \Big) \ , \\
S_{\rm cos} &=& 4 \int\rd^{3|4}z \,E \, \cV G
\eea
\esubeq
and $\l$ is the cosmological constant. 
Integrating by parts leads to the following equivalent form for 
$S_{\rm SG}$:
\bea
S_{\rm SG} &=& 
4\int\rd^{3|4}z \,E \, \cV \mathbb G 
= - 16 \int\rd^{3|4} z \,E \, \cV G \mathscr S \ .
\eea
The corresponding component action may be derived by putting $\bm G \rightarrow \mathbb G$ into the $BF$ action \eqref{BFN=2}. 
It is straighforward to show that the component field $\bm h$ contains a term proportional to $\frac{1}{g} \Box g$. Making use of the 
gauge conditions \eqref{compTypeIIgauge}
and the results \eqref{usefulResultsTypeIIcomp}, one can see that it gives rise
to the Einstein-Hilbert term in the component action. The cosmological term comes from the 
U(1) Chern-Simons term described by the invariant $S_{\rm cos}$. The full component action 
for supergravity with a cosmological term was analysed in detail in \cite{KLRST-M}.


\subsubsection{The ${R} h^n$ invariants}

One can construct an invariant containing a ${R} h^n$ term with $n > 1$ as follows
\bea
S_{{R} h^n} &=&\int\rd^{3|4}z \,E \,  \Big( \frac{\mathbb G}{G} \Big)^{n} G 
= \int \rd^{3|4}z \,E \,   
\Big( \frac{\mathbb G}{G} \Big)^{n-1} \mathbb G \non\\
&=& (- 4)^n \int\rd^{3|4}z \,E \, G \mathscr S^n  \ .
\label{Rhn_0}
\eea
Upon integrating by parts one finds the equivalent forms
\bea
S_{{R} h^n} &=&\int\rd^{3|4} z \,E \,  \cV \mathbb G_{n} 
= \int\rd^{3|4}z \,E \,   \mathbb G_{n-1} \ln \Big( \frac{G}{\Phi \bar \Phi} \Big) \ ,
\eea
where
\be \mathbb G_n = \ri \nabla^\g \bar \nabla_\g \Big( \frac{\mathbb G}{G} \Big)^n \ .
\ee
The component form of the above action can be obtained from the $BF$ action \eqref{BFN=2} by putting 
for instance
\be 
\bm G \rightarrow \mathbb G_{n}
\ee
into the $BF$ action \eqref{BFN=2}.
It is straightforward to check that the component action contains the term
\bea
\frac{3n(n-1)}{16}
{R} h^{n} 
\eea
upon imposing the gauge conditions \eqref{compTypeIIgauge}.

For $n = 1$ the invariant \eqref{Rhn_0} vanishes and so we have to consider the $n=1$ case separately.
A locally supersymmetric invariant containing a $Rh$ term is described by
\bea
S_{{R} h} &=&
-4\int\rd^{3|4}z \,E \, \mathbb G \ln \Big( \frac{G}{\Phi \bar \Phi} \Big) 
= 16 \int\rd^{3|4}z \,E \, G \mathscr S \ln \Big( \frac{G}{\Phi \bar \Phi} \Big)\ .
\eea
The component action can be worked out by putting $G \rightarrow \mathbb G$ and $\bm G \rightarrow \mathbb G$ 
into the $BF$ action \eqref{BFN=2}. It gives rise to a term proportional to $\frac{1}{g^2} h \Box g$, which upon gauge fixing 
leads to the ${R} h$ term in the component action. The bosonic action was explictly given in \cite{BOS14}.

It is important to note that in the $n = 2$ case 
the action also contains a scalar curvature squared term. However, as was discussed in \cite{BOS14}, an independent invariant containing 
a curvature squared term is not known to exist. This can be attributed to
the fact that only a real scalar composite $\mathscr S$ can be constructed from the linear multiplet, while for the 
type I case one can construct a complex scalar composite $\mathscr R$ leading to an extra invariant.


\subsubsection{The Ricci curvature squared invariant}

An invariant containing a Ricci squared term is given by
\be S_{\rm Ricci^2} = 4 \int\rd^{3|4}z \,E \, G \mathscr C^{\a\b} \mathscr C_{\a\b} \ . \label{TypeIIRicci2first}
\ee
The action also contains a scalar curvature squared term contribution. This can be checked by using the 
results of \cite{KLRST-M}. It is natural to introduce the one parameter family of invariants
\be
S_\z = 4 \int\rd^{3|4}z \,E \, G (\mathscr C^{\a\b} \mathscr C_{\a\b} + \z \mathscr S^2) \ ,
\ee
where $\z$ parametrizes the scalar curvature squared contribution.

The invariant $S_{\rm Ricci^2} = S_0$ can be seen to correspond to the one given in \cite{BOS14} 
with the gauge conditions \eqref{gaugeFixN=21} and \eqref{gaugeFixN=22}. 
In this gauge we find $\cR = 
0$ and 
$\cC_{\a\b}$ satisfies a constraint 
reminiscent of a $\cN = 2$ Yang-Mills multiplet
\be \cD^2 \cC_{\a\b} 
= 0 \ .
\ee
The above constraints mean that the supersymmetry transformations of $\cC_{\a\b}$ 
can be put into one-to-one correspondence with a Yang-Mills multiplet. The action 
with the gauge conditions \eqref{gaugeFixN=21} and \eqref{gaugeFixN=22} reads
\be S_{\rm Ricci^2} = 4 \int\rd^{3|4}z \,E \, \cC^{\a\b} \cC_{\a\b} \ .
\ee
However, we can identity this action up to some multiplicative constant 
as a special case of the Yang-Mills action
\be
\int\rd^{3|4}z \,E \, \tr \ \cG^2 \ ,
\ee
where $\cG = \cC^{\a\b} M_{\a\b}$ and we are tracing over the Lorentz group. Therefore we can equally construct the 
action in the gauge $G = 1$ using the correspondence with the Yang-Mills multiplet. This provides a geometric 
explanation for the procedure employed in \cite{BOS14} at the component level. However, in our approach it is 
not necessary to work in the gauge \eqref{compTypeIIgauge} since one can just use the action \eqref{TypeIIRicci2first}.

Finally, It is worth mentioning that a pure Ricci curvature squared invariant is given by
\bea
S_{\rm Ricci^2}^{(\rm pure)}
=
S_{-40/3}
= 
4\int\rd^{3|4}z \,E \, G\Big(
\mathscr C^{\a\b}\mathscr  C_{\a\b}
-\frac{40}{3}\mathscr S^2
\Big) \ .
\eea
In the gauge \eqref{compTypeIIgauge} it gives rise to a 
${R}^{ab}{R}_{ab}$ term in the component action.

It should be mentioned that although we have restricted our attention to curvature squared terms. Similarly to the type I case, 
one can always consider higher derivative and locally supersymmetric actions by considering 
other instances of the action \eqref{generalTypeIIaction}. One should mention the following important 
descendent of the 
primary composites \eqref{compositesN=2}:
\be \mathscr W_{\a\b} = - \frac{\ri}{4} [\mathscr D^\g , \bar{\mathscr D}_\g] \mathscr C_{\a\b} 
+ \hf [\mathscr D_{(\a} , \bar{\mathscr D}_{\b)}] \mathscr S
+ 2 \mathscr S \mathscr C_{\a\b} \ .
\ee
It is related to the super-Cotton tensor $W_{\a\b}$ as
\be W_{\a\b} = G^2 \mathscr W_{\a\b} \ .
\ee
It should be kept in mind that the super-Cotton tensor is actually independent of the compensator $G$.


\subsection{Models for massive supergravity}

In this section, in analogy to type I supergravity, we analyse the equations of motion for a general 
supergravity model and determine the conditions in which we have massive theories of supergravity.

We consider the following type II supergravity model
\be
S =
 \l S_{\rm cos} 
 + \frac{1}{\kappa} S_{\rm SG} 
+ \frac{1}{\tilde{\mu}} S_{\rm CSG}
+ \mu_1 S_{{R} h} 
+ \mu_2 S_{{R} h^2}
+ \mu_3 S_{\rm Ricci^2} 
~,
\label{N=2_SG_model-II}
\ee
where $\kappa$, $\tilde{\mu}$, $\mu_1$, $\mu_2$ and $\mu_3$ are real, and we
make use of the invariants defined in subsection \ref{subsection4.3TypeII}. Here $S_{\rm CSG}$ denotes the 
$\cN = 2$ conformal supergravity action given in our conventions in \cite{BKNT-M2}. Here we assume 
$\mu_3 \geq 0$ as in \cite{BOS14}.

The equations of motion corresponding to the theory with action \eqref{N=2_SG_model-II} can 
be derived by varying the action with respect to the prepotential $\cV$ of the real linear compensator $G$. 
One finds the equation of motion on the compensator to be
\bea \label{TypeIIcompensatorEoM}
0&=&
\ri{\mathscr D}^\a{\bar{\mathscr D}}_\a\Big{[}
8\mu_1 \mathscr S
-2\mu_2\Big(\ri {\mathscr D}^\b {\bar{\mathscr D}}_\b\mathscr S +2\ri  {\mathscr S}^2\Big)
+ \mu_3\Big(2\ri{\mathscr D}^\b{\bar{\mathscr D}}_\b \mathscr S-{\mathscr C}^{\a\b}{\mathscr C}_{\a\b}\Big)\Big{]}
\non\\
&&
+2 \l
-\frac{4}{\kappa}\mathscr S
~.~~~~~~
\eea

The equation of motion for the conformal supergravity prepotential is
\bea
\frac{1}{\mu} W_{\a\b} + T_{\a\b} = 0 \ ,
\label{111112}
\eea
where $\mu$ is related to $\tilde{\mu}$ by a multiplicative constant. 
One can find the supergravity equation of motion \eqref{111112} by making use of
the results for the deformation of the prepotential in appendix \ref{prepotentialN=2}, which imply
\be 
\d S_{\rm CSG} = \int \rd^{3|4} z E \d H^{\a\b} W_{\a\b}
~,~~~~~~\d \Big{[}S-\frac{1}{\tilde\mu}S_{\rm CSG}\Big{]} = \int \rd^{3|4} z E \d H^{\a\b} T_{\a\b}
\ .
\ee
It can be checked that the supercurrent 
$T_{\a\b}$ is given by
\be T_{\a\b} = G^2 \mathbb T_{\a\b} \ ,
\ee
where
\be 
\mathbb T_{\a\b} =
 \l \mathbb T^{(\rm cos)}{}_{\a\b}
 + \frac{1}{\kappa}\mathbb  T^{(\rm SG)} {}_{\a\b}
+ \mu_1 \mathbb T^{({\rm R} h)}{}_{\a\b}
+ \mu_2 \mathbb T^{({\rm R} h^2)}{}_{\a\b}
+ \mu_3 \mathbb T^{(\rm Ricci^2)}{}_{\a\b}
~,
\ee
and
\bsubeq
\bea
\mathbb T^{(\rm cos)}{}_{\a\b} 
&=&
2{[} {\mathscr D}_{(\a},{\bar{\mathscr D}}_{\b)}{]}  \cV
~,
\\
\mathbb T^{(\rm SG)}{}_{\a\b} 
&=&
2 {\mathscr C}_{\a\b}
-4{\mathscr S}{[} {\mathscr D}_{(\a},{\bar{\mathscr D}}_{\b)}{]}  \cV
~,
\\
\mathbb T^{(\cR h)}{}_{\a\b}
&=&
4\big({[}{\mathscr D}_{(\a},{\bar{\mathscr D}}_{\b)}{]} +4{\mathscr C}_{\a\b}\big){\mathscr S}
+8([{\mathscr D}_{(\a} , \bar{\mathscr D}_{\b)}] \cV)   \ri{\mathscr D}^\g{\bar{\mathscr D}}_\g{\mathscr S}
~,
\\
\mathbb T^{(\cR h^n)}{}_{\a\b} 
&=&
-(-4)^{n}\Big\{\,
 \frac{n}{32} \big( {[}{\mathscr D}_{(\a},  {\bar{\mathscr D}}_{\b)}{]}+4{\mathscr C}_{\a\b}\big)
 \ri{\mathscr D}^\g {\bar{\mathscr D}}_\g {\mathscr S}^{(n-1)} 
\non\\
&&~~~~~~~~~~~~~
+\frac{n(n-1)}{2}{\mathscr S}^{n-2} ({\mathscr D}_{(\a}{\mathscr S}){\bar{\mathscr D}}_{\b)}{\mathscr S}
\Big\}
\non\\
&&
-\frac{(-4)^{n}}{16}
   ([{\mathscr D}_{(\a} ,{\bar{\mathscr D}}_{\b)}] \cV)
\ri{\mathscr D}^\g {\bar{\mathscr D}}_\g\Big(
n\ri{\mathscr D}^\g {\bar{\mathscr D}}_\g{\mathscr S}^{(n-1)}
+4(n-1) {\mathscr S}^{n} 
\Big)
~,~~~~~~~~~
\\
\mathbb T^{(\rm Ricci^2)}{}_{\a\b} &=&
 - \frac{1}{4} {[}{\mathscr D}^\g, {\bar{\mathscr D}}_\g{]} {[}{\mathscr D}^\d, {\bar{\mathscr D}}_\d{]}{\mathscr C}_{\a\b}
- 4  {\mathscr D}^a {\mathscr D}_a{\mathscr C}_{\a\b}
+ \ri  {\mathscr D}_{(\a}{}^{\g}{[}{\mathscr D}^\d, {\bar{\mathscr D}}_{|\d|}{]}{\mathscr C}_{\b)\g}
\non\\
&&
+24 \bar{{\mathscr C}}_{\a\b\g}{\mathscr D}^{\g}{\mathscr S}
- 24 {\mathscr C}_{\a\b\g}{\bar{\mathscr D}}^{\g}{\mathscr S}
+  {[}{\mathscr D}_{(\a},{\bar{\mathscr D}}_{\b)}{]}{\mathscr C}^{\g\d}{\mathscr C}_{\g\d} 
\non\\
&&
-\frac{16}{3}  {\mathscr C}_{\a\b}\,\ri{\mathscr D}^\g {\bar{\mathscr D}}_\g {\mathscr S}
+ 4 {\mathscr C}^{\g\d}\big(
\ri{\bar{\mathscr D}}_{(\a}{\mathscr C}_{\b\g\d)}
+\ri{\mathscr D}_{(\a} \bar{{\mathscr C}}_{\b\g\d)}
\big)
+ 8{\mathscr C}_{(\a}{}^{\g}{\mathscr D}_{\b)\g}{\mathscr S}
\non\\
&&
- 4 \ri{\mathscr S}{[}{\mathscr D}^\g,{\bar{\mathscr D}}_\g{]}{\mathscr C}_{\a\b}
- 8 {\mathscr S}\ve^{cab}(\g_c)_{\a\b}{\mathscr D}_a{\mathscr C}_b
- 32 {\mathscr S}^2{\mathscr C}_{\a\b}
\non\\
&&
+ ([{\mathscr D}_{(\a} ,{\bar{\mathscr D}}_{\b)}] \cV)\,\ri{\mathscr D}^\g {\bar{\mathscr D}}_\g
\Big(
2\ri{\mathscr D}^\g{\bar{\mathscr D}}_{\g}{\mathscr S}
-{\mathscr C}_{\a\b}{\mathscr C}^{\a\b}
\Big)
~.
\eea
\esubeq
Making use of the compensator equation of motion, the above expression 
becomes
\bea \label{TypeIISUGRAEoMs}
0&=&
\frac{1}{\mu} {\mathscr W}_{\a\b} 
 + \frac{2}{\kappa} {\mathscr C}_{\a\b}
+4 \mu_1\big({[}{\mathscr D}_{(\a},{\bar{\mathscr D}}_{\b)}{]} +4{\mathscr C}_{\a\b}\big){\mathscr S}
\non\\
&&
-\mu_2\Big{[}
\big( {[}{\mathscr D}_{(\a},  {\bar{\mathscr D}}_{\b)}{]}+4{\mathscr C}_{\a\b}\big)
 \ri {\mathscr D}^\g {\bar{\mathscr D}}_\g {\mathscr S} 
+16 ({\mathscr D}_{(\a}{\mathscr S}){\bar{\mathscr D}}_{\b)}{\mathscr S}
\Big{]}
\non\\
&&
- \frac{\mu_3}{4} 
\Big{[}\,
 {[}{\mathscr D}^\g, {\bar{\mathscr D}}_\g{]} {[}{\mathscr D}^\d, {\bar{\mathscr D}}_\d{]}{\mathscr C}_{\a\b}
+16  {\mathscr D}^a {\mathscr D}_a{\mathscr C}_{\a\b}
-4\ri  {\mathscr D}_{(\a}{}^{\g}{[}{\mathscr D}^\d, {\bar{\mathscr D}}_{|\d|}{]}{\mathscr C}_{\b)\g}
\non\\
&&~~~~~~
-96\bar{{\mathscr C}}_{\a\b\g}{\mathscr D}^{\g}{\mathscr S}
+96{\mathscr C}_{\a\b\g}{\bar{\mathscr D}}^{\g}{\mathscr S}
-4 {[}{\mathscr D}_{(\a},{\bar{\mathscr D}}_{\b)}{]}{\mathscr C}^{\g\d}{\mathscr C}_{\g\d} 
\non\\
&&~~~~~~
+\frac{64}{3}  {\mathscr C}_{\a\b}\,\ri{\mathscr D}^\g {\bar{\mathscr D}}_\g {\mathscr S}
-16{\mathscr C}^{\g\d}\big(
\ri{\bar{\mathscr D}}_{(\a}{\mathscr C}_{\b\g\d)}
+\ri{\mathscr D}_{(\a} \bar{{\mathscr C}}_{\b\g\d)}
\big)
-32{\mathscr C}_{(\a}{}^{\g}{\mathscr D}_{\b)\g}{\mathscr S}
\non\\
&&~~~~~~
+16\ri{\mathscr S}{[}{\mathscr D}^\g,{\bar{\mathscr D}}_\g{]}{\mathscr C}_{\a\b}
+32{\mathscr S}\ve^{cab}(\g_c)_{\a\b}{\mathscr D}_a{\mathscr C}_b
+128{\mathscr S}^2{\mathscr C}_{\a\b}
\Big{]}
~.
\eea
The supercurrent $T_{\a\b}$ obeys the conservation equation
\be \nabla^\b T_{\a\b} = 0
\ee
when the compensator obeys its equation of motion \eqref{TypeIIcompensatorEoM}.

We are interested in  maximally supersymmetric solutions 
of the supergravity equations of motion. In type II supergravity
backgrounds, all maximally supersymmetric backgrounds \cite{KLRST-M,Kuzenko15}
are characterised by 
 dimension-1 torsion superfields under the following constraints 
 \be \cR = 0 \ , \quad \cD_A \cS = 0 \ , \quad \cD_\a \cC_b = 0
 \quad \implies \quad \cD_a \cC_b = 2 \eps_{abc} \cC^c \cS \ , \quad \cC^a \cC_a = {\rm const} \ . \label{genCondTypeIIback}
 \ee
 The corresponding algebra of covariant derivatives is
\bsubeq \label{4.544}
\bea
\{ \cD_\a  , \cD_\b\} &=& 0 \ , \\
\{ \cD_\a , \bar\cD_{\b} \} &=& - 2 \ri (\g^c)_{\a\b} \big( \cD_c - 2 \cS M_c - \ri \cC_c J \big)
+ 4 \eps_{\a\b} \Big( \cC^c M_c - \ri \cS J \Big) \ , \\
\left[ \cD_a , \cD_\b \right] &=& \ri \eps_{abc} (\g^b)_\b{}^\g \cC^c \cD_\g
+ (\g_a)_\b{}^\g \cS \cD_\g \ , \\
\left[ \cD_a , \cD_b \right]
&=& 4 \eps_{abc} \big(
\cC^c \cC_d + \d^c_d \cS^2
\big) M^d \ .
\eea
\esubeq

The equations of motion \eqref{TypeIIcompensatorEoM} and \eqref{TypeIISUGRAEoMs}
simplify significantly for maximally supersymmetric backgrounds where 
we have the conditions
\be
\cC_{a} = 0 \ , \quad 
\cS = \cS_0 = {\rm const} \ . 
\ee
In this case the equation on the compensator reduces to
\bea
\cS_0=\frac{\kappa}{2}\l
~,
\eea
and the supercurrent vanishes,
\be T_{\a\b} = W_{\a\b} = 0 \ .
\ee

The supergravity equations of motion are by construction 
super-Weyl invariant. One can fix this super-Weyl invariance by 
imposing the gauge
\be 
G= 1 \ .
\ee
Keep in mind that the super-Weyl invariance can be restored by replacing
$\cD_\a \rightarrow \mathscr D_\a$, 
$\cC_{a} \rightarrow \mathscr C_a$ and 
$\cS \rightarrow \mathscr S$ everywhere.
We will assume the above gauge condition in what follows.

We consider perturbations in the model \eqref{N=2_SG_model-II} about the 
maximally supersymmetric solution:
\be 
\cS = \cS_0 + {\D\cS} \ , \quad \cC_{\a\b} = {\D \cC}_{\a\b} \ .
\ee
The equation of motion on the compensator becomes
\bea
0&=&
\frac{2}{\kappa}{\D \cS}
- 4 \big(\mu_1-\mu_2 \cS_0\big)\ri\cD^\a\cDB_\a{\D \cS}
+ \big(\mu_2 - \mu_3\big)(\ri\cD^\a\cDB_\a)^2 {\D \cS}
 \ . \label{typeIIEoMperturb}
\eea
Thus we see that $\D \cS$ is propagating in general. However one can simplify the equation 
of motion \eqref{typeIIEoMperturb} by turning it into an algebraic one by setting
\bea
\mu_1=\mu_2 \cS_0
~,~~~~~~
\mu_3= \mu_2
~.
\eea
In this case, if $\frac{1}{\kappa} \neq 0$ we must require $\D\cS = 0$. 

By linearizing the Bianchi identity \eqref{BItypeIC} about the background chosen 
one can show that 
$\D \cC_{\a\b}$ is divergenceless, $\cD^\b \D \cC_{\a\b} = 0$. Using this condition one 
can write the supergravity equation of motion in the form
\bea \label{TypeIIbackgroundSUGRAEoM}
0&=&
\mu_3(\ri\cD^\g\cDB_\g)^2{\D \cC}_{\a\b}
-\Big(8\mu_3\cS_0+\frac{1}{2\mu}\Big)\ri\cD^\g\cDB_\g{\D \cC}_{\a\b}
\non\\
&&
+2\Big(
 \frac{1}{\kappa} 
 +\frac{1}{\mu}\cS_0
-32\mu_3\cS^2_0\Big)
{\D \cC}_{\a\b}
~.
\eea

Consistency of the previous equation may be checked by making use 
of a general property of symmetric divergenceless superfields in the 
background chosen. 
Specifically, given a symmetric real spinor $T_{\a_1 \cdots \a_n} = T_{(\a_1 \cdots \a_n)}$ such that $\cD^\a T_{\a \a_1 \cdots \a_{n-1}} = 0$, 
one can check that the following holds
\be 
\cD^\a \cD^\g \bar\cD_\g T_{\a \a_1 \cdots \a_{n-1}} =
 0 \ .
\ee
This implies that the operator $\frac{\ri}{2} \cD^\g \bar\cD_\g$ preserves the divergenceless 
condition of the superfield $\D \cC_{\a\b}$. Thus one can check that \eqref{TypeIIbackgroundSUGRAEoM} is 
consistent.

It should be mentioned that \eqref{TypeIIbackgroundSUGRAEoM} can also be rewritten in terms of 
vector covariant derivatives as follows
\bea
0&=&
\mu_3\cD^{a} \cD_{a}{\D \cC}_{\a\b}
+\Big(4\mu_3\cS_0-\frac{1}{4\mu}\Big)\ve^{cab}(\g_c)_{\a\b}\cD_a{\D \cC}_b
\non\\
&&
+\hf\Big(
 \frac{1}{\kappa} 
-\frac{1}{\mu}\cS_0
 -16\mu_3\cS^2_0
\Big){\D \cC}_{\a\b}
~.
\eea

In the case $\mu_3 > 0$ we may factorize \eqref{TypeIIbackgroundSUGRAEoM} as follows
\be 
\Big(\frac{\ri}{2}\cD^\g\cDB_\g + m_-\Big)\Big(\frac{\ri}{2}\cD^\g\cDB_\g-m_+\Big){\D \cC}_{\a\b}=0
~,
\ee
where
\bsubeq
\bea m_+ - m_- &=& 
4\cS_0
+\frac{1}{4\mu_3\mu}
\ , \\
m_+ m_- &=&- \hf\Big(
 \frac{1}{\kappa\mu_3} 
 +\frac{1}{\mu\mu_3}\cS_0
-32\cS^2_0\Big)
\ .
\eea
\esubeq
The constants $m_+$ and $m_-$ are real for
\be
\Big(4\mu_3\cS_0+\frac{1}{4\mu}\Big)^2
 \geq 
2\mu_3\Big(
 \frac{1}{\kappa} 
 +\frac{1}{\mu}\cS_0
-32\mu_3\cS^2_0\Big) \label{TypeIIinequalitymassive}
~.
\ee

Massive supergravity may be described by the model \eqref{N=2_SG_model-II} 
with various choices of parameters. We see from eq. \eqref{TypeIIinequalitymassive} that in a 
Minkowski background ($\cS_0 = 0$)
with $\mu_3 > 0$ and $\mu = \infty$ it is necessary to have a
negative Einstein Hilbert term, $\kappa < 0$ or no Einstein-Hilbert term $\kappa \rightarrow \infty$. 
About a non-Minkowski background, $\cS_0 \neq 0$, 
the presence of a $R h^2$ in the action is problematic for ghost freedom \cite{BOS14}.
The choice $\l = 0$ and $\mu_1 = 0$ recovers the $\cN = 2$ generalised massive supergravity model 
discussed in \cite{BOS14}.

It is worth noting that maximally supersymmetric backgrounds are characterised by the more 
general conditions \eqref{genCondTypeIIback}. In this case
\be \cS = \cS_0 = {\rm const} \ , \quad \cC^a = \cC^a_0= {\rm const}
\ee
and the equation on the compensator 
reduces to
\be  \cS_0 = \frac{\kappa}{2} \l \ ,
\ee
while the  equation for the gravitational superfield  becomes
\be 0 = \Big{[}\frac{1}{\kappa} + \Big(\frac{1}{\mu} + 8 \mu_1 \Big) \cS_0 - 16 \mu_3 \cS_0^3 \Big{]} \cC_0^a \ .
\ee
We have already studied the case $\cC_0^a = 0$. If $\cC_0^a \neq 0$ we have the condition
\be \frac{1}{\kappa} + \Big(\frac{1}{\mu} + 8 \mu_1 \Big) \cS_0 - 16 \mu_3 \cS_0^3 = 0 \ .
\ee
We do not discuss linearization about this background here.


\section{$\cN = 2$ supergravity models with a complex linear compensator} 
\label{section5}

In the previous sections we have constructed $\cN=2$ supergravity models by using 
 a chiral and a real linear compensator.
In complete analogy to four-dimensional $\cN=1$ supergravity, see \cite{GGRS,WB,Ideas} for detailed reviews,
3D $\cN=2$ off-shell Poincar\'e and AdS supergravities can be realised 
by using a complex linear superfield coupled to conformal supergravity.
These non-minimal 3D $\cN=2$ models were introduced in \cite{KLT-M11,KT-M11}.
In this section we aim to
show that massive supergravity can be constructed
in the non-minimal case.

\subsection{Non-minimal supergravity in conventional superspace}

To describe non-minimal supergravity one makes use of a complex linear compensator $\S$ 
that obeys the constraint
\bea 
(\cDB^2-4\cR)\S=0
\label{complex-linear}
\eea
and is subject to no reality condition. 
By definition,  the compensator $\S$ is chosen to be nowhere vanishing 
and transform as a primary field of weight $w$  under the super-Weyl group.
The   U$(1)_R$  charge of $\S$ 
 is uniquely determined 
  \cite{KLT-M11}, 
\bea
\d_\s \S=w\s\S \quad \Longrightarrow \quad {J}\S=(1-w) \S
~.
\label{complex-linear2}
\eea

For every value of $w\ne0,1$ the following action
\bea
S_{\text{non-minimal}}= \frac{4w}{1-w}
\int \rd^{3|4}z
\,E\,
\big(\bar \S \S\big)^{\frac{1}{2w}} 
\label{NM4}
\eea
describes off-shell non-minimal Poincar\'e supergravity
providing a supersymmetric extension of the Einstein-Hilbert term.
On the other hand, it turns out that the complex linear superfield $\S$ is not suitable to construct a cosmological 
constant term and describe AdS supergravity. 
The way around this limitation 
was found in the four-dimensional case in \cite{BK11dual} and applied to three dimensions in \cite{KT-M11}.
The core of the idea is that when $w=-1$
the complex linear constraint \eqref{complex-linear}  admits non-trivial deformations. 

Consider a new conformal compensator $\G$ that has  the transformation properties
\bea
\d_\s \G= -\s\G~,\qquad
{J}\G=2\G
\eea
and  obeys  the {\it improved} linear constraint \cite{KT-M11}
\begin{align}
-\frac{1}{4} (\bar \cD^2 - 4 \cR) \Gamma = \l ={\rm const} \ .
\label{mCLc}
\end{align}
This constraint is super-Weyl invariant and
the complex parameter $\l \neq 0$ turns out to play the role of a cosmological constant.
In fact, the action
\bea
S_{\text{AdS}} = -2 \int \rd^{3|4}z
\,E\,
{ 
{ (\bar \G  \G)} 
}^{-1/2}
\label{NM-AdS}
\eea
describes AdS supergravity.
We can prove this statement by showing that the action
\eqref{NM-AdS} is dual to the type I minimal supergravity action \eqref{Type-I_AdS}.
Consider the first-order action
\bea
S_{\text{first-order}} = \int \rd^{3|4}z
\,E\,
\Big( -4  \bar \F \F 
+ \G\,\F^4 + \bar \G \,\bar\F^4 \Big) ~,
\label{NM7}
\eea
where $\F$ is {\it complex unconstrained}, and $\G$ obeys the constraint (\ref{mCLc}).
Varying $S_{\text{first-order}} $ with respect to $\G$ yields $\bar \cD_\a \F =0$, 
and then \eqref{NM7} reduces to the supergravity matter action \eqref{Type-I_AdS}
where for simplicity we have set $\k=1$. 
On the other hand, we can integrate out the fields $\F$ and $\bar \F$ 
to end up with the action \eqref{NM-AdS}.
In the following we will focus only on non-minimal supergravity where the compensator satisfies \eqref{mCLc}.

Let us now  discuss some geometrical properties of the $w=-1$ non-minimal supergravity
within conventional superspace.
The super-Weyl and local U$(1)_R$ symmetries can be used to impose the gauge condition
\bea
\G=1~.
\label{4.50}
\eea
In this gauge, 
some restrictions on the geometry occur \cite{KLT-M11}.
To describe them, it is useful to  split the covariant derivatives as 
\bea
\cD_\a={\rm D}_\a+\ri T_\a{J}
~,~~~
\cDB_\a=\bar{{\rm D}}_\a+\ri \bar{T}_\a{J}~,
\eea
where $T_\a$ is related to 
the original complex U$(1)_R$ connection $\F_\a$ as $T_\a = - \Phi_\a$.
In the gauge (\ref{4.50}), the  constraint $({\bar \cD}^2 -4\cR)\G=-4\l$ turns into 
\bea
\cR&=&\l+\frac{\ri}{2}\Big(
\bar{{\rm D}}_\a\bar{T}^\a
+\ri\bar{T}_\a\bar{T}^\a
\Big)~.
\label{4.52}
\eea
Evaluating explicitly $\{ \cD_\a, \cD_\b\} \G $ and  $\{ \cD_\a, {\bar \cD}_\b\} \G $
and then setting $\G=1$
gives
\bsubeq
\bea
&&{\rm D}_{(\a}T_{\b)}=0~,~~~~~~
\cS=\frac{1}{8}\Big(
{\bar{{\rm D}}}^\a T_\a
-{\rm D}^\a\bar{T}_\a
+2\ri T^\a\bar{T}_\a
\Big)
~,
\label{SDT}
\\
&&~~~~~~
\F_{\a\b}
=
\cC_{\a\b}
+\frac{\ri}{2}{\rm D}_{(\a}\bar{T}_{\b)}
+\frac{\ri}{2}{\bar{{\rm D}}}_{(\a} T_{\b)}
+T_{(\a}\bar{T}_{\b)}
~.
\label{FCDT}
\eea
\esubeq
If we define a new vector covariant derivative ${\rm D} _a$ by $ {\rm D}_a:=\cD_a -\ri \F_a {J}$, 
then the algebra of the covariant derivatives ${\rm D}_A =({\rm D}_a, {\rm D}_\a , {\bar {\rm D}}^\a)$
proves to be 
\bsubeq
\label{NMalg}
\bea
\{{\rm D}_\a,{\rm D}_\b\}&=&
-2\ri T_{(\a}{\rm D}_{\b)}
-4\bar{\cR}M_{\a\b}
~,
\\
\{{\rm D}_\a,{\bar{{\rm D}}}_\b\}&=&
-2\ri{\rm D}_{\a\b}
-\ri\bar{T}_{(\a} {\rm D}_{\b)}
+\ri T_{(\a}\bar{ {\rm D}}_{\b)}
+\frac{\ri}{2}\ve_{\a\b}\big(\bar{T}^\g {\rm D}_\g+T^{\g}\bar{{\rm D}}_{\g}\big)
\non\\
&&
-2\ve_{\a\b}\cC^{\g\d}M_{\g\d}
+4\ri\cS M_{\a\b}
~.
\eea
\esubeq
Note that here the independent curvature tensor superfields  $T_\a$ and $\cC_{\a\b}$ are
 of mass dimension $1/2$ and one, respectively, while 
both $\cS$ and $\cR$ are now descendants of the torsion superfields $T_\a$ and $\bar T_\a$.

Note that the Bianchi identities of the non-minimal algebra imply the following constraints
\begin{subequations}
\bea
\bar{{\rm D}}_\a \cR&=&2\ri \bar{T}_\a\cR~,
\\
{\rm D}^{\b}\cC_{\a\b}
&=&
-\frac{1 }{ 2}\Big[
\big( \bar{\rm D} _{\a}
+2\ri\bar{T}_{\a}\big)\bar{\cR}
+4\ri {\rm D}_{\a}{\cS}
\Big]~.
\eea
\end{subequations}
Moreover, the constraint defining a real linear superfield becomes
\bea
\big({\rm D}^2
+\ri {T}^\a{\rm D}_\a
-4\bar{\cR}\big)G
=0
~.
\eea
Such a constraint also holds for $\cS$.
The expression for the super-Cotton tensor expressed in terms of the non-minimal 
covariant derivatives ${\rm D}_A$ is
\bea
\cW_{\a\b}
&=&
-\frac{\ri}{4}{[}{\rm D}^\g,\bar{\rm D}_\g{]}\cC_{\a\b}
+\hf{[}{\rm D}_{(\a},\bar{\rm D}_{\b)}{]}\cS
-\frac{1}{4}  T^\g\bar{\rm D}_{(\a}\cC_{\b\g)}
+\frac{1}{4}\bar{T}^\g{\rm D}_{(\a}\cC_{\b\g)}
\non\\
&&
-\frac{1 }{ 12} T_{(\a}{\rm D} _{\b)}\cR
+\frac{1 }{ 12}\bar{T}_{(\a}\bar{\rm D} _{\b)}\bar{\cR}
-\frac{\ri }{6}\big( T_{(\a}\bar{\rm D}_{\b)}+\bar{T}_{(\a}{\rm D}_{\b)}\big){\cS}
+2\cS\cC_{\a\b}
~,
\eea
or, equivalently,
\bea
\cW_{\a\b}
=
-\hf{[}{\rm D}_{(\a},\bar{\rm D}_{\b)}{]}\cS
+\frac{\ri}{2} T_{(\a}\bar{\rm D}_{\b)}\cS
+\frac{\ri}{2} \bar{T}_{(\a}{\rm D}_{\b)}\cS
-\ve^{cab}(\g_c)_{\a\b}{\rm D}_{a}\cC_{b}
-2\cS\cC_{\a\b}
~.~~~~~~
\eea


\subsection{The superconformal setting}

In conformal superspace the conformal compensator $\G$ is a primary superfield
of dimension -1 and U$(1)_{R}$ weight 2,
\be \mathbb D \G = - 1 \ , \quad {J} \G = 2 \G \ , \quad K_A \G = 0 \ ,
\ee
and satisfies the constraint 
\be 
- \frac{1}{4} \bar\nabla^2 \G = \l = {\rm const.}
\label{CLc}
\ee

In complete analogy to the type I and II cases,
using $\G$ one can introduce new covariant derivatives that take primary  superfields 
to primary  superfields. We define the new covariant derivatives 
$\mathscr D_A=(\mathscr D_a,\mathscr D_\a,\bar{\mathscr D}^\a)$
as follows:
\bsubeq
\bea
\mathscr D_\a
&:=&
\bar\G^{\frac{1}{2}}\Big\{
\de_\a
+\hf\de^{\g}\log{\big(\G\bar{\G}\big)}M_{\a\g}
-{\frac{1}{4}}\de_{\a}\log{\Big(\frac{\G}{\bar{\G}} \Big)}{J}
+\frac{1}{2}}\de_{\a}\log{\big(\G\bar\G\big)\mathbb D
\Big\} \ , 
~~~~~~
\\
\mathscr D_{\a\b}
&:=&
(\G\bar\G)^{\hf}\Big\{
\de_{\a\b}
+\frac{\ri}{2}\bar\de_{(\a}\log{(\G\bar\G)}\de_{\b)}
+\frac{\ri}{2}\de_{(\a}\log{(\G\bar\G)}\bar{\de}_{\b)}
\non\\
&&~~~~~~~~~
+\hf\de_{(\a}{}^{\d}\log{(\G\bar\G)}M_{\b)\d}
-\frac{\ri}{8}\big(\de^{\g}\log{(\G\bar\G)}\big)\bar\de_{\g}\log{(\G\bar\G)}M_{\a\b}
\non\\
&&~~~~~~~~~
-\frac{\ri}{8}\Big{[}
\big(\de_{(\a}\log{(\G\bar\G)}\big)\bar\de_{\b)}\log{\Big(\frac{\G}{\bar{\G}} \Big)} 
+\big(\bar\de_{(\a}\log{(\G\bar\G)}\big)\de_{\b)}\log{\Big(\frac{\G}{\bar{\G}} \Big)} 
\Big{]}{J}
\non\\
&&~~~~~~~~~
-\frac{1}{4}\de_{\a\b}\log{\Big(\frac{\G}{\bar{\G}} \Big)}{J}
+\hf\de_{\a\b}\log{\big(\G\bar\G\big)}\mathbb D
\Big\} \ .
\eea
\esubeq
These covariant derivatives are such that 
\bea
{[}\mathbb D,\mathscr D_A{]}=0
~,~~~
{[}J,\mathscr D_A{]}=0
~,~~~
[K_A,\mathscr D_B\}=0
~,
\eea
 and satisfy the algebra:
\bsubeq
\label{confNMalg}
\bea
\{\mathscr D_\a,\mathscr D_\b\} &=&
-2\ri \mathscr T_{(\a}\mathscr D_{\b)}
-4\bar{\mathscr R}M_{\a\b} \ , \\
\{\mathscr D_\a,\bar{\mathscr D}_\b\}&=&
-2\ri\mathscr D_{\a\b}
-\ri\bar{\mathscr T}_{(\a}\mathscr D_{\b)}
+\ri \mathscr T_{(\a}\bar{\mathscr D}_{\b)}
+\frac{\ri}{2}\ve_{\a\b}\big(\bar{\mathscr T}^\g\mathscr D_\g+\mathscr T^{\g}\bar{\mathscr D}_{\g}\big)
\non\\
&&
-2\ve_{\a\b}\mathscr C^{\g\d}M_{\g\d}
+4\ri\mathscr SM_{\a\b} \ .
\eea
\esubeq
Here we have introduced the following primary dimensionless, and U(1)$_R$ chargeless superfields
\bsubeq
\bea
\mathscr T_\a
&=&
\frac{\ri}{4}\bar\G^{\frac{1}{2}}
\de_{\a}
\log{\big(\G\bar{\G}^{3}\big)}
~,
\\
\mathscr S&=&
\frac{\ri}{8}(\G\bar\G)^{\hf}
\de^\a\bar{\de}_{\a}\log{\big(\G\bar{\G}\big)}
\ , \\
\mathscr R&=&
-\frac{1}{4} (\G^{3}\bar{\G})^{\frac{1}{2}}\bar{\nabla}^2 (\G\bar\G)^{-\hf} 
\ ,\\
\mathscr C_{\a\b}
&=&
-\frac{1}{4}{[}\de_{(\a},\bar\de_{\b)}{]}(\G\bar\G)^{\hf} \ ,
\eea
\esubeq
together with their complex conjugates.
Note also that $\G$ is covariantly constant with respect to the derivatives $\mathscr D_A$,
\bea
\mathscr D_A\G=0
~.
\eea
The  algebra of covariant derivatives \eqref{confNMalg} formally coincides with
that of
$w=-1$ non-minimal supergravity, \eqref{NMalg}.

By using the compensator $\G$ and the torsion superfield $\mathscr T_\a$, together with its descendants,
we can construct general actions of the form
\be
S_{\rm NM} = \int \rd^{3|4}z \, E \, (\G \bar \G)^{-\hf} \frak{L}(\mathscr T , \mathscr D \mathscr T \ , \cdots ) \ ,
\label{general_non-minimal}
\ee
where  $\frak{L}$ is a primary 
dimensionless superfield constructed out of the torsion components
$\mathscr T$ and its covariant derivatives.

It is worth underlining the peculiarity of the presence of a primary, spinorial torsion $\mathscr T_\a$ 
in $\cN=2$ non-minimal supergravity.
Its presence is ultimately related to the fact that non-minimal supergravity has 4+4 extra auxiliary fields compared
to the minimal type I and type II cases.
Consider the following independent components of the complex linear compensator
\bsubeq\label{NM-auxiliaries}
\bea
&B=\G|~,~~~
\rho_\a=\de_\a\G|~,~~~
\z_\a=\de_\a\bar\G|~,
\\
&H=\de^2\G|~,~~~
p=\bar \de^\a \de_\a\G|~,~~~
p_{\a\b}=\bar \de_{(\a} \de_{\b)}\G|~,~~~
\Bar{\b}_\a={1\over 2}\de^\b\bar \de_\a \de_\b\G|~,
~~~~~~~~~
\eea
\esubeq
together with their complex conjugates.
The gauge choice
\bea
\G=
1~,~~~~~~
B_A=0~,
\eea
that fixes dilatation, U(1)$_R$ and special conformal symmetry,
at the component level corresponds to setting
\bea
B=
1~,~~~
\big(\rho_\a+\z_\a\big)
=0
~,~~~
b_m=0~.
\eea
The first condition fixes the dilatation and U(1)$_R$ symmetries.
 The second condition fixes $S$-supersymmetry. The last condition fixes the 
conformal boosts.
It is clear that only half of the eight spinor components of $\z_\a$, $\bar{\z}_\a$  $\rho_\a$ and $\bar{\rho}_\a$
are used to fix $S$-supersymmetry while the remainder fit in the non-minimal spinor auxiliary field
\bea
T_{\a}|=\frac{1}{4}\big(\rho_\a-\z_\a\big)
~.
\eea

\subsection{Massive supergravity}

In this section we use the non-minimal formulation to construct a 
novel consistent theory of massive supergravity.

Consider the action
\be
 S =
  \frac{1}{\tilde\mu} S_{\rm CSG}
+S_{\text{first-order}}  \ ,
\label{final_first-order}
\ee
where $S_{\rm CSG}$ is the conformal supergravity action
while $S_{\text{first-order}} $ is given by
\bea
S_{\text{first-order}} = \int \rd^{3|4}z
\,E\,
\Big( \bar \F \F\, \frak{L}(\F,\bar\F)
+ \G\,\F^4 + \bar \G \,\bar\F^4 \Big) ~,
\label{dualMassive-1}
\eea
and the dimensionless primary superfield $\frak{L}$ is given by
\bea
\frak{L}:=-4
- 4  \mu_2 \mathscr R \bar {\mathscr R}
+ 4 \mu_3  \mathscr C^{\a\b} \mathscr C_{\a\b} 
+\big(
 \mu_0 \mathscr R^3
+ \mu_1 \mathscr R^4
+{\rm c.c.}
\big)
~.
\label{frakLNM}
\eea
The superfields $\G$ and $\bar\G$ are the complex linear compensators defined in the previous sections and
satisfy the constraint \eqref{CLc} and its conjugate, respectively.
The superfields $\F$ and $\bar\F$ here are unconstrained complex primaries such that
\bea
\mathbb D \F=\hf\F~,~~~~~~
J \F=-\hf\F
~.
\eea
The dimensionless primary superfields 
$\mathscr R $, $\bar{\mathscr R} $ and $\mathscr C_{\a\b}$ 
are functionals of $\F$ and $\bar\F$ defined as (compare with \eqref{GeocompositeN=2})
\bea
\mathscr R :=- \frac{1}{4 \Phi^4} \bar \nabla^2 (\F\bar \Phi) \ , 
\quad \mathscr C_{\a\b} := - \frac{1}{4} [\nabla_{(\a} , \bar \nabla_{\b)}] \frac{1}{\Phi \bar \Phi}
\ .~~~~~~
\eea

An important feature of the action \eqref{final_first-order} is that
if we vary it with respect to $\G$ the Lagrange multiplier term $\G\F^4$ yields the chiral constraint,
 $\bar \de_\a \F =0$.
 Then 
\eqref{final_first-order} with \eqref{frakLNM} reduces to the general type I massive 
action \eqref{N=2_SG_model}
(where for simplicity we have set $\k=1$). 
On the other hand, we can formally integrate out the fields $\F$ and $\bar \F$, which 
upon imposing their equations of motion become functionals of $\G$ and $\bar\G$.
If one then plugs  the expressions for $\F(\G,\bar\G)$ and $\bar\F(\G,\bar\G)$ into eq.~\eqref{dualMassive-1},
the resulting model describes massive supergravity in the $\cN=2$ non-minimal case.
On-shell the resulting dynamical system is equivalent to the model described by eq.~\eqref{N=2_SG_model}.

We have just demonstrated that, by dualizing type I models, 
massive supergravity can be constructed also in the non-minimal case.
An open question remains whether the models obtained by dualizing the type I 
models are the most general massive supergravity theories one can construct 
in the non-minimal formulation.
We leave the investigation of this question for future work.


\section{$\cN = 3$ supergravity with a compensating vector multiplet} 
\label{section6}

The conventional superspace formalism of \cite{KLT-M11} 
offers the ability to construct 
the most general $\cN = 3$ supergravity models. In this section we focus on $\cN = 3$ 
supergravity with a nowhere vanishing off-shell vector multiplet and construct 
various supergravity invariants up to and including curvature squared invariants.  
Furthermore, we will demonstrate 
how such invariants may be constructed 
within the superconformal framework. It should also be mentioned 
that in the context of projective superspace the $\cN = 3$ vector multiplet 
is often referred to as the $\cO(2)$ multiplet. In what follows we use either 
name interchangeably.


\subsection{Conventional superspace} \label{subsection6.1}

The curved superspace $\cM^{3|6}$ is parametrized by real bosonic $(x^m)$ and 
real fermionic $(\q^\mu_{\rm I})$ coordinates $z^M = (x^m , \q_{\rm I}^\mu)$, where $m = 0, 1, 2$, $\mu = 1, 2$, and $\rm I = 1,2,3$. 
The superspace geometry \cite{KLT-M11} is described in terms of 
covariant derivatives of the form
\be \cD_A = E_A{}^M \pa_M - \hf \Omega_A{}^{ab} M_{ab} - \hf \Phi_A{}^{PQ} N_{PQ} \ .
\ee
Here $E_A = E_A{}^M(z) \partial_M$ is the inverse supervielbein, $\Omega_A{}^{ab}$ is the Lorentz connection
and $\Phi_A{}^{PQ}$ is the $\rm SO(3)$ connection. The SO(3) generators $N_{IJ} = - N_{JI}$ act on the 
spinor covariant derivatives as follows
\be [N^{IJ} , \cD_\a^K] = 2 \d^{K[I} \cD_\a^{J]} \ .
\ee
The supergravity spinor covariant derivatives obey the following anti-commutation relations:
\bea
\{ \cD_\a^I , \cD_\b^J \} &=& 
2 \ri \d^{IJ} (\g^c)_{\a\b} \cD_c 
- 2 \ri \eps_{\a\b} \cC^{\g\d IJ} M_{\g\d} 
- 4 \ri (\cS^{IJ} + \d^{IJ} \cS) M_{\a\b} \non\\ 
&&
+\Big{[}
- 4 \ri \eps_{\a\b} \cS^{K[ I} \d^{J] L} 
- 4 \ri \eps_{\a\b} \d^{K[I} \d^{J] L} \cS
+ \ri \cC_{\a\b}{}^{KL} \d^{IJ} 
\non\\
&&~~~~
- 4 \ri \cC_{\a\b}{}^{K(I} \d^{J)L} \Big{]} N_{KL}\ ,
\eea
where $\cS^{IJ}$ and $\cC_{\a\b}{}^{IJ}$ satisfy the symmetry properties
\be \cS^{IJ} = \cS^{(IJ)} \ , \quad \cS^{I}{}_I = 0 \ , \quad \cC_{\a\b}{}^{IJ} = \cC_{\a\b}{}^{[IJ]} = \cC_{(\a\b)}{}^{IJ} \ ,
\ee
as well as the Bianchi identities
\bsubeq
\bea
\cD_{\a}^{I} C_{\b\g}{}^{JK}
&=&
{2\over 3}\ve_{\a(\b}\Big(
\cC_{\g)}{}^{IJK}
+3\cT_{\g)}{}^{JKI}
+4(\cD_{\g)}^{[J} \cS)\d^{K]I}
-\frac{1}{3}\cS_{\g)}{}^{[J}\d^{K]I}
\Big)
\non\\
&&
+\cC_{\a\b\g}{}^{IJK}
-2 \cC_{\a\b\g}{}^{[J}\d^{K]I}
\ ,
\\
\cD_\a^I \cS^{JK}
&=&
2\cT_{\a}{}^{I(JK)}
+\cS_\a{}^{(J}\d^{K)I}
-\frac{1}{3}\cS_\a{}^{I}\d^{JK}
\ .
\eea
\esubeq
The symmetry properties of $\cC_{\a}{}^{IJK}$, $\cC_{\a\b\g}{}^{IJK}$ and $\cC_{\a\b\g}{}^I$ are
\bea 
\cC_{\a}{}^{IJK}=\cC_{\a}{}^{[IJK]}
~,~~~
\cC_{\a\b\g}{}^{IJK} = \cC_{\a\b\g}{}^{[IJK]} = \cC_{(\a\b\g)}{}^{IJK} ~,~~~
\cC_{\a\b\g}{}^I = \cC_{(\a\b\g)}{}^I ~,
\eea
while the superfield $\cT_{\a}{}^{IJK}$ is such that
\bea
\cT_{\a}{}^{IJK}=\cT_{\a}{}^{[IJ]K}~, \qquad
\d_{JK}\cT_{\a}{}^{IJK}=
\cT_{\a}{}^{[IJK]}=0
~.
\eea
The remaining covariant derivative commutation relations follow from the spinor covariant derivative 
anti-commutator, see \cite{KLT-M11} for more details.

The superspace geometry describes conformal supergravity because it admits super-Weyl 
transformations of
the form \cite{1205.4622}
\bsubeq
\bea
\cD'{}_\a^I &=& \re^{\hf \s} \Big( 
\cD_\a^I
+ \cD^{\b I} \s M_{\a\b}
+ \cD_{\a J} \s N^{IJ}
\Big) \ , \\
\cD'_a &=& \re^\s \Big(
\cD_a
+ \frac{\ri}{2} (\g_a)^{\a\b} (\cD_\a^K \s) \cD_{\b K}
+ \eps_{abc} (\cD^b \s) M^c
+ \frac{\ri}{4} (\cD^\r_K \s) \cD_\r^K \s M_a 
\non\\
&&~~~~~~~
+ \frac{\ri}{48} (\g_a)^{\a\b} \re^{- 3 \s} [\cD_\a^K , \cD_\b^L] \re^{3 \s}
\Big) \ .
\eea
\esubeq
The corresponding transformations of the torsion superfields are
\bsubeq
\bea
\cS' &=& - \frac{\ri}{6} \re^{\hf \s} (\cD^\g_K \cD_\g^K + 6 \ri \cS) \re^{\hf \s}
\ , \\
\cS'^{IJ} &=& \frac{\ri}{4} \re^{2 \s} \Big(
\cD^{\g I} \cD_\g^J
- \frac{1}{3} \d^{IJ} \cD^\g_K \cD_\g^K
- 4 \ri \cS^{IJ}
\Big) \re^{- \s}
\ , \\
\cC'_{\a\b}{}^{IJ} &=& \frac{\ri}{2} \Big( \cD_{(\a}^{[I} \cD_{\b)}^{J]} - 2 \ri \cC_{\a\b}{}^{IJ} \Big) \re^\s
\ .
\eea
\esubeq
All supergravity-matter invariants can be made super-Weyl invariant 
with the use of the vector multiplet compensator field strength $G^I$, which 
satisfies the Bianchi identity
\be \cD_\a^{(I} G^{J)} = \frac{1}{3} \d^{IJ} \cD_\a^K G_K \ .
\ee
One only needs to make use of its magnitude $G$, 
\be G^2 := G^I G_I \ ,
\ee
which transforms homogeneously under super-Weyl transformations,
\be G' = \re^\s G \ .
\ee

It should be mentioned that the abelian vector multiplet is also known as 
the $\cO(2)$ multiplet. The reason for this is made evident by making use of the isomorphism 
$\rm SO(3) \cong SU(2)/ {\mathbb Z}_2$ and replacing any SO(3) vector index by a 
symmetric pair of SU(2) spinor indices. For the isovector $G^I$, one can instead 
work with the symmetric spinor $G^{ij}$ defined by
\bea
G^I := (\S^I)_{ij} G^{ij}~, \quad G_{ij} = (\S_I)_{ij} G^I \ , \label{G^{ij}intr}
\eea
where the sigma-matrices are given by 
\bea
(\S_I)_{i j}:=({\mathbbm 1},\ri\s_1, \ri\s_3) =(\S_I)_{ji}~.
\eea
The $\cO(2)$ multiplet can be shown to satisfy the analyticity constraint
\be \cD_\a^{(ij} G^{kl)} =  0  \ ,
\ee
where $\cD_\a^{ij} := (\S_I)^{i j} \cD_\a^I$.
The isospinor representation of isotensors is used in the projective superspace 
formulation \cite{KLT-M11}.

General off-shell matter couplings in $\cN=3$ supergravity were constructed in 
\cite{KLT-M11} by using projective superspace.
  Given a supergravity-matter system, its 
dynamics can be described by a Lagrangian $\cL^{(2)}(v)$ 
which is a real weight-two covariant projective supermultiplet,\footnote{In what follows, 
we suppress explicit $z$-dependence of $\cN=3$ superfields.} 
with $v^i$ the homogeneous  coordinates for ${\mathbb C}P^1$. 
We refer the reader to \cite{KLT-M11} for the definition and notations of $\cN=3$ covariant projective superspace
and details about the supersymmetric action principle in that context.
See also \cite{KN14} for the generalization to the $\cN=3$ conformal superspace of \cite{BKNT-M1}.

It was shown in \cite{KN14} that in the presence of an Abelian vector multiplet with nowhere 
vanishing gauge invariant field strength $G^{ij}$, $G:=\sqrt{G^{ij}G_{ij}}\neq 0$, 
the action functional can be rewritten as a $BF$ term for a 
composite real $\cO (2) $ multiplet $\bm G^{ij}$. In this sense the $BF$ action may be used as 
a \emph{universal} action principle for $\cN = 3$. We will discuss the $BF$ action 
in more detail below in the superconformal context.


\subsection{The superconformal setting} \label{subsection6.2}

Since our actions can be made to be superconformally invariant 
it is natural to work with a manifestly superconformal framework. In this 
subsection we introduce such a framework and provide the ingredients 
for the construction of supergravity invariants.

\subsubsection{The abelian vector multiplet}

Here we make use of $\cN = 3$ conformal superspace described in appendix \ref{geometry}. The abelian vector multiplet 
superfield strength $G^I$ is a dimension 1 primary,
\be \mathbb D G^I = G^I \ , \quad K_A G^I = 0 \ ,
\ee
and satisfies the Bianchi identity
\be \nabla_\a^{(I} G^{J)} = \frac{1}{3} \d^{IJ} \nabla_\a^K G_K \ . 
\ee
The superfield $G^{I}$ naturally appears in the components of the gauge-invariant 
field strength $F = \rd V$,
\bea
 F = 
\hf E^\b_J \wedge E^\a_I \, F_{\a}^I{}_{ \b}^J 
 + E^\b_J \wedge E^a\,  F_{a}{}_{\b}^J  
 + \hf E^b \wedge E^a\,  F_{ab}~, 
\eea 
where
\bsubeq \label{FSCompsN=3}
\bea
F_{\a}^I{}_\b^J &=&  -2 \ri\ve_{\a\b} \eps^{IJK} G_K \ , \\
F_{a}{}_\a^I&=&
\frac{1}{ 2}(\g_a)_\a{}^{\b} \eps^{IJK} \nabla_{\b J}  G_K
~,
\\
F_{ab}&=&
-\frac{\ri}{ 24}\ve_{abc}(\g^c)^{\a\b} \eps_{IJK} [\nabla_{\a}^{ I}, \nabla_{\b}^{ J}] G^K
~.
\eea
\esubeq

Using $G = \sqrt{G^I G_I}$ one can construct dimensionless covariant derivatives 
that take primary superfields to primary superfields. The covariant derivatives are
\bsubeq \label{DGcovDerN=3}
\bea \mathscr D_\a^I &=& G^{- \hf} \Big(\nabla_\a^I - (\nabla^{\b I} \ln G) M_{\a\b} - (\nabla_{\a J} \ln G) N^{IJ} 
- (\nabla_\a^I \ln G) \mathbb D \Big) \ ,  \\
\mathscr D_a &=& \frac{\ri}{12} (\g_a)^{\a\b} \{ \mathscr D_\a^K , \mathscr D_{\b K} \}
+ 2 \mathscr S M_a + \frac{1}{6} \mathscr C_a{}^{KL} N_{KL} \ ,
\eea
\esubeq
where we have introduced the dimensionless primary superfields
\bsubeq
\bea
\mathscr C_{a}{}^{IJ} 
&:=& - \frac{\ri}{4} (\g_a)^{\a\b} \nabla_\a^{[I} \nabla_\b^{J]} \frac{1}{G} \equiv \eps^{IJK} \mathscr C_{a K} 
\ , \\
\mathscr S &:=& - \frac{\ri}{6 G^{\hf}} \nabla^\g_K \nabla^K_\g G^{- \hf} \ .
\eea
\esubeq
It can be checked that the covariant derivatives annihilate $G$, $\mathscr D_A G = 0$. On primary superfields one can 
verify that the covariant derivatives \eqref{DGcovDerN=3} satisfy the algebra
\bea
\{ \mathscr D_\a^I , \mathscr D_\b^J \} &=& 
2 \ri \d^{IJ} (\g^c)_{\a\b} \mathscr D_c 
- 2 \ri \eps_{\a\b} \mathscr C^{\g\d IJ} M_{\g\d} 
- 4 \ri (\mathscr S^{IJ} + \d^{IJ} \mathscr S) M_{\a\b} \non\\ 
&&+\Big(- 4 \ri \eps_{\a\b} \mathscr S^{K[ I} \d^{J] L} 
- 4 \ri \eps_{\a\b} \d^{K[I} \d^{J] L} \mathscr S
+ \ri \mathscr C_{\a\b}{}^{KL} \d^{IJ} \non\\
&&\qquad \qquad- 4 \ri \mathscr C_{\a\b}{}^{K(I} \d^{J)L} \Big) N_{KL}\ ,  
\eea
where we have introduced
\be \mathscr S^{IJ} := \frac{\ri}{4 G^2} \Big( \nabla^{\g (I} \nabla_\g^{J)} - \frac{1}{3} \d^{IJ} \nabla^\g_K \nabla_\g^K \Big) G \ .
\ee
The algebra of covariant derivative of $\mathscr D_A$ can be seen to formally agree with the one of 
$\rm SO(3)$ superspace. It is worth mentioning that in the gauge $B_A= 0$ the superfields $\mathscr S^{IJ}$, 
$\mathscr S$ and $\mathscr C_{\a\b}{}^{IJ}$ become
\bsubeq
\bea
\mathscr C_{\a\b}{}^{IJ} &=& \frac{\ri}{2} \Big( \cD_{(\a}^{[I} \cD_{\b)}^{J]} - 2 \ri \cC_{\a\b}{}^{IJ} \Big) \frac{1}{G} \ , \\
\mathscr  S &=& - \frac{\ri}{6 G^{\hf}} \Big( \cD^\g_K \cD_\g^K + 6 \ri \cS \Big) \frac{1}{G^{\hf}} \ , \\
\mathscr S^{IJ} &=& \frac{\ri}{4 G^2} \Big( 
\cD^{\g I} \cD_{\g}^J - \frac{1}{3} \d^{IJ} \cD^\g_K \cD_\g^K - 4 \ri \cS^{IJ}
\Big) G
\ .
\eea
\esubeq
Similar to the previous cases the superfields $\mathscr S^{IJ}$, 
$\mathscr S$ and $\mathscr C_{\a\b}{}^{IJ}$ can be seen to degauge to $\cS^{IJ}$, $\cS$ and $C_{\a\b}{}^{IJ}$ upon 
imposing the gauge condition $G = 1$.

Instead of working with $G^I$, one can also equivalently make use of the 
prescription \eqref{G^{ij}intr} and introduce
$G^{(2)}$ defined such that
\be \nabla_\a^{(2)} G^{(2)} = 0 \quad \Longleftrightarrow \quad \nabla_\a^{(ij} G^{kl)} = 0 \ , 
\quad G^{(2)}(v) := G_{ij} v^i v^j \ , \quad \nabla_\a^{(2)} := v_i v_j \nabla_\a^{ij} \ , \label{O(2)constraint}
\ee
where $v^i$ are homogeneous  coordinates for ${\mathbb C}P^1$.\footnote{Refer to \cite{KLT-M11,KN14} for details 
on the formalism of projective superspace.} In this form the $\cO(2)$ multiplet can be 
given the prepotential realisation \cite{KLT-M11}
\bea
G^{(2)}(v) := G_{ij} v^i v^j 
= \D^{(4)} \oint_{\hat \g} \frac{ ({\hat v},\rd {\hat v})}{  2\p(v, {\hat v})^2}\,\cV( {\hat v})
~,
\label{C.9}
\eea
where $\cV(v)$ is the tropical prepotential for the vector multiplet,  
$\nabla^{(2)}_\a \cV =0$, and we have introduced the analytic projection operator \cite{KLT-M11}
\be \D^{(4)} := \frac{\ri}{4} \nabla^{\a(2)} \nabla_\a^{(2)} \ .
\ee
Here the prepotential $\cV$ possesses the gauge transformations 
\bea
\d \cV = \l + \breve{\l} ~,\qquad \nabla^{(2)}_\a \l =0~, \label{N=3GT}
\label{7.3}
\eea
where the gauge parameter $\l$ is an arbitrary weight-0 arctic multiplet
and $\breve \l$ is its smile-conjugate, see \cite{KLT-M11} for more details. 

A noteworthy and useful composite $\cO(2)$ multiplet $\mathbb G^{(2)}$ 
may be constructed using the prepotential realisation \eqref{C.9} 
with the prepotential \cite{KLT-M11}
\be \cV = \ln \Big( \frac{G^{(2)}}{\ri \U^{(1)} \breve \U^{(1)}} \Big) \ ,
\ee
where $\U^{(1)}$ is a weight-1 arctic multiplet and $\breve \U^{(1)}$ its smile 
conjugate \cite{KLT-M11}. 
The composite $\mathbb G^I$, which was used in the 
description of (2,1) AdS supergravity \cite{KN14}, can be expressed in 
terms of the composites $\mathscr S^{IJ}$ and $\mathscr S$ as follows:
\bea \mathbb G^{I} &=& 4 (\mathscr S^{IJ} + \d^{IJ} \mathscr S) G_J \ , \non\\
&=& \frac{\ri}{G^2} G_J (\nabla^{\g (I} \nabla_\g^{J)} - \frac{1}{3} \d^{IJ} \nabla^\g_K \nabla_\g^K ) G
- \frac{2\ri}{3 G^{\hf}} G^I \nabla^\g_K \nabla_\g^K G^{-\hf} \ .
\eea
It will be a useful ingredient in the construction of supergravity invariants.


\subsubsection{The $BF$ action}

General off-shell matter couplings in $\cN=3$ supergravity were constructed in 
\cite{KLT-M11} and it was shown in \cite{KN14} that the general action functional there
can be rewritten as a $BF$ term
\bea
S_{\rm LM}&=&
\frac{1}{2\pi\ri} \oint_\g (v, \rd v)
\int \rd^{3|6} z
\,E\, 
C^{(-4)}
\cL^{(2)} \non\\
 &=& 
\frac{1}{2\pi\ri} \oint_\g (v, \rd v)
\int \rd^{3|6} z \,\,E\, 
C^{(-4)}
 \cV \,\bm G^{(2)}~, 
\label{7.1} 
\quad  \rd^{3|6} z = \rd^3 x \,{\rm d}^6\q \ ,
\eea
where $\cV(v)$ is the tropical prepotential for the $\cO(2)$ multiplet $G^{(2)}$,  
while $\bm G^{(2)}$ is some composite $\cO(2)$ multiplet. Here the {\it model-independent}  
isotwistor superfield $C^{(-4)} (v)$ of weight $-4$  
is required to be conformally primary and of dimension $-1$.
The action \eqref{7.1} is invariant under the gauge transformations \eqref{N=3GT}. 
The action \eqref{7.1} is also called the $\cN=3$ linear multiplet action. The component form of the 
action \eqref{7.1} is \cite{KN14}
\begin{align}
S_{\rm LM} =&\, \hf \int \rd^3 x \,e \, \Big(
	\eps^{abc} v_a \bm f_{bc}
	- 2 \ri \c^\g \bm \c_\g  - \frac{\ri}{4} \l^{\g IJ} \bm \l_{\g IJ} + g^I \bm h_I + \bm g^I h_I \non\\
	&- \hf (\g^a)_{\g\d} \psi_a{}^\g_I (\l^{\d IJ} \bm g_J + \bm \l^{\d IJ} g_J  + \ri \c^\d \bm g^I + \ri \bm \c^\d g^I) \non\\
	&+ \frac{\ri}{2} \eps^{abc} (\g_a)_{\g\d} \psi_b{}^\g_K \psi_c{}^{\d}_L (\d^{KL} g^P \bm g_P - 2 g^K \bm g^L)  \Big) \ ,
\label{7.8}
\end{align}
where the component fields are defined as follows:
\bsubeq
\begin{align}
g_I &= G_I| \ , \quad \l_{\a}{}^{IJ} = 2 \nabla_{\a}^{[I} G^{J]}| \ , \quad \c_\a = \frac{\ri}{3}  \nabla_\a^I G_I | \ , \quad h_I = - \ri  \nabla^{\g J}  \nabla_{\g J} G_I | \ , \\
v_a &= e_a{}^m V_m| = V_a| + \hf \psi_a{}^\a_I V_\a^I | \ , \\
f_{ab} &=  F_{ab}| - \hf (\psi_{[a}{}^K \g_{b]}  \l_K) + \frac{\ri}{2} \psi_a{}^{\g K} \psi_b{}_\g^L  g_{KL} 
\non\\
&= 
- \frac{\ri}{12} \eps_{abc} (\g^c)^{\a\b} \eps^{IJK} 
\nabla_{\a I} \nabla_{\b J} G_K |
- \frac{1}{4} \eps^{IJK} (\psi_{[a I} \g_{b]} \l_{JK}) + \frac{\ri}{2} \eps^{IJK} \psi_a{}^{\g}_I \psi_b{}_{\g J}   g_{K} \ .
\end{align}
\esubeq
The same definitions hold for the component fields of $\bm G$.

The component fields of the Weyl multiplet were defined in \cite{BKNT-M1}. The 
vielbein $e_m{}^a$, the gravitino $\psi_m{}^\a_I$, the $\rm SO(3)$ gauge field 
$V_m{}^{IJ}$ and the dilatation gauge field $b_m$, are defined as the lowest components of 
their corresponding superforms,
\begin{align}\label{eq:GaugeFields}
e_m{}^a &:= E_m{}^a| \ , \qquad \psi_m{}^\a_I := 2 E_m{}^\a_I| ~, \qquad
V_m{}^{IJ} := \Phi_m{}^{IJ}| \ ,\qquad b_m := B_m| ~.
\end{align}
There also exists an additional component field $w_\a$ which is defined as the 
component projection of the super-Cotton tensor $W_\a$, $w_\a = W_\a|$. 
At the component level, we will be mostly interested in the bosonic sector of locally supersymmetric invariants and therefore 
will not be concerned with fermionic fields.

To fix the dilatations, $S$-supersymmetry, special conformal symmetry and 
$\rm SO(3)$ symmetry in our invariants one can make use of the 
gauge conditions
\be G^{I} = (0 , 0 , G) \ , \quad G = 1 \ , \quad B_A = 0 \ ,
\ee
which leads to the following conditions at the component level:
\be g^I = (0,0, g) \ , \quad g := \sqrt{g^I g_I} = 1 \ , \quad \c_\a = \l_\a{}^{IJ} = 0 \ , \quad b_m := B_m| = 0 \ . \label{compGFN=3}
\ee
The first breaks the SO(3) symmetry, the second breaks dilatations, the third fixes $S$-supersymmetry transformations and 
the last fixes the special conformal transformations. Note that the gauge conditions 
\eqref{compGFN=3} also imply that
\be h^I \equiv (0, 0 , h) \ .
\ee
In the gauge \eqref{compGFN=3} one can verify the following useful results:
\bsubeq \label{N=3compResults}
\bea
\Box g^I &=& \frac{1}{4} {R} g^I + {\rm fermion \ terms} \ , \\
\Box^2 g^I &=& {R}^{ab} {R}_{ab} g^I - \frac{1}{8} {R}^2 g^I +\frac{1}{4}({D}^a{D}_a {R}) g^I + {\rm fermion \ terms} \ ,
\eea
\esubeq
where we have defined
\bsubeq
\bea 
\Box g^I &:=& \nabla^a \nabla_a G^I| \ , \\
\Box^2 g^I &:=& \nabla^a \nabla_a \nabla^b \nabla_b G^I| \ ,
\eea
and
\esubeq
\be {D}_a = e_a{}^m \Big( \partial_m - \hf \omega_m{}^{bc} M_{bc} - V_m{}^{KL} N_{KL} - b_m \mathbb D \Big) \ .
\ee
The above results 
will be useful in identifying the curvature terms in the invariants at the component level.


\subsection{Supergravity invariants}

In this subsection we construct various supergravity invariants in superspace using the composites 
$\mathscr S^{IJ}$ and $\mathscr S$.


\subsubsection{The supergravity action}

The $\cN = (2,1)$ supergravity action with a cosmological term is described by the 
Lagrangian  \cite{KLT-M11}
\be \cL^{(2)} = \frac{1}{\kappa} \cL^{(2)}_{\rm SG} + \l \cL^{(2)}_{\rm cos} \ , \label{N=3SUGRA}
\ee
where
\bsubeq
\bea \cL_{\rm SG}^{(2)} &=& G^{(2)} \ln \Big( \frac{G^{(2)}}{\ri \U^{(1)} \breve \U^{(1)}} \Big) \ , \\
\cL_{\rm cos}^{(2)} &=& \cV G^{(2)} \ .
\eea
\esubeq
The component action corresponding to $\cL_{\rm SG}^{(2)}$
can be constructed by letting $\bm G^I \rightarrow \mathbb G^I$ in the component $BF$ action \eqref{7.8}.
After doing so, one can verify that the component field $\bm h^I$ contains a term proportional to $\frac{1}{g} \Box g^I$. Imposing the 
gauge condition \eqref{compGFN=3} and using the results \eqref{N=3compResults} recovers the Einstein-Hilbert 
term in the action. The cosmological term coming from 
$\cL_{\rm cos}^{(2)}$
corresponds to a U(1) Chern-Simons term. The equations of 
motion for the supergravity model was given in \cite{KN14}.


\subsubsection{The ${R} h^n$ invariants}

An invariant can be constructed using
\be \cL^{(2)}_{{R} h^n} = \Big( \frac{\mathbb G^{(2)}}{G^{(2)}} \Big)^n G^{(2)} \ .
\ee
Using integration by parts one can show that the corresponding action can be equivalently constructed from 
\be \cL^{(2)} = \cV \mathbb G_{n}^{(2)} \ ,
\ee
where we have defined \cite{KN14}
\bea
\mathbb G^{(2)}_n := \D^{(4)} \oint_{\hat \g} \frac{ ({\hat v},\rd {\hat v})}{  2\p(v, {\hat v})^2}\,
\Big( \frac{\mathbb G^{(\hat{2})}}{G^{(\hat{2})}} \Big)^n \ .
\eea
The component action corresponding to $\cL^{(2)}_{{R} h^n}$ may be obtained  
by letting $\bm G^I \rightarrow \mathbb G^I$ in the component $BF$ action \eqref{7.8}.
For $n = 1$ the action can be seen to vanish since upon integrating by parts it coincides with
\be \cL^{(2)} = \mathbb G^{(2)} \ ,
\ee
which may be thought of as a $BF$ action with one of the vector multiplets set to zero. 
For $n \geq 2$ it can be shown to contain a term proportional to ${R} h^n$ upon imposing the gauge 
conditions \eqref{compGFN=3}.

For the $n=1$ case one can use the following independent invariant 
\be \cL^{(2)}_{{R} h} = \mathbb G^{(2)} \ln \Big( \frac{G^{(2)}}{\ri \U^{(1)} \breve \U^{(1)}} \Big) \ .
\ee
The component action can be analysed by putting $G^I \rightarrow \mathbb G^I$ and 
$\bm G^I \rightarrow \mathbb G^I$
into the $BF$ action. A term of the form
\be \frac{1}{g^2} h_I \Box g^I
\ee
will arise in the component action in the contribution $g^I \bm h_I + \bm g^I h_I$. Upon gauge fixing it gives rise to a ${R} h$ term. 
In analogy with the $\cN = 2$ case with a vector multiplet  
we call the invariant a ${R} h$ invariant.

It is also worth mentioning that upon integrating by parts one can show that 
the supergravity invariant corresponding to $\cL^{(2)}_{{R} h}$ can 
be described by the Lagrangian
\be \cL^{(2)} = V \mathbb G^{(2)}_0 \ ,
\ee
where
\be \mathbb G^{(2)}_0 
= 
\D^{(4)} \oint_{\hat \g} \frac{ ({\hat v},\rd {\hat v})}{  2\p(v, {\hat v})^2}\,
\frac{{\mathbb G}^{(\hat{2})}}{G^{(\hat{2})}} \ln \Big( \frac{G^{(\hat{2})}}{\ri \U^{(\hat{1})} \breve \U^{(\hat{1})}} \Big) \ .
\ee


\subsubsection{Ricci squared invariants}

In the previous subsections it was straightforward to write down the superspace Lagrangian $\cL^{(2)}$
for various invariants. To construct a Ricci squared invariant 
we will instead adopt a different approach and  first 
construct an invariant in full superspace. To construct locally supersymmetric invariants one 
can use the conventional $\cN = 3$ locally supersymmetric action
\be S = \ri \int \rd^{3|6} z \, E \,\cL \ , \label{fullSuperspaceInvN=3}
\ee
where the Lagrangian $\cL$ is a dimensionless primary scalar superfield
\be
\mathbb D \cL = 0 \ , \quad K_A \cL = 0 \ .
\ee
It was shown in \cite{KN14} that the above action can be recast in the form \eqref{7.1} with
\be \cL^{(2)} = 2 \D^{(4)} \frac{G \cL}{G^{(2)}} \ . \label{KNresult-1}
\ee
and
\bea
{\bm G}^{(2)} 
=\D^{(4)}
\oint_{\hat \g} \frac{({\hat v},\rd {\hat v})}{ 2\pi (v, {\hat v} )^2}  \frac{\cL^{(2)} ( {\hat v}) }
{G^{(2)}  ( {\hat v})} ~. \label{KNresult-2}
\label{C15.b}
\eea

One can consider the general actions of the form
\be 
S = \ri \int \rd^{3|6} z \, E \,\frak{L}(\mathscr T , \mathscr D \mathscr T  , \mathscr D^2 \mathscr T  , \cdots ) \ , \label{genActionN=3}
\ee
where $\frak{L}$ is a dimensionless superfield constructed out of the torsion 
components $\mathscr T$ of the covariant derivatives $\mathscr D_A$.
One can in principle construct general higher derivatives couplings beyond those of the 
previous subsections using the action \eqref{genActionN=3}. As in the previous sections we will focus 
our attention on actions containing at most curvature squared terms and thus search for an invariant 
containing a Ricci squared term.

To construct a Ricci squared term we will 
consider the invariant
\be S = \ri \int\rd^{3|6}z \, E \ \cL_{\z , \r} 
\ ,
\ee
with
\be \cL_{\z , \r} = \frac{\z}{G^2} \mathscr S^{IJ} G_I G_J + \r \mathscr S \ ,
\ee
and $\zeta$ and $\r$ are arbitrary constants. Using eq. \eqref{KNresult-1} and \eqref{KNresult-2} we can rewrite the full superspace action in 
terms of a $BF$ action with the composite $\cO(2)$ multiplet
\bea
{\bm G}^{(2)}(v) 
=  2 \D^{(4)}
 \oint_{\hat \g} \frac{({\hat v},\rd {\hat v})}{ 2\pi (v, {\hat v} )^2}  
 \frac{1}{\big(G^{(2)}( {\hat v})\big)^2}
 \D^{(\hat{4})} (\hat v) 
 (G \cL_{\z , \r})
 \ .
\eea
It can be shown that the above composite for $\z = 0$ or $\r = 0$ leads to a Ricci curvature squared term in the action 
since the action will contain a term proportional to $\frac{1}{g^3}g_I\Box^2g^{I}$. Furthermore, one can show that the 
Ricci curvature squared contributions cancel for $\r = 4 \z$. However, the corresponding invariant does not 
coincide with a linear combination of the previous invariants.
Thus we expect that by using the action corresponding to $\cL_{1 , 4}$ one can construct an invariant containing a 
scalar curvature squared contribution independent of the ${R} h^2$ invariant already constructed.

It should be mentioned that since $\cL$ is dimensionless we can let $\cL = 1$ without 
breaking the dilatation symmetry. Upon doing so it is not immediately obvious whether 
the invariant \eqref{fullSuperspaceInvN=3} gives another non-vanishing invariant, 
which is a full superspace volume. However, it is not difficult to verify that it does indeed vanish. To see this we 
construct the action using the superspace Lagrangian
\be \cL^{(2)} = 2 \D^{(4)} \frac{G}{G^{(2)}} \ .
\ee
One can check that it is proportional to the composite $\mathbb G^{(2)}$,
\be \cL^{(2)} = 2 \mathbb G^{(2)} \ .
\ee
From which it is clear that the full superspace volume must vanish, 
\bea
 \int \rd^{3|6}z \, E =0 \ .
 \label{6.611}
 \eea

It would be interesting to re-derive our result  \eqref{6.611} using the normal coordinates techniques of \cite{Kuzenko:2008ry}. 
Eq.  \eqref{6.611} is actually quite remarkable.
The point is that there exist only three conformal supergravity theories for which 
one can define a full superspace volume (without use of any compensator), specifically: 2D $\cN=(2,2)$, 
3D $\cN=3$ and 4D $\cN=2$. It is only in these cases that the superspace measure 
is dimensionless, and therefore invariant under the super-Weyl transformations.
In the  2D $\cN=(2,2)$ case, the corresponding full superspace volume is vanishing, 
as follows from eq. (4.1) in \cite{Grisaru:1995dr}. 
In the  4D $\cN=2$ case, the full superspace volume also vanishes, 
as follows from eqs. (3.22) and (3.23) in \cite{Kuzenko:2008ry}.
Eq.  \eqref{6.611} tells us that this property holds in the 
remaining case, and is therefore generic. 
It should be remarked that the 3D $\cN=2$ property
\bea
S = \int \rd^{3|4}z \, E \,G =0
\eea
can also be interpreted as the vanishing superspace volume in type II
supergravity provided one makes use of the covariant derivatives \eqref{4.111}, 
which is analogous to  the new minimal formulation for 4D $\cN=1$ supergravity
\cite{Howe:1981et}. However, the latter result holds in the presence of a conformal 
compensator and thus corresponds to Poincar\'e supergravity.


\section{Concluding comments} \label{section7}

Using the off-shell formulations for 3D $\cN$-extended conformal supergravity
\cite{KLT-M11,BKNT-M1} and the results in \cite{KT-M11,Kuzenko12, KLRST-M, BKNT-M2, KN14}, 
in this paper we have developed the geometric superspace settings to construct arbitrary higher derivative couplings (including $R^n$ terms)
in supergravity theories with $\cN\leq 3$.  We have concentrated on the 
explicit construction of all supersymmetric invariants with up to and including four derivatives, 
since these invariants are used in the models for $\cN=1$ and $\cN=2$ massive supergravity
advocated in \cite{Andringa:2009yc,BHRST10,BOS14}. 

All four-derivative invariants in $\cN=1$ supergravity were constructed in 
\cite{Andringa:2009yc,BHRST10} using component techniques. 
However, these papers did not provide tools to generate arbitrary higher order invariants.  
The novelty of our superspace approach is twofold: (i) our construction 
is geometric; (ii) it allows one to generate supersymmetric invariants of arbitrary order 
in powers of curvature and its covariant derivatives.

There are three off-shell formulations for 3D $\cN=2$ Poincar\'e and
AdS supergravity theories \cite{KLT-M11,KT-M11}:
(i) type I minimal; (ii) type II minimal; and (iii) non-minimal. 
For the minimal $\cN=2$ supergravity theories, four-derivative invariants were
derived in \cite{BOS14} using the component superconformal tensor calculus. 
In the present paper, we have developed an alternative approach which is 
not only geometric but also possesses, unlike the one of \cite{BOS14}, 
the following key properties: 
(i) it allows the generation of 
supersymmetric invariants of arbitrary order 
in powers of the curvature and its covariant derivatives; 
(ii) it keeps manifest the local superconformal symmetry; 
and (iii)  it does not make use of any gauge choice in deriving the curvature squared 
invariants. 
The important point of our constructions is that 
they
provide a 
complete description of the fourth order invariants for the minimal supergravity theories. 
In particular, we have described the Ricci curvature squared invariant in type I 
supergravity beyond the bosonic level originally given in \cite{BOS14}. 
For the case of 
type II supergravity we have provided a geometric explanation 
for the gauge-dependent 
procedure used in \cite{BOS14} 
to construct the Ricci curvature squared invariant.
We have also given a simple geometric reason 
for the non-existence of two independent invariants containing $R h^2$ terms. 
Finally,  we have provided for the first time a geometric 
setup to construct arbitrary higher derivative invariants within non-minimal supergravity.

Using the supergravity invariants given in the 
$\cN=1$ and minimal $\cN=2$ supergravity cases,  we have 
constructed general supergravity models with curvature-squared 
and lower order terms in order to study models for massive supergravity. 
In these cases we have derived the superfield equations 
of motion, linearised them about maximally  supersymmetric backgrounds and 
obtained restrictions on the parameters that have lead to models for massive supergravity. 
To derive the superfield equations of motion, we have worked out in appendices 
\ref{prepotentialN=1} and \ref{prepotentialN=2} 
the response of all geometric objects to an infinitesimal prepotential deformation. 
For type I supergravity we have identified a new massive supergravity model 
which does not propagate any degrees of freedom associated with the 
component field $M$ about any AdS background satisfying \eqref{Rcubic_eq}. 
In the the non-minimal formulation for $\cN = 2$ supergravity 
we have constructed for the first time  
a novel consistent theory of massive  supergravity. 

In the case of $\cN = 3$ supergravity we have considered the 
off-shell formulation with a compensating vector multiplet. 
For this supergravity theory the AdS action was given in \cite{KLT-M11}, 
the off-shell $\cN=3$ conformal supergravity action 
was constructed in \cite{BKNT-M2} and the model for topologically 
massive supergravity was studied in \cite{KN14}. In the present paper 
we have constructed for the first time new higher derivative invariants 
with four and less derivatives.
The new $\cN=3$ invariants constructed appear to be analogous to 
the invariants in the type II case in $\cN = 2$ supergravity. However, by using the vector multiplet as compensator it does not seem possible to 
remove one of the torsion components $\cS$, $\cS^{IJ}$ or $\cC_a{}^{IJ}$ by gauge fixing. Thus 
it appears that more invariants are possible in the $\cN = 3$ case. Perhaps the most interesting point raised in 
our $\cN=3$ analysis is that there appears to exist an independent scalar curvature squared invariant in contrast to the the type II case.
Our results provide the building blocks for the construction of general  
massive $\cN = 3$ supergravity theories.

The off-shell $\cN = 3$ formulation with a 
compensating vector multiplet corresponds to 
$(2 , 1)$ AdS supergravity. So far the appropriate 
compensator for $(3,0)$ AdS supergravity is 
not known. 
However, one expects the compensator to be described by a 
Lorentz and $\rm SO(3)$ scalar primary superfield $Y$, while
without loss of generality we can take this compensator 
to be have dimension 1, $\bbD Y = Y$. It is expected that $(3,0)$ AdS superspace is 
a solution to the equations of motion for 
$\cN = 3$ supergravity coupled to the compensator. 
On the other hand one can describe $(3,0)$ AdS superspace by imposing some 
differential constraints on $Y$. The appropriate constraints are\footnote{
In the isospinor notation, 
the constraint \eqref{constraints(3,0)b} can be written in 
the compact form $\de^{\a (ij}\de_\a^{kl)} Y = 0$. The constraint \eqref{constraints(3,0)a} 
implies the vanishing of the super-Cotton tensor. 
In the super-Poincar\'e case, 
the constraint \eqref{constraints(3,0)b} describes one of the two $\cN=3$ multiplets 
obtained by reducing the $\cN=4$ supercurrent 
to $\cN=3$ Minkowski superspace \cite{BKS}, with the second 
multiplet being the $\cN=3$ supercurrent.
}
\bsubeq \label{constraints(3,0)}
\bea
\nabla^{\g (I} \nabla_\g^{J)} Y &=& \frac{1}{3} \d^{IJ} \nabla^\g_K \nabla_\g^K Y \label{constraints(3,0)b} \ , \\
\nabla_{(\a}^{[I} \nabla_{\b)}^{J]} Y^{-1} &=& 0 \label{constraints(3,0)a} \ .
\eea
\esubeq
To see this, one just degauges to conventional superspace \cite{BKNT-M1} 
and imposes the gauge condition $Y = 1$.
We find the following constraints on the torsion superfields:
\be 
\cS^{IJ} = 0 
\ ,~~~~~~
\cC_a{}^{IJ}=0 \ ,
\ee
which defines $(3,0)$ AdS superspace \cite{1205.4622}. It is also interesting to note that
the constraints \eqref{constraints(3,0)} implies the equation
\be  \nabla^\g_K \nabla^K_\g Y^{- \hf} - 6 \ri \mu Y^{\hf} = 0 \ , \quad \mu = {\rm const} \ ,
\ee
where $\mu$ coincides with $\cS$ in the gauge $Y = 1$.
There appears to be a striking similarity between the constraint \eqref{constraints(3,0)b} and the one 
defining the $\cN = 2$ off-shell vector multiplet suggesting that perhaps one should treat it as 
an off-shell condition for the compensator. We thus suggest that the constraint \eqref{constraints(3,0)b} defines 
an off-shell multiplet while \eqref{constraints(3,0)a} is an on-shell condition derived from 
an appropriate action which is currently unknown.

In this paper we have restricted our attention to supergravity with $\cN \leq 3$.
Keeping in mind certain similarities between the general $\cN=3$ and $\cN=4$ supergravity-matter systems, 
see \cite{KLT-M11} for more details,
it is natural to expect that techniques analogous to those used 
for the $\cN = 3$ case can be applied to the $\cN = 4$ case in order 
to construct higher derivative couplings.

In conclusion, we give a bi-product of our analysis and present superfield 
equations for massive higher spin multiplets in (1,0),  (1,1) and (2,0) anti-de Sitter superspaces.
A massive higher spin multiplet in $\cN=1$ AdS superspace, eq.   \eqref{N=1AdSSS},
is described by a real symmetric rank-$n$ spinor, 
$T_{\a_1 \cdots \a_n} = T_{(\a_1 \cdots \a_n)}$,  
constrained by 
\begin{subequations}
\bea
\cD^\b  T_{\a_1 \cdots \a_{n-1} \b} &=& 0~,  \\
\Big( \frac{\ri}{2} \cD^2 + m \big)  T_{\a_1 \cdots \a_n} &=&0~,
\eea
\end{subequations}
with $m$ a real mass parameter. It can be shown that
\bea
\Big( \frac{\ri}{2} \cD^2 \Big)^2 T_{\a_1 \cdots \a_n} 
=
\Big( \cD^a \cD_a
 - \ri (n + 2) \cS \cD^2 
 - n (n + 2) \cS^2 \Big) T_{\a_1 \cdots \a_n} \ ,
\eea
where the second term on the right can be rewritten as follows:
\be 
\frac{\ri}{2} \cD^2 T_{\a_1 \cdots \a_n} = 
-
\cD_{(\a_1}{}^\b T_{\a_2 \cdots \a_n ) \b} 
-
(n + 2) \cS T_{\a_1 \cdots \a_n} \ .
\ee

A massive higher spin multiplet in (1,1) AdS superspace, 
which corresponds to the algebra \eqref{4.37} with $\cC_{\a\b}=0$, 
is described by a real symmetric rank-$n$ spinor, 
$T_{\a_1 \cdots \a_n} = T_{(\a_1 \cdots \a_n)}$ 
constrained by 
\begin{subequations} \label{7.77}
\bea
\cD^\b T_{ \a_1 \cdots \a_{n-1} \b} 
= \bar \cD^\b T_{ \a_1 \cdots \a_{n-1} \b}&=&0~, \\ 
\Big( \frac{\ri}{2} \cD^\g\bar \cD_\g + m \Big) T_{\a_1 \cdots \a_n} &=&0~.
\eea
\end{subequations}
It can be shown that 
\bea \label{7.8'}
\Big( \frac{\ri}{2} \cD^\g\bar \cD_\g\Big)^2 T_{\a_1 \cdots \a_n}  = 
\Big( \cD^a \cD_a + 2 (n + 2) |\cR|^2 \Big) T_{\a_1 \cdots \a_n} \ .
\eea

In the case of (2,0) AdS superspace, which corresponds to the algebra 
 \eqref{4.544} with $\cC_{a} =0$,  
 massive higher spin multiplets are also  described 
 by the equations
 \eqref{7.77}. However, the identity \eqref{7.8'} turns into  
\bea
\Big(\frac{\ri}{2}\cD^\g\cDB_\g\Big)^2 T_{\a_1\cdots\a_n}
&=&
\Big( \cD^{a} \cD_{a} 
+(n+2) \ri \cS\cD^\g\cDB_\g 
-n(n+2)\cS^2 \Big) T_{\a_1\cdots\a_n}
~,~~~~~~~~~~~~
\eea
where the second term on the right can be rewritten as follows:
\bea
\frac{\ri}{2}\cD^\g\cDB_\g T_{\a_1\cdots\a_n}
&=&
\cD_{(\a_1}{}^{\g} T_{\a_2\cdots\a_n)\g}
+(n+2)\cS T_{\a_1\cdots\a_n} \ .
\eea

$~$\\
\noindent
{\bf Acknowledgements:}\\
This work was supported
in part by the Australian Research Council (ARC) Discovery Project DP140103925.
The work of GT-M and JN was supported by the ARC Discovery Early Career Award (DECRA), project No. DE120101498.
GT-M is grateful to the Instituut voor Theoretische Fysica, KU Leuven, for kind hospitality during January--May 2015.
From July 2015 the work of GT-M is supported by the Interuniversity Attraction Poles Programme
initiated by the Belgian Science Policy (P7/37)
and in part by COST Action MP1210 
``The String Theory Universe.''


\appendix


\section{Geometry of 
conformal superspace} \label{geometry}

Here we collect the essential details of the $\cN$-extended 
superspace geometry developed in \cite{BKNT-M1} for the cases
$\cN=1,2,3$.

We begin with a curved three-dimensional $\cN$-extended superspace
$\cM^{3|2 \cN}$ parametrized by
local bosonic $(x^m)$ and fermionic coordinates $(\theta^\m_I)$:
\be z^M = (x^m, \q^\mu_I) \ ,
\ee
where $m = 0, 1, 2$, $\mu = 1, 2$ and $I = 1, \cdots , \cN$. The 
structure group is chosen to be ${\rm{OSp}}(\cN|4, {\mathbb R})$ and 
the covariant derivatives have the form
\be
\nabla_A = E_A{}^M \pa_M - \o_A{}^{\underline b} X_{\underline b} 
= E_A{}^M \pa_M - \hf \Omega_A{}^{ab} M_{ab} - \hf \Phi_A{}^{PQ} N_{PQ} - B_A \mathbb D - \mathfrak{F}_A{}^B K_B \ .
\ee
Here $E_A = E_A{}^M(z) \partial_M$ is the inverse supervielbein, $M_{ab}$ are the Lorentz generators, $N_{IJ}$ are generators of the 
$\rm SO(\cN)$ group, $\mathbb D$ is the dilatation generator and $K_A = (K_a , S_\a^I)$ are the special superconformal 
generators.\footnote{As usual, we refer to $K_a$ as the special conformal generator and $S_\a^I$ as the $S$-supersymmetry 
generator.} The supervielbein $E^A = \rd z^M E_M{}^A$ is defined such that
\be E_M{}^A E_A{}^N = \d_M^N \ , \quad E_A{}^M E_M{}^B = \d^B_A \ .
\ee

The Lorentz generators obey
\bsubeq
\bea
&[M_{ab} , M_{cd}] = 2 \eta_{c[a} M_{b] d} - 2 \eta_{d [a} M_{b] c} \ , \\
&[M_{ab} , \nabla_c ] = 2 \eta_{c [a} \nabla_{b]} \ , \quad [M_{\a\b} , \nabla_\g^I] = \eps_{\g(\a} \nabla_{\b)}^I \ .
\eea
\esubeq
The $\rm SO(\cN)$ and dilatation generators obey
\bsubeq
\bea
&[N_{KL} , N^{IJ}] = 2 \d^I_{[K} N_{L]}{}^J - 2 \d^J_{[K} N_{L]}{}^I \ , \quad [N_{KL} , \nabla_\a^I] = 2 \d^I_{[K} \nabla_{\a L]} \ ,  \\
&[\mathbb D , \nabla_a] = \nabla_a \ , \quad [\mathbb D , \nabla_\a^I] = \nabla_\a^I \ .
\eea
\esubeq
The special conformal generators $K_A$ transform under Lorentz and $\rm SO(\cN)$ transformations as
\begin{align}
[M_{ab} , K_c] = 2 \eta_{c[a} K_{b]} \ , \quad [M_{\a\b} , S_\g^I] = \eps_{\g(\a} S_{\b)}^I \ , \quad
[N_{KL} , S_\a^I] = 2 \d^I_{[K} S_{\a L]} \ ,
\end{align}
while under dilatations as
\begin{align}
[\mathbb D , K_a] = - K_a \ , \quad [\mathbb D, S_\a^I] &= - \hf S_\a^I \ .
\end{align}
Among themselves, the generators $K_A$ obey the algebra
\begin{align}
\{ S_\a^I , S_\b^J \} = 2 \ri \d^{IJ} (\g^c)_{\a\b} K_c \ .
\end{align}
Finally, the algebra of $K_A$ with $\nabla_A$ is given by
\bsubeq
\begin{align}
[K_a , \nabla_b] &= 2 \eta_{ab} \mathbb D + 2 M_{ab} \ , \\
[K_a , \nabla_\a^I ] &= - \ri (\g_a)_\a{}^\b S_\b^I \ , \\
[S_\a^I , \nabla_a] &= \ri (\g_a)_\a{}^\b \nabla_{\b}^I \ , \\
\{ S_\a^I , \nabla_\b^J \} &= 2 \eps_{\a\b} \d^{IJ} \mathbb D - 2 \d^{IJ} M_{\a \b} - 2 \eps_{\a \b} N^{IJ} \ .
\end{align}
\esubeq
All other (anti-)commutators vanish.

The covariant derivatives obey the (anti-)commutation relations of the form
\begin{align}
[ \nabla_A , \nabla_B \}
	&= -T_{AB}{}^C \nabla_C
	- \frac{1}{2} \RM_{AB}{}^{cd} M_{cd}
	- \frac{1}{2} \RN_{AB}{}^{PQ} N_{PQ}
	\non \\ & \quad
	- \RD_{AB} \mathbb D
	- \RS_{AB}{}^\g_K S_\g^K
	- \RK_{AB}{}^c K_c~,
\end{align}
where $T$ is the torsion, and $\RM$, $\RN$, $\RD$, $\RK$ are the curvatures.

The covariant derivatives transform under the conformal supergravity gauge group as follows
\bea 
\d_\cG \nabla_A &=& [\cK , \nabla_A] \ ,
\label{TransCD}
\eea
where $\cK$ denotes the first-order differential operator
\bea
\cK = \xi^C \nabla_C + \hf \L^{ab} M_{ab} + \hf \L^{IJ} N_{IJ} + \s \mathbb D + \L^A K_A ~.
\eea
Covariant (or tensor) superfields transform as
\bea 
\d_{\cG} T &=& \cK T~.
\eea

The algebra of covariant derivatives are constrained entirely in terms of a single 
primary superfield, the super-Cotton tensor. It is used 
in the construction of the component fields of the 3D Weyl multiplet \cite{BKNT-M2}. 
The super-Cotton tensor takes different forms for the $\cN=1,2,3$ cases. We 
summarise these cases below.


\subsection{The $\cN = 1$ case}

The $\cN=1$ super-Cotton tensor $W_{\a\b\g}$ is a symmetric primary superfield of dimension-$5/2$
\be S_\d W_{\a\b\g} = 0 \ , \quad \mathbb D W_{\a\b\g} = \frac{5}{2} W_{\a\b\g} ~.
\ee
The algebra of covariant derivatives is given by
\begin{subequations}
\begin{align} \{ \nabla_\a , \nabla_\b \} &= 2 \ri \nabla_{\a\b} \ , \\
[ \nabla_a , \nabla_\a ] &= \frac{1}{4} (\g_a)_\a{}^\b W_{\b \g\d} K^{\g\d} \ , \\
[\nabla_a , \nabla_b] &= - \frac{\ri}{8} \eps_{abc} (\g^c)^{\a\b} \nabla_\a W_{\b\g\d} K^{\g\d} - \frac{1}{4} \eps_{abc} (\g^c)^{\a\b} W_{\a\b\g} S^\g \ ,
\end{align}
\end{subequations}
The Bianchi identities imply an additional constraint on $W_{\a\b\g}$,
\be \nabla^\a W_{\a \b\g} = 0 \ .
\ee


\subsection{The $\cN = 2 $ case}

Here we make use of the complex basis for the $\cN=2$ covariant derivatives $\de_{A}=(\de_\a,\deb^\a,\de_a)$,
see \cite{BKNT-M1} for more details. 
The complex spinor covariant derivatives have definite U(1) charges:
\be [{J} , \nabla_\a] = \nabla_\a \ , \quad [{J} , \bar{\nabla}_\a] = - \bar{\nabla}_\a \ ,
\ee
with the U(1) generator defined by
\be 
{J} := - \frac{\ri}{2} \eps^{KL} N_{KL} \ .
\ee
The $\rm SO(2)$ connection and curvature take the form
\be \hf \Phi_A{}^{KL} N_{KL} = \ri \Phi_A {J} \ , \quad \hf \RN_{AB}{}^{KL} N_{KL} = \ri \RJ_{AB} {J} \ .
\ee
The conjugation rule in the complex basis is
\be (\nabla_\a F)^* = (-1)^{\eps(F)} \bar{\nabla}_\a \bar{F} \ ,
\ee
where $F$ is a complex superfield and $\bar{F} = (F)^*$ is its complex conjugate.

The 
super-Cotton tensor $W_{\a\b}$ is a symmetric primary superfield of dimension $2$
\be S_\g W_{\a\b} = 
0 \ , \quad \mathbb D W_{\a\b} = 2 W_{\a\b} ~.
\ee
As in the $\cN=1$ case, its spinor divergence vanishes,
\be \nabla^{\a} W_{\a\b} = 0 \ .
\ee

In the complex basis $(\nabla_\a , \bar{\nabla}_\a)$, the covariant derivative algebra 
takes the form
\begin{subequations} \label{N=2AlgebraCB}
\begin{align} 
\{ \nabla_\a , \nabla_\b \} &= 0 \ , 
\\
\{ \nabla_\a , \bar{\nabla}_\b \} &= - 2 \ri \nabla_{\a\b} - \eps_{\a\b} W_{\g\d} K^{\g\d}\ , \\
[\nabla_a , \nabla_\b] &= \frac{\ri}{2} (\g_a)_\b{}^\g \nabla_\g W^{\a\d} K_{\a\d} - (\g_a)_{\b\g} W^{\g\d} \bar{S}_\d \ , \\
[\nabla_a , \nabla_b] &= - \frac{\ri}{8} \eps_{abc} (\g^c)^{\g\d} \Big( \ri [\nabla_\g , \bar{\nabla}_\d] W_{\a\b} K^{\a\b} + 4 \bar{\nabla}_\g W_{\d\b} \bar{S}^\b + 4 \nabla_\g W_{\d \b} S^\b \\
&\qquad - 8 W_{\g\d} {J} \Big) \ .
\end{align}
\end{subequations}
Here the generators $M_{ab}, J, \mathbb{D}, S_\a,\bar S^\a, K_a$ and the covariant derivatives 
$\de_{A}$
satisfy the following algebraic relations 
{\allowdisplaybreaks\begin{subequations} \label{gensCB}
\begin{gather}
[M_{\a\b} , \nabla_\g] = \eps_{\g(\a} \nabla_{\b)} \ , 
\quad [\mathbb D, \nabla_\a] = \hf \nabla_\a \ , 
\\
\{ S_\a , S_\b \} = 0 \ , 
\quad 
\{ S_\a , \bar{S}_\b \} = 2 \ri K_{\a\b} \ , \\
[S_\a , K_b] = 0 \ , \\
[M_{\a\b} , S_\g] = \eps_{\g(\a} S_{\b)}\ ,
\quad 
[{J} , S_\a] = - S_\a
 \ , 
\quad 
[\mathbb D, S_\a] = - \hf S_\a \ , 
\\
[K_a , \nabla_\a ] = - \ri (\g_a)_\a{}^\b \bar{S}_\b \ , 
\quad
[S_\a , \nabla_a] = - \ri (\g_a)_\a{}^\b \bar{\de}_{\b} \ , \\
\{ \bar{S}_\a , \nabla_\b \} = 0 \ , 
\quad
\{ S_\a , \nabla_\b \} = 2 \eps_{\a\b} \mathbb D -
2 M_{\a\b} - 2 \eps_{\a\b} {J} \ ,
\end{gather}
\end{subequations}}
together with their complex conjugates.


\subsection{The $\cN = 3$ case}

The $\cN = 3$ super-Cotton tensor $W_{\a}$ is a primary superfield of dimension $3/2$
with vanishing spinor divergence,
\be S_\b^I W_{\a} = 0 \ , \qquad \mathbb D W_{\a} = \frac{3}{2} W_{\a}~, \qquad
\nabla^{\alpha I} W_\alpha = 0~.
\ee
The algebra of covariant derivatives is
\begin{subequations} \label{N=3Algebra}
\begin{align} 
\{ \nabla_\a^I , \nabla_\b^J \} &= 2 \ri \d^{IJ} \nabla_{\a\b} - 2 \eps_{\a\b} \eps^{IJL} W^\g S_{\g L} + \ri \eps_{\a\b} (\g^c)^{\g\d} \eps^{IJK} (\nabla_{\g K} W_\d) K_c \ , \\
[\nabla_a , \nabla_\b^J ] &= \ri \eps^{JKL} (\g_a)_{\b\g} W^\g N_{KL} + \ri \eps^{JKL} (\g_a)_{\b\g} (\nabla^\g_K W^\d) S_{\d L} \non\\
&\qquad + \frac{1}{4} \eps^{JKL} (\g_a)_{\b\g} (\g^c)_{\d\rho} (\nabla^\g_K \nabla^\d_L W^\rho) K_c \ , \\
[\nabla_a , \nabla_b] &= \eps_{abc} (\g^c)_{\a\b} \Big[ - \hf \eps^{IJK} (\nabla^\a_I W^\b) N_{JK} - \frac{1}{4} \eps^{IJK} (\nabla^\a_I \nabla^\b_J W^\g ) S_{\g K} \non\\
&\qquad +\frac{\ri}{24} \eps^{IJK} (\g^d)_{\g\d} (\nabla^\a_I \nabla^\b_J \nabla^\g_K W^\d) K_d \Big] \ .
\end{align}
\end{subequations}


\section{Results for $\cN = 1$ prepotential deformation} \label{prepotentialN=1}

In order to compute the equations of motion corresponding to the supergravity action 
\eqref{N=1_SG_model}, it is necessary to know how the functionals listed in subsection 
\ref{subsection2.3} depend on the unconstrained prepotential for $\cN=1$ conformal supergravity, which is a real symmetric rank-3 spinor $\J_{\a\b\g}$ in accordance  
with the prepotential formulation for 3D $\cN =1$ supergravity sketched 
in \cite{GGRS}. Here we will build on the ideas put forward in 
the classic papers by Grisaru and Siegel \cite{GrisaruSiegel} devoted to the background field method
in 4D $\cN=1$ supergravity (see \cite{Ideas} for a pedagogical review and applications).

In $\cN=1$ conformal supergravity, the  gauge group consists of 
(i) the super-Weyl transformations \eqref{N=1sW}; 
and (ii) the superspace general coordinate 
and local Lorentz transformations, which have the infinitesimal form 
\bea
\d_\cK\cD_A=[\cK,\cD_A]~,~~~~~~
\cK=\x^B (z) \cD_B+\hf K^{bc} (z) M_{bc}
~,
\label{SUGRA-gauge-group1}
\eea
with the gauge parameters $\x^B$ and $K^{bc}$ obeying natural reality conditions 
but otherwise arbitrary.

Let ${\mathbb D}_A$ be some other set of covariant derivatives 
which differ from $\cD_A$ by finite deformations 
but satisfy the same (anti-)commutation relations \eqref{cdaN=1-con} as 
the covariant derivatives $\cD_A$. We can represent 
\bea
{\mathbb D}_A = {\mathbb E}_A{}^B \cD_B -\hf {\bm \O}_A{}^{bc} M_{bc}~, 
\label{B.2}
\eea
for some tensor superfields ${\mathbb E}_A{}^B  $ and  ${\bm \O}_A{}^{bc} $.
In such a setting, the gauge transformations \eqref{N=1sW} and 
\eqref{SUGRA-gauge-group1} can be realised in two different incarnations: 
(i) as ``background'' transformations; and (ii) as  ``quantum'' transformations 
\cite{GrisaruSiegel}. The latter gauge freedom associated with 
super-Weyl ($\s$), coordinate ($\x^\b$) and local Lorentz ($K^{bc})$ 
parameters may be used to bring the operator ${\mathbb D}_\a$ to
the form: 
\bea
\mathbb{D}_\a
&=&
{\cD}_\a
+\ri\Psi_{\a\g\d} \cD^{\g\d}
-\hf {\bm \O}_{\a}{}^{\g\d}M_{\g\d} \ ,\qquad \J_{\a\b\g} = \J_{(\a\b\g)} \ . 
\label{B4}
\eea
The deformed connection
${\bm \O}_{\a}{}^{\g\d}$ may be determined as a function of  
the prepotential $ \J_{\a\b\g}$
by requiring the spinor derivatives $\mathbb D_\a$ to satisfy the 
same algebra 
as that of conventional superspace \eqref{cdaN=1-con}. Specifically, we 
require
\bsubeq\label{deformedAlgebraCovDerN=1}
\bea
\{ \mathbb D_\a , \mathbb D_\b \} &=& 2 \ri \mathbb D_{\a\b} - 4 \ri \mathbb S M_{\a\b} \ , \\
\left[ \mathbb D_{\a \b} , \mathbb D_\g \right] &=& 
- 2 \eps_{\g(\a} \mathbb S \mathbb D_{\b)} + 2 \eps_{\g(\a} \mathbb C_{\b) \d \r} M^{\d\r}
\non\\
&&
+ \frac{2}{3} \big( \cD_\g \mathbb S M_{\a\b} 
- 4 \cD_{(\a} \mathbb S M_{\b) \g} \big)
\ ,
\eea
\esubeq
where we define the deformed vector derivative by
\be 
\mathbb D_a = \frac{\ri}{4} (\g_a)^{\a\b} \{ \mathbb D_{\a} , \mathbb D_{\b} \}
+ 2 \mathbb S M_a \ .
\ee
The above torsion superfields $\mathbb C_a$ and $\mathbb S$ are some functions of the prepotential $\Psi_{\a\b\g}$ 
and its covariant derivatives.

Requiring the algebra \eqref{deformedAlgebraCovDerN=1} fixes ${\bm \O}_{\a}{}^{\g\d}$ 
as a function of $\Psi_{\a\b\g}$ and its covariant derivatives.
The deformed spinor covariant derivative is
\bea
\mathbb{D}_\a
&=&
\cD_\a
+\ri\Psi_{\a\g\d}\cD^{\g\d}
-\frac{1}{4}\cD^2\Psi_{\a\b\g} M^{\b\g}
-\frac{\ri}{2} \cD_{(\a}{}^{\d}\Psi_{\b\g)\d} M^{\b\g} \non\\
&&
- \frac{2\ri}{3} \cD_{\b \g} \Psi^{\b\g\d} M_{\d\a}
+ \frac{3\ri}{2}\cS\Psi_{\a\b\g}M^{\b\g}~+~\cO(\J^2)
~.
\eea
Here we have omitted all the terms of second and higher order in $\J_{\a\b\g}$, 
for these terms are not necessary for our goals in the present paper. 
All the results below also hold modulo terms quadratic in $\J_{\a\b\g}$, 
but we do not write explicitly $\cO(\J^2)$.
Since $\mathbb{D}_\a$ has been determined, 
requiring the  (anti-)commutation relations \eqref{deformedAlgebraCovDerN=1} fixes the 
torsion superfields as follows:
\bsubeq
\bea
{\mathbb S}
&=&
\cS
-\frac{1}{8}\Big(\cD^{(\a}\cD^{\b\g)}+2\cC^{\a\b\g}\Big)\Psi_{\a\b\g} \ ,\\
{\mathbb C}_{\a\b\g}
&=&
\cC_{\a\b\g}
+\frac{1}{4}\cD_{(\a}{}^{\d}\cD^2\Psi_{\b\g)\d}
+\frac{\ri}{2}\cD_{(\a}{}^{\d}\cD_{\b}{}^{\r}\Psi_{\g)\d\r}
- \frac{1}{4} \cS\cD^2\Psi_{\a\b\g} \non\\
&&
-\frac{5}{3}\cC_{(\a}{}^{\r\t}\cD_{\b}\Psi_{\g\r\t)}
- \frac{4}{3}(\cD^\d\cS)\cD_{(\d}\Psi_{\a\b\g)}
+\frac{1}{2}\cC_{(\a\b}{}^\r\cD^{\t}\Psi_{\g)\r\t}
\non\\
&&+ \frac{1}{3}(\cD_{(\a}\cS)\cD^{\t}\Psi_{\b\g)\t}
+\frac{\ri}{2}(\cD_{(\a}{}^{\r}\cS)\Psi_{\b\g)\r}
-\frac{\ri}{2} \cS^2\Psi_{\a\b\g} \ .
\eea
\esubeq
Finally, the deformed vector covariant derivative is given by
\begin{align}
\mathbb{D}_{\a\b}
=&~
\cD_{\a\b}
+ \cD_{(\a}\Psi_{\b\g\d)}\cD^{\g\d}
+\frac{1}{2} \cD^{\d}\Psi_{\d\r(\a} \cD_{\b)}{}^{\r}
\non\\
&
+\Big{[}
-\frac{\ri}{4}\cD^2\Psi_{\a\b\g}
+\frac{1}{2}\cD_{(\a}{}^{\d}\Psi_{\b\g)\d}
-\frac{1}{3} \cD^{\r \d} \Psi_{\r \d(\a} \ve_{\b)\g}
+\hf\cS\Psi_{\a\b\g}
\Big{]}\cD^\g \non\\
&
+\Big{[}
-\frac{5}{4}\cD_{(\a}{}^{\r}\cD_{\b} \Psi_{\g\d\r)}
+\frac{1}{4}\cD_{(\a\b}\cD^\t \Psi_{\g\d)\t}
+\frac{1}{4}\cD^{\r\t}\Big(
\cD_{(\t}\Psi_{\g\r\a)}\ve_{\b\d}
+\cD_{(\t}\Psi_{\g\r\b)}\ve_{\a\d}
\Big)
\non\\
&~~~~~
-\frac{1}{12} \ve_{\g(\a}\ve_{\b)\d}\cD^{ \r\e} \cD^{\t}\Psi_{\r\t\e} 
+\frac{1}{8} 
\Big(
\cD_{(\a}{}^{ \r}\ve_{\b)(\g} \cD^{\t}\Psi_{ \d)\r\t} 
- \cD_{(\g}{}^{ \r}\ve_{\d)(\a} \cD^{\t}\Psi_{ \b)\r\t} 
\Big)
\non\\
&~~~~~
-\cC_{(\a\b}{}^{\t}\Psi_{\g\d)\t}
+\frac{1}{2}\cC^{\r\t}{}_{(\a}\ve_{\b)(\g}\Psi_{\d)\r\t}
-\frac{1}{2}\cC^{\r\t}{}_{(\g}\ve_{\d)(\a}\Psi_{\b)\r\t}
-\frac{1}{6}\ve_{\g(\a}\ve_{\b)\d}\cC^{\r\t\e}\Psi_{\r\t\e}
\non\\
&~~~~~
+ \frac{4}{3} (\cD_{(\a}\cS)\Psi_{\b\g\d)}
-\frac{2}{3}(\cD^{\r}\cS)\Psi_{\r\g(\a}\ve_{\b)\d}
\Big) 
\Big{]} M^{\g\d} \ .
\end{align}

For the derivation of the above results and the relations \eqref{2.522},
a number of identities prove to be useful. 
The most important identities are:
\bsubeq
\bea
\cD^2\cC_{\a\b\g}
&=&
2\ri\cD_{(\a}{}^\d \cC_{\b\g)\d}
+\frac{8\ri}{3}\cD_{(\a\b}\cD_{\g)}\cS
+10\ri \cS \cC_{\a\b\g}
~,
\\
\cD^{\b\g}\cC_{\a\b\g}
&=&
-\frac{4}{3}\cD_{\a\b}\cD^\b\cS
-4\cS\cD_\a\cS
~,
\\
\cD_{\a}\cD^2
&=&
-2\ri\cD_{\a\b}\cD^\b
+2\ri\cS\cD_\a
+4\ri\cS\cD^\b M_{\a\b}
+2\ri \cC_{\a\r\t}M^{\r\t}
\non\\
&&
+\frac{8\ri}{3}(\cD^{\d}\cS) M_{\a\d}
~,
\\
\cD_{(\a}{}^\d \cD_{\b}{}^\r \cC_{\g) \d\r}
&=&
\cD^a \cD_a \cC_{\a\b\g}
- \frac{4}{3} \cD_{(\a\b} \cD_{\g\d)} \cD^\d \cS
+2\ri  \cC_{\a\b\g} \cD^2 \cS
- 6 (\cD_{(\a\b} \cS) \cD_{\g)} \cS
\non\\
&&
- 4 \cS \cD_{(\a\b} \cD_{\g)} \cS 
+ 3 K_{\d\r(\a\b} \cC_{\g)}{}^{\d\r}
+ 10 \cS^2 \cC_{\a\b\g}
~,
\eea
\bea
{[}\cD_{(\a}{}^{\r},\cD^2{]}\cC_{\b\g)\r}
&=&
12\ri \cC_{\r(\a\b}\cD^\r{}_{\g)}\cS
-\frac{10\ri}{3}(\cD^\d\cS)K_{\a\b\g\d}
+\frac{10\ri}{3}(\cD_{(\a}\cS)\cD_{\b\g)}\cS
\non\\
&&
+\frac{5}{3}(\cD^2\cS)\cC_{\a\b\g}
~,
\\
\ve^{abc} (\g_c)_{(\a\b} [\cD_a , \cD_b ] \cD_{\g)} \cS
&=&
-\ri \cC_{\a\b\g}\cD^2 \cS
-2 \cC_{(\a\b}{}^{\r}\cD_{\g)\r} \cS
+{8\over 3}(\cD_{(\a}\cS)\cD_{\b\g)} \cS
\non\\
&&
-2K_{\a\b\g\d}\cD^{\d} \cS
~,
\\
\ve^{abc} (\g_c)_{(\a}{}^\d [\cD_a , \cD_b] \cC_{\b\g) \d}
&=&
6 K_{\d\r(\a\b} \cC_{\g)}{}^{\d\r}
- \frac{4}{3} (\cD^\d \cS) K_{\a\b\g\d}
- \frac{20}{9} (\cD_{(\a} \cS) \cD_{\b\g)} \cS
\non\\
&&
- \frac{4}{3} \cC_{(\a\b}{}^\d \cD_{\g) \d} \cS 
+ \frac{10 \ri}{3} (\cD^2 \cS) \cC_{\a\b\g}
+ 20 \cS^2 \cC_{\a\b\g}
~,
\eea
\esubeq
where we have denoted $K_{\a\b\g\d}:=\ri\cD_{(\a}\cC_{\b\g\d)}$.

In the gauge \eqref{B4}, 
there remains some residual gauge freedom. 
It is described by certain transformations of the form
\bea
\d\mathbb D_A:={[}\cK,\mathbb D_A{]}+\d_\s\mathbb D_A~,~~~~~~
\cK=\x^b\mathbb D_b +\x^\a\mathbb D_\a
+\hf K^{bc}M_{bc}~,
\eea
where $\d_\s\mathbb D_A$ denotes an infinitesimal  super-Weyl transformation parametrised by $\s$, which is obtained from 
eq. \eqref{N=1sW} by replacing $\cD_A \to {\mathbb D}_A$. 
In order to preserve the condition \eqref{B4}, it may be shown that the parameters
$\x^\a$, $K^{bc}$ and $\s$ should be functions of $\x^a$ and its covariant derivatives, 
with $\x^a$ being real unconstrained. 
Modulo $\J$-dependent terms, 
the parameters $\x^\a$, $K^{bc}$ and $\s$ have the explicit form 
\bsubeq
\bea
\x_\a&=&
-\frac{\ri}{6}\cD^\b\x_{\b\a}
~,
\\
K_{\a\b}
&=&
2\cD_{(\a}\x_{\b)}
-2\cS\x_{\a\b}
~,
\\
\s
&=&
\cD_\a\x^{\a}
~.
\eea
\esubeq
The gauge transformation of the prepotential is 
\bea
\d\Psi_{\a\b\g}
=
\hf\cD_{(\a}\x_{\b\g)}~+~ \cO(\J)
~.
\label{B.12}
\eea

Let $S=S[\cD_A, \vf]$ be a supergravity action such as \eqref{N=1_SG_model},
with $\vf$ being the compensator. The action has to be invariant under the supergravity gauge transformations \eqref{N=1sW} and \eqref{SUGRA-gauge-group1}. 
Assuming that the compensator obeys its equation of motion, 
$\d S / \d \vf =0$, we consider the variation of the action induced by an 
infinitesimal deformation of the gravitational superfield $\J_{\a\b\g}$, 
\bea 
\d  S[\cD_A, \vf] = \ri \int \rd^{3|2} z \,E \,\d \Psi^{\a\b\g} {\mathbb T}_{\a\b\g}~, 
\eea
for some superfield ${\mathbb T}_{\a\b\g}$. This variation must vanish 
if $\d \Psi^{\a\b\g} $ is the gauge transformation \eqref{B.12}. Since the gauge parameter
$\x_{\b\g} $ in \eqref{B.12} is an arbitrary superfield, we conclude that 
\bea 
\cD^\g  {\mathbb T}_{\a\b\g} = 0~.
\eea


\section{Results for $\cN = 2$ prepotential deformation} \label{prepotentialN=2}

The prepotential formulation for $\cN=2$ conformal supergravity was given in 
\cite{Kuzenko12} as a generalisation of the prepotential solution in 4D $\cN = 1$ supergravity \cite{Siegel78}.
 Modulo purely gauge degrees of freedom, 
the 3D $\cN=2$ Weyl multiplet is described by a real vector superfield $H^a$.
In order to derive the equations of motion for the supergravity actions
\eqref{N=2_SG_model} and \eqref{N=2_SG_model-II}, we have to know 
the dependence of these actions on $H^a$. The necessary technical tools 
are given in this appendix. 

In $\cN=2$ conformal supergravity, the  gauge group consists of 
(i) the super-Weyl transformations \eqref{super-WeylN=2}; 
and (ii) the superspace general coordinate 
and local Lorentz and U$(1)_R$ transformations, which have the infinitesimal form 
\bea
 \d_\cK  \cD_A = \big[ \cK , \cD_A \big]~,
\qquad \cK = \x^B(z) \cD_B +\hf K^{b c}(z) M_{bc}
+\ri  \t (z) J  ~,
\label{C1}
\eea
where the gauge parameters $\x^B$, $K^{bc}$ and $\t$ obey natural reality conditions 
but are otherwise arbitrary.

Let ${\mathbb D}_A$ be some other set of covariant derivatives 
which differ from $\cD_A$ by finite deformations 
but satisfy the same (anti-)commutation relations \eqref{algebra-final} as 
the covariant derivatives $\cD_A$. 
By analogy with background-quantum splitting in 
4D $\cN=1$ supergravity \cite{GrisaruSiegel,Ideas}, the operators
${\mathbb D}_\a$ and $\bar {\mathbb D}_\a$ may be represented 
in the form 
\begin{subequations} \label{C.2}
\bea
{\mathbb D}_\a &=& \re^{\cW } \Big[ \cF \cD_\a -\hf \D \O_\a{}^{bc}M_{bc} 
- \ri \D \F_\a J \Big] \re^{-\cW}~, \\
\bar {\mathbb D}_\a &=& \re^{\bar \cW } \Big[ \bar \cF \cD_\a -\hf \D \bar \O_\a{}^{bc}M_{bc} 
- \ri \D \bar \F_\a J \Big] \re^{-\bar \cW}~, 
\eea
for some complex first-order operator $\cW$ of the form
\bea
\cW= \cW^B \cD_B  -\hf {\frak W}^{bc}M_{bc} 
- \ri {\frak W} J~.
\eea
\end{subequations}
The introduction of representation \eqref{C.2} is accompanied by the appearance 
of a new gauge invariance that acts on $\cW$ and $\bar \cW$ by 
\bea
\re^{\cW'} = \re^\cW \re^{- \bar \L}~, \qquad
\re^{\bar \cW'} = \re^{\bar \cW} \re^{- \L}~, \qquad 
\L= \L^B \cD_B  +\hf {\L}^{bc}M_{bc} + \ri {\L} J~.
\label{C-lambda}
\eea
This transformation should be accompanied by certain transformations of 
$\cF$, $\D \O_\a{}^{bc}$, $ \D \F_\a $ and their conjugates 
such that ${\mathbb D}_\a$ and $\bar {\mathbb D}_\a$ remain unchanged,
which leads to some restrictions on the superfield parameters in $\L$.
In such a setting, the supergravity gauge transformations 
\eqref{super-WeylN=2} and \eqref{C1}
can be realised in two different incarnations: 
(i) as ``background'' transformations; and (ii) as  ``quantum'' transformations.
The quantum gauge transformations and the  $\L$-transformations \eqref{C-lambda} 
may be used to choose a {\it quantum chiral representation} 
in which the operators  
 $\bar {\mathbb D}_\a$ and ${\mathbb D}_\a$  
 take the form:
\bsubeq \label{C.3} 
\bea
\bar {\mathbb D}_\a &=&  {\bar \cD}_\a + \cdots \ , \\
{\mathbb D}_\a &=& \re^{-2 \ri H} (\cN_\a{}^\b  \cD_\b + \cdots ) \re^{2 \ri H} \ , 
\qquad \det \cN = 1\ , \quad \cN \bar \cN = {\mathbbm 1}_2
\ ,
\eea
\esubeq
where we have introduced the $2\times 2$ matrix $\cN = (\cN_\a{}^\b) $, 
its complex conjugate $\bar \cN$, as well as  the differential operator 
\be H = H^{a} \cD_{a} ~, \qquad {\bar H}^a = H^a~.
\ee
The ellipses in \eqref{C.3} denote all terms with the Lorentz and U$(1)_R$ generators. 
The above steps are analogous to the background-quantum splitting 
in 4D $\cN=1$ supergravity \cite{GrisaruSiegel} described in detail in \cite{Ideas}.
The novel feature of the 3D $\cN=2$ case is the appearance of the matrix $\cN$, 
its origin is explained in  \cite{Kuzenko12}.

All the building blocks in \eqref{C.3}, as well as the torsion tensors for ${\mathbb D}_A$
can be expressed in terms of the gravitational superfield $H^a$ by requiring 
these covariant derivatives to obey the (anti-)commutation relations 
for the  conventional superspace. 
In this paper we are only interested in explicit expressions for these objects
at first-order in $H^a$. 
In this approximation the covariant derivatives \eqref{C.3} are 
\bsubeq
\bea {\mathbb D}_\a &=& \cD_\a - \ri \cD_\a H^{\g\d} \cD_{\g\d}
+ {N}_\a{}^\b \cD_\b + 2  \bar \cR H_\a{}^\g  {\bar \cD}_\g
- \hf {\bm \Omega}_\a{}^{\g\d} M_{\g\d} - \ri {\bm \Phi}_\a {J} \ , \\
{\bar {\mathbb D}}_\a
&=& {\bar \cD}_\a - \hf \tilde{\bm \Omega}_\a{}^{\g\d} M_{\g\d} - \ri \tilde{\bm \Phi}_\a {J} \ ,
\eea
\esubeq
where ${N}_\a{}^\b$ is traceless and is related to $\cN_\a{}^\b$ as follows:
\be 
 \cN_\a{}^\b - N_\a{}^\b = 
 \d_\a^\b 
+ H_\a{}^\g \cC^\b{}_\g
+ H^{\b\g} \cC_{\a\g}
+ 2 \ri H_\a{}^\b \cS \ .
\ee
All the superfields ${N}_\a{}^\b$,  
${\bm \Omega}_\a{}^{\g\d}$, 
$\tilde{\bm \Omega}_\a{}^{\g\d}$, ${\bm \Phi}_\a$ and $\tilde{\bm \Phi}_\a$ 
may be expressed in term of $H^a$ and its covariant derivatives. These 
are fixed by requiring the following  algebra to be satisfied:
\bsubeq
\bea
\{ {\mathbb D}_\a , {\mathbb D}_\b \} &=& - 4 \bar {\mathbb R} M_{\a\b} \ , \quad
\{ \bar{\mathbb D}_\a , \bar{\mathbb D}_\b \} = 4 {\mathbb R} M_{\a\b} \ , \\
\{ {\mathbb D}_\a , \bar{\mathbb D}_\b \} &=& - 2 \ri {\mathbb D}_{\ab}
- 2 {\mathbb C}_{\a\b} {J} - 4 \ri \eps_{\a\b} {\mathbb S} {J}
+ 4 \ri {\mathbb S} M_{\a\b} - 2 \eps_{\a\b} {\mathbb C}^{\g\d} M_{\g\d} \ .
\eea
\esubeq
Here we have defined the deformed vector derivative by
\be \mathbb D_a = - \frac{\ri}{4} (\g_a)^{\a\b} \{ \mathbb D_{(\a} , \bar {\mathbb D}_{\b)} \}
+ \ri \mathbb C_a {J} + 2 \mathbb S M_a~.
\ee
It should be noted that in the chiral representation 
the torsion superfields $\mathbb C_a$ and $\mathbb S$ are no longer real; 
instead they obey some modified reality conditions.
Similarly, 
$\bar{\mathbb R}$ is no longer conjugate to ${\mathbb R}$. 
Direct calculations lead to the following 
expressions: 
\bsubeq
\bea
{N}_\a{}^\b &=& - \hf \bar \cD_\g \cD^\g H_{\a}{}^{\b} \ , \\
{\bm \Omega}_{\a, \b\g} &=&  4 (\cD_\a H_{(\b}{}^\d) \cC_{\g)\d}
+  4 \ri \cS \cD_\a H_{\b\g} + 2 \cD_\a {N}_{\b\g} - 4 \ri \eps_{\a(\b} \Phi_{\g)} \ , \\
\tilde{\bm \Omega}_{\a , \b\g} &=&  4 \ri \eps_{\a (\b} \tilde{\Phi}_{\g)} \ , \\
{\bm \Phi}_\a &=&  \frac{\ri}{4} \cD_\a \cD^\b {\bar \cD}^\g H_{\b\g}
+ \ri (\cD_\a \cC_{\b\g}) H^{\b\g} + 4 (\cD^\b \cS) H_{\a\b} \ , \\
\tilde{\bm \Phi}_\a &=&  \frac{\ri}{4} \bar \cD_\a \bar \cD^\b \cD^\g H_{\b\g} \ .
\eea
\esubeq
The deformed torsion superfields are 
\bsubeq
\bea
\mathbb R &=& \cR + \frac{\ri}{2} \bar \cD^\a \tilde{\bm \Phi}_\a \ , \\
\bar {\mathbb R} &=& \bar \cR 
+ \frac{\ri}{2} \cC^{\a\b\g} \cD_\a H_{\b\g}
+ \frac{4 \ri}{3} (\cD_\b \cS) \cD_\a H^{\a\b}
- \frac{1}{6} ({\bar \cD}_\a \bar \cR) \cD_\b H^{\a\b}
\non\\
&&+ 2 \bar \cR \cC^{\a\b} H_{\a\b} + \frac{\ri}{2} \cD^\a {\bm \Phi}_\a
\ , \\
\mathbb S &=&
\cS 
- \frac{1}{8} \bar \cC^{\a\b\g} \cD_\a H_{\b\g}
- \frac{1}{24} (8 \bar\cD_\a \cS - \ri \cD_\a \cR) \cD_\b H^{\a\b}
- \frac{\ri}{4} {N}^{\a\b} \cC_{\a\b} \non\\
&&+ \frac{1}{8} \cD^\a \tilde{\bm \Phi}_\a
- \frac{1}{8} \bar \cD^\a {\bm \Phi}_\a
\ , \\
\mathbb C_{\a\b} &=& 
\cC_{\a\b}
- \frac{1}{4} \bar \cD^\g \cD_\g {N}_{\a\b}
- 2 \ri \cS {N}_{\a\b}
- 2 H_{\a\b} \cR \bar \cR
+ \frac{\ri}{2} \cD_{(\a} \tilde{\bm \Phi}_{\b)}
+ \frac{\ri}{2} \bar \cD_{(\a} {\bm \Phi}_{\b)}
- {N}_{(\a}{}^\d \cC_{\b) \d} \non\\
&&- \frac{\ri}{2} (\cD_\g H^{\g\d}) \bar \cC_{\a\b\d}
+ \frac{\ri}{2} (\cD^\g H^\d{}_{(\a}) \bar \cC_{\b)\g\d}
+ \frac{\ri}{12} (5 \ri \cD^\g \cR - 4 \bar\cD^\g \cS) \cD_{(\a} H_{\b) \g} \non\\
&&+ \frac{\ri}{12} (\ri \cD_{(\a} \cR + 4 \bar\cD_{(\a} \cS) \cD^\g H_{\b)\g}
- \frac{\ri}{12} (\ri \cD^\g \cR + 4 \bar\cD^\g \cS) \cD_\g H_{\a\b}
\ .
\eea
\esubeq

In conclusion, a few comments are in order regarding the chiral representation
used above. In order to switch from the original real to the chiral representation,
every scalar superfield $T$ has to be transformed to
\be \mathbb T = \re^{- \ri H^a \cD_a} T \ .
\ee
This tells us that $\d T = - \ri H^a \cD_a T =  \mathbb T - T  + \cO(H^2) $
is the complete variation in the case that $T$ is unconstrained. 
For instance, the prepotential $\cV$ for the $\cN = 2$ linear multiplet 
varies  as
\be 
\d \cV =  - \ri H^a \cD_a \cV ~.
\ee
However constrained superfields transform in a more complicated fashion since their constraints 
must be preserved under shifts in the prepotential. For instance, for a covariantly 
chiral superfield $\Psi$  of U$(1)_R$ charge $-\hf $ and its conjugate
$\bar \Psi$ we obtain
\bsubeq
\bea
\d \Psi &=& - \frac{1}{16} \J ([\cD^\a , \bar \cD^\b] H_{\a\b} - 4 \ri \cD^a H_a )  \ , \\
\d \bar \Psi &=& - \frac{1}{16} \bar \J ([\cD^\a , \bar \cD^\b] H_{\a\b} + 4 \ri \cD^a H_a 
)  -2\ri H^a \cD_a \bar \Psi \ .
\eea
\esubeq
Similarly, it can be shown that the $\cN = 2$ linear superfield varies as
\bea
\d G &=& 
\frac{\ri}{4} (\bar \cD^\a \cD_\a H^{\g\d}) [\cD_\g , \bar\cD_\d] \cV
- \hf H^{\g\d} \bar\cD^\a \cD_\a \cD_{\g\d} \cV
- \hf (\bar \cD^\a H^{\g\d}) \cD_\a \cD_{\g\d} \cV
\non\\
&&
+ \hf (\cD^\a H^{\g\d}) \bar \cD_\a \cD_{\g\d} \cV 
- 2 \ri (\bar \cD_\a H^{\a\g}) \bar \cR \bar\cD_\g \cV
- 2 \ri (\cD^\a H_{\a\g}) \cR \cD^\g \cV
\non\\
&&
- 2 \ri H^{\a\b} (\bar\cD_\a \bar \cR) \bar\cD_\b \cV \ ,
\eea
which can be rewritten in the following form:
\bea
\d G &=&  - \ri H^{a}\cD_{a} G
+\frac{\ri}{8} ({[}\cD^\a, \cDB_\a{]} H^{\g\d}) [\cD_\g , \bar\cD_\d] \cV
\non\\
&&
+ H^{\a\b}\Big(
2\cC_{(\a}{}^\g\cD_{\b)\g} 
-\ri(\cD_{(\a}\cR)\cD_{\b)}
-\ri(\cDB_{(\a}\bar{\cR})\cDB_{\b)}  
\Big) \cV 
\non\\
&&
+ \hf (\cD^\g H^{\a\b}) \Big( \cD_{\a\b} \cDB_\g   
+\ri \cC_{\g\a}\cDB_{\b} \Big) \cV 
- \hf (\bar \cD^\g H^{\a\b}) \Big( \cD_{\a\b} \cD_\g  
-\ri \cC_{\g\a}\cD_{\b} \Big)\cV
\non\\
&&
+(\bar \cD_\a H^{\a\b})\Big(
 \frac{\ri}{2}  \cC_{\b\d}\cD^{\d}
+\cS\cD_{\b}
-\ri\bar{R}\cDB_{\b}
\Big)\cV \non \\
&&+(\cD_\a H^{\a\b})\Big(
\frac{\ri}{2} \cC_{\b\d}\cDB^{\d}
-\cS\cDB_{\b}
-\ri R\cD_{\b}
\Big) 
\cV ~.
\eea


\begin{footnotesize}

\end{footnotesize}


\begin{thebibliography}{66}

\bibitem{UDW} 
  R.~Utiyama and B.~S.~DeWitt,
  ``Renormalization of a classical gravitational field interacting with quantized matter fields,''
  J.\ Math.\ Phys.\  {\bf 3}, 608 (1962).
  
\bibitem{DeWitt} 
  B.~S.~DeWitt,
 {\it Dynamical Theory of Groups and Fields}, Gordon and Breach, New York, 1965. 
 
\bibitem{Stelle} 
K.~S.~Stelle,
``Renormalization of higher derivative quantum gravity,''
  Phys.\ Rev.\ D {\bf 16}, 953 (1977).
  
\bibitem{Starobinsky} 
  A.~A.~Starobinsky,
  ``A new type of isotropic cosmological models without singularity,''
  Phys.\ Lett.\ B {\bf 91}, 99 (1980).
  

\bibitem{DJT} 
  S.~Deser, R.~Jackiw and S.~Templeton,
   ``Topologically massive gauge theories,''
  Annals Phys.\  {\bf 140}, 372 (1982)
  [Erratum-ibid.\  {\bf 185}, 406 (1988)].
  
\bibitem{BHT1} 
  E.~A.~Bergshoeff, O.~Hohm and P.~K.~Townsend,
  ``Massive gravity in three dimensions,''
  Phys.\ Rev.\ Lett.\  {\bf 102}, 201301 (2009)
  [arXiv:0901.1766 [hep-th]];
  
\bibitem{BHT2} 
  E.~A.~Bergshoeff, O.~Hohm and P.~K.~Townsend,
  ``More on massive 3D gravity,''
  Phys.\ Rev.\ D {\bf 79}, 124042 (2009)
  [arXiv:0905.1259 [hep-th]].
  
\bibitem{Nakasone:2009bn} 
  M.~Nakasone and I.~Oda,
  ``On unitarity of massive gravity in three dimensions,''
  Prog.\ Theor.\ Phys.\  {\bf 121}, 1389 (2009)
  [arXiv:0902.3531 [hep-th]].


\bibitem{Deser:2009hb} 
  S.~Deser,
  ``Ghost-free, finite, fourth order D=3 (alas) gravity,''
  Phys.\ Rev.\ Lett.\  {\bf 103}, 101302 (2009)
  [arXiv:0904.4473 [hep-th]].
  
\bibitem{Ohta12} 
  N.~Ohta,
  ``A Complete Classification of Higher Derivative Gravity in 3D and Criticality in 4D,''
  Class.\ Quant.\ Grav.\  {\bf 29}, 015002 (2012)
  [arXiv:1109.4458 [hep-th]].

\bibitem{Oda:2009ys} 
  I.~Oda,
  ``Renormalizability of massive gravity in three dimensions,''
  JHEP {\bf 0905}, 064 (2009)
  [arXiv:0904.2833 [hep-th]].
  
\bibitem{MuneyukiOhta12}
  K.~Muneyuki and N.~Ohta,
  ``Unitarity versus Renormalizability of Higher Derivative Gravity in 3D,''
  Phys.\ Rev.\ D {\bf 85}, 101501 (2012)
  [arXiv:1201.2058 [hep-th]].
 
\bibitem{Deser}
  S.~Deser,
  ``Cosmological topological supergravity,''
 in {\it Quantum Theory of Gravity}, S. M. Christensen (Ed.), 
 Adam Hilger, Bristol, 1984, pp. 374-381. 


\bibitem{LSS} 
  W.~Li, W.~Song and A.~Strominger,
 ``Chiral gravity in three dimensions,''
  JHEP {\bf 0804}, 082 (2008)
  [arXiv:0801.4566 [hep-th]].
  
\bibitem{CDWW} 
  S.~Carlip, S.~Deser, A.~Waldron and D.~K.~Wise,
  ``Topologically massive AdS gravity,''
  Phys.\ Lett.\ B {\bf 666}, 272 (2008)
  [arXiv:0807.0486 [hep-th]];
 ``Cosmological topologically massive gravitons and photons,''
  Class.\ Quant.\ Grav.\  {\bf 26}, 075008 (2009)
  [arXiv:0803.3998 [hep-th]].
 
\bibitem{DK} 
S.~Deser and J.~H.~Kay,
``Topologically massive supergravity,''
Phys.\ Lett.\ B {\bf 120}, 97 (1983).
 
  \bibitem{KLRST-M} 
  S.~M.~Kuzenko, U.~Lindstr\"om, M.~Ro\v{c}ek, I.~Sachs 
  and G.~Tartaglino-Mazzucchelli,
  ``Three-dimensional N=2 supergravity theories: From superspace to components,''
  Phys.\ Rev.\ D {\bf 89}, 085028 (2014)
  [arXiv:1312.4267 [hep-th]].
  
\bibitem{KN14} 
  S.~M.~Kuzenko and J.~Novak,
  ``Supergravity-matter actions in three dimensions and Chern-Simons terms,''
  JHEP {\bf 1405}, 093 (2014)
  [arXiv:1401.2307 [hep-th]].
  

\bibitem{Andringa:2009yc} 
  R.~Andringa, E.~A.~Bergshoeff, M.~de Roo, O.~Hohm, E.~Sezgin and P.~K.~Townsend,
  ``Massive 3D supergravity,''
  Class.\ Quant.\ Grav.\  {\bf 27}, 025010 (2010)
  [arXiv:0907.4658 [hep-th]].

\bibitem{BHRST10} 
  E.~A.~Bergshoeff, O.~Hohm, J.~Rosseel, E.~Sezgin and P.~K.~Townsend,
  ``More on massive 3D supergravity,''
  Class.\ Quant.\ Grav.\  {\bf 28}, 015002 (2011)
  [arXiv:1005.3952 [hep-th]].
  
\bibitem{Bergshoeff:2014ida} 
  E.~Bergshoeff and M.~Ozkan,
 ``3D Born-Infeld gravity and supersymmetry,''
  JHEP {\bf 1408}, 149 (2014)
  [arXiv:1405.6212 [hep-th]].
  
  
  \bibitem{BOS14} 
  G.~Alkac, L.~Basanisi, E.~A.~Bergshoeff, M.~Ozkan and E.~Sezgin,
 ``Massive $ \mathcal{N} $ = 2 supergravity in three dimensions,''
  JHEP {\bf 1502}, 125 (2015)
  [arXiv:1412.3118 [hep-th]].
  
\bibitem{BG}
M.~Brown and S.~J.~Gates Jr.,
``Superspace Bianchi identities and the supercovariant derivative,''
 Annals Phys.\  {\bf 122}, 443 (1979).


\bibitem{Siegel}
  W.~Siegel,  ``Unextended superfields in extended supersymmetry,''
  Nucl.\ Phys.\  B {\bf 156}, 135 (1979).
  


\bibitem{GGRS}
S. J.~Gates Jr., M.~T.~Grisaru, M.~Ro\v{c}ek and W.~Siegel,
{\it Superspace, or One Thousand and One Lessons in Supersymmetry},
Benjamin/Cummings (Reading, MA),  1983, hep-th/0108200.  

\bibitem{HT}
  P.~S.~Howe and R.~W.~Tucker,
  ``Local supersymmetry in (2+1) dimensions. 1. Supergravity and differential forms,''
  J.\ Math.\ Phys.\  {\bf 19}, 869 (1978);
  ``Local supersymmetry in (2+1) dimensions. 2. An action for a spinning membrane,''
  J.\ Math.\ Phys.\  {\bf 19}, 981 (1978).
  
\bibitem{KLT-M11} 
 S.~M.~Kuzenko, U.~Lindstr\"om and G.~Tartaglino-Mazzucchelli,
  ``Off-shell supergravity-matter couplings in three dimensions,''
 JHEP {\bf 1103}, 120 (2011)
  [arXiv:1101.4013 [hep-th]].
  
\bibitem{BKNT-M1} 
  D.~Butter, S.~M.~Kuzenko, J.~Novak and G.~Tartaglino-Mazzucchelli,
  ``Conformal supergravity in three dimensions: New off-shell formulation,''
  JHEP {\bf 1309}, 072 (2013)
  [arXiv:1305.3132 [hep-th]].
 


\bibitem{BKNT-M2}
D.~Butter, S.~M.~Kuzenko, J.~Novak and G.~Tartaglino-Mazzucchelli,
 ``Conformal supergravity in three dimensions: Off-shell actions,''
  JHEP {\bf 1310}, 073 (2013)
  [arXiv:1306.1205 [hep-th]].


\bibitem{KNT-M} 
  S.~M.~Kuzenko, J.~Novak and G.~Tartaglino-Mazzucchelli,
  ``N=6 superconformal gravity in three dimensions from superspace,''
  JHEP {\bf 1401}, 121 (2014)
  [arXiv:1308.5552 [hep-th]].
  

\bibitem{KT-M11}
S.~M.~Kuzenko and G.~Tartaglino-Mazzucchelli,
 ``Three-dimensional N=2 (AdS) supergravity and associated supercurrents,''
JHEP {\bf 1112}, 052 (2011)
[arXiv:1109.0496 [hep-th]].

\bibitem{Theisen} 
  S.~Theisen,
  ``Fourth-order supergravity,''
  Nucl.\ Phys.\ B {\bf 263}, 687 (1986)
  [Nucl.\ Phys.\ B {\bf 269}, 744 (1986)].


\bibitem{RvN}
M.~Ro\v{c}ek and P.~van Nieuwenhuizen,
``${\rm N}   \geq 2$ supersymmetric Chern-Simons terms as D = 3 extended conformal
supergravity,''
Class.\ Quant.\ Grav.\  {\bf 3}, 43 (1986).
       

\bibitem{ZP88}
  B.~M.~Zupnik and D.~G.~Pak,
   ``Superfield formulation of the simplest three-dimensional gauge theories and
  conformal supergravities,''
  Theor.\ Math.\ Phys.\  {\bf 77} (1988) 1070
  [Teor.\ Mat.\ Fiz.\  {\bf 77} (1988) 97]. 


\bibitem{NG93} 
  H.~Nishino and S.~J.~Gates Jr.,
  ``Chern-Simons theories with supersymmetries in three-dimensions,''
  Int.\ J.\ Mod.\ Phys.\ A {\bf 8}, 3371 (1993).
  


\bibitem{HIPT}
  P.~S.~Howe, J.~M.~Izquierdo, G.~Papadopoulos and P.~K.~Townsend,
``New supergravities with central charges and Killing spinors in 2+1 dimensions,''
  Nucl.\ Phys.\  B {\bf 467}, 183 (1996).
  [arXiv:hep-th/9505032].
  
\bibitem{dWNT}
 B.~de Wit, H.~Nicolai and A.~K.~Tollsten,
 ``Locally supersymmetric D = 3 nonlinear sigma models,''
 Nucl.\ Phys.\  B {\bf 392}, 3 (1993)
 [arXiv:hep-th/9208074].


\bibitem{IT}
 J.~M.~Izquierdo and P.~K.~Townsend,
 ``Supersymmetric space-times in (2+1) adS supergravity models,''
 Class.\ Quant.\ Grav.\  {\bf 12}, 895 (1995)
 [arXiv:gr-qc/9501018].

\bibitem{DKSS} 
 N.~S.~Deger, A.~Kaya, E.~Sezgin and P.~Sundell,
 ``Matter coupled AdS(3) supergravities and their black strings,''
 Nucl.\ Phys.\ B {\bf 573}, 275 (2000)
 [hep-th/9908089].

\bibitem{old}
J.~Wess and B.~Zumino,
``Superfield Lagrangian for supergravity,''
Phys.\ Lett.\  B {\bf 74}, 51 (1978);
K.~S.~Stelle and P.~C.~West,
``Minimal auxiliary fields for supergravity,''
Phys.\ Lett.\  B {\bf 74},  330 (1978);
S.~Ferrara and P.~van Nieuwenhuizen,
``The auxiliary fields of supergravity,''
Phys.\ Lett.\  B {\bf 74}, 333 (1978).

\bibitem{Ideas} 
  I.~L.~Buchbinder and S.~M.~Kuzenko,
  ``Ideas and methods of supersymmetry and supergravity: 
  Or a walk through superspace,''
  IOP, Bristol, 1995 (Revised Edition: 1998),   656 p.
  
\bibitem{WB}
J.~Wess and J.~Bagger,
{\it Supersymmetry and Supergravity},
Princeton University Press, Princeton, 1992.


\bibitem{new}
M.~F.~Sohnius and P.~C.~West,
``An alternative minimal off-shell version of N=1 supergravity,''
Phys.\ Lett.\  B {\bf 105}, 353 (1981).

\bibitem{non-min}
P.~Breitenlohner,
 ``A geometric interpretation of local supersymmetry,''
  Phys.\ Lett.\  B {\bf 67}, 49 (1977);
``Some invariant Lagrangians for local supersymmetry,''
Nucl.\ Phys.\ {\bf B124}, 500 (1977).


\bibitem{SG}
W.~Siegel and S.~J.~Gates Jr.
 ``Superfield supergravity,''  Nucl.\ Phys.\  B {\bf 147}, 77 (1979).
 
\bibitem{BK11dual} 
  D.~Butter and S.~M.~Kuzenko,
  ``A dual formulation of supergravity-matter theories,''
  Nucl.\ Phys.\ B {\bf 854}, 1 (2012)
  [arXiv:1106.3038 [hep-th]].
\bibitem{AT}
  A.~Ach\'ucarro and P.~K.~Townsend,
  ``A Chern-Simons action for three-dimensional anti-de Sitter supergravity
 theories,''
  Phys.\ Lett.\  B {\bf 180}, 89 (1986).
  

\bibitem{Deser70} 
S.~Deser, 
``Scale invariance and gravitational coupling,''
Annals Phys.\  {\bf 59}, 248 (1970).

\bibitem{Zumino} B. Zumino, 
``Effective Lagrangians and broken symmetries," 
in {\it Lectures on Elementary Particles and Quantum Field Theory,
Vol. 2}, S. Deser, M. Grisaru and H. Pendleton (Eds.),
Cambridge, Mass. 1970, pp. 437-500.
  
\bibitem{ZP89} 
B.~M.~Zupnik and D.~G.~Pak,
``Differential and integral forms in supergauge theories and supergravity,''
Class.\ Quant.\ Grav.\  {\bf 6}, 723 (1989).
  

\bibitem{LR-brane}
  U.~Lindstr\"om and M.~Ro\v{c}ek,
  ``A super-Weyl-invariant spinning membrane,''
  Phys.\ Lett.\  B {\bf 218}, 207 (1989).


 
\bibitem{vanN85}
P.~van Nieuwenhuizen,
``D = 3 conformal supergravity and Chern-Simons terms,''
Phys.\ Rev.\  D {\bf 32}, 872 (1985).




\bibitem{KT-M12} 
  S.~M.~Kuzenko and G.~Tartaglino-Mazzucchelli,
  ``Conformal supergravities as Chern-Simons theories revisited,''
  JHEP {\bf 1303}, 113 (2013)
  [arXiv:1212.6852 [hep-th]].
   
  
\bibitem{WZ2}
J.~Wess and B.~Zumino,
``The component formalism follows from 
the superspace formulation of supergravity,''
Phys.\ Lett.\ B {\bf 79},  394 (1978).

 
\bibitem{KugoU}
  T.~Kugo and S.~Uehara,
 ``N=1 superconformal tensor calculus: Multiplets with external indices and derivative operators,''
  Prog.\ Theor.\ Phys.\  {\bf 73}, 235 (1985).
   
\bibitem{BKO89} 
 I.~L.~Buchbinder, S.~M.~Kuzenko and O.~A.~Soloviev,
``One-loop counterterms of Wess-Zumino model in the N=1 non-minimal 
supergravity background,'' Nucl.\ Phys.\ B {\bf 322}, 277 (1989).
  
\bibitem{Butter09} 
  D.~Butter,
  ``Background field formalism for chiral matter and gauge fields conformally coupled to supergravity,''
  Nucl.\ Phys.\ B {\bf 828}, 233 (2010)
  [arXiv:0909.4901 [hep-th]].

\bibitem{BN12} 
  D.~Butter and J.~Novak,
``Component reduction in N=2 supergravity: the vector, tensor, 
and vector-tensor multiplets,''  JHEP {\bf 1205}, 115 (2012) [arXiv:1201.5431 [hep-th]].
  
\bibitem{BCGLMP04} 
  M.~Becker, D.~Constantin, S.~J.~Gates Jr., W.~D.~Linch, 
  W.~Merrell and J.~Phillips,
  ``M-theory on Spin(7) manifolds, fluxes and 3D, N=1 supergravity,''
  Nucl.\ Phys.\ B {\bf 683}, 67 (2004)
  [hep-th/0312040].


\bibitem{Kuzenko15} 
  S.~M.~Kuzenko,
  ``Supersymmetric spacetimes from curved superspace,''
  PoS CORFU {\bf 2014} (2014)
  [arXiv:1504.08114 [hep-th]].
       
  \bibitem{BK11} 
  D.~Butter and S.~M.~Kuzenko,
  ``New higher-derivative couplings in 4D N = 2 supergravity,''
  JHEP {\bf 1103}, 047 (2011)
  [arXiv:1012.5153 [hep-th]].
  
\bibitem{BKT} 
  I.~L.~Buchbinder, S.~M.~Kuzenko and A.~A.~Tseytlin,
  ``On low-energy effective actions in N=2, N=4 superconformal theories 
  in four-dimensions,''
  Phys.\ Rev.\ D {\bf 62}, 045001 (2000)
  [hep-th/9911221].
      
\bibitem{Kuzenko12} 
S.~M.~Kuzenko,
``Prepotentials for N=2 conformal supergravity in three dimensions,''
JHEP {\bf 1212}, 021 (2012)  [arXiv:1209.3894 [hep-th]].
  
\bibitem{1205.4622} 
  S.~M.~Kuzenko, U.~Lindstr\"om and G.~Tartaglino-Mazzucchelli,
  ``Three-dimensional (p,q) AdS superspaces and matter couplings,''
  JHEP {\bf 1208}, 024 (2012)
  [arXiv:1205.4622 [hep-th]].
  
\bibitem{Kuzenko:2008ry} 
S.~M.~Kuzenko and G.~Tartaglino-Mazzucchelli, 
``Different representations for the action principle in 4D N = 2 supergravity,''
JHEP {\bf 0904}, 007 (2009) [arXiv:0812.3464 [hep-th]].  
  
\bibitem{Grisaru:1995dr} 
  M.~T.~Grisaru and M.~E.~Wehlau,
  ``Superspace measures, invariant actions, and component projection formulae for (2,2) supergravity,''
  Nucl.\ Phys.\ B {\bf 457}, 219 (1995)
  [hep-th/9508139].
  
\bibitem{Howe:1981et} 
  P.~S.~Howe, K.~S.~Stelle and P.~K.~Townsend,
  ``The vanishing volume of $N=1$ superspace,''
  Phys.\ Lett.\ B {\bf 107}, 420 (1981).
  
\bibitem{BKS}
E.~I.~Buchbinder, S.~M.~Kuzenko, I.~B.~Samsonov,
 ``Superconformal field theory in three dimensions:
 Correlation functions of conserved currents,'' 
 JHEP {\bf 1506}, 138 (2015)
 [arXiv: 1503.04961 [hep-th]].

\bibitem{GrisaruSiegel} 
  M.~T.~Grisaru and W.~Siegel,
 ``Supergraphity (I). Background field formalism,''
  Nucl.\ Phys.\ B {\bf 187}, 149 (1981);
 ``Supergraphity (II). Manifestly covariant rules and higher loop finiteness,''
  Nucl.\ Phys.\ B {\bf 201}, 292 (1982)

\bibitem{Siegel78} 
 W.~Siegel,
 ``Solution to constraints in {Wess-Zumino} supergravity formalism,''
 Nucl.\ Phys.\ B {\bf 142}, 301 (1978).




\end{thebibliography}
\end{document}